\documentclass[11pt,a4paper]{article}
\usepackage[utf8x]{inputenc}
\usepackage[T1]{fontenc}

\usepackage[margin=25truemm]{geometry}
\usepackage{graphicx} 
\usepackage{calc} 
\usepackage{authblk}
\usepackage{bm}
\usepackage{enumitem} 
\usepackage{braket,mathtools,amsmath,amssymb,subcaption}
\usepackage{authblk}
\usepackage[all]{nowidow} 
\usepackage{url}

\usepackage{lipsum} 

\usepackage{framed} 
\usepackage[framed]{ntheorem}
\newframedtheorem{frm-thm}{Theorem}
\usepackage{tikz}
\usetikzlibrary{backgrounds}
\usepackage{subcaption}
\def\tht{'t Hooft}

\def\p{\partial}

\def\hx{\hat{x}}
\def\hy{\hat{y}}

\def\mb{\mathbb{Z}_2}

\usepackage{cite}
\usepackage{hyperref}
\usepackage{tikz}  
\usetikzlibrary{arrows.meta,positioning,calc}
\hypersetup{colorlinks,linkcolor=blue,citecolor=cyan}
\begin{document} 
\title{Type-IV ’t Hooft Anomalies on the Lattice: Emergent Higher-Categorical Symmetries and Applications to LSM Systems}
\author{Tsubasa Oishi$^1$, Hiromi Ebisu$^{2}$~\footnote{hiromi.ebisu@riken.jp}}
\affil{$^1$Yukawa Institute for Theoretical Physics, Kyoto University, Kyoto 606-8502, Japan}
\affil{$^2$Interdisciplinary Theoretical and Mathematical Sciences Program (iTHEMS)~RIKEN, Wako 351-0198, Japan}

\maketitle
\thispagestyle{empty}

\begin{abstract}
’t Hooft anomalies impose fundamental constraints on quantum matter and often lead to emergent symmetry structures upon gauging.
We analyze a lattice model with four global symmetries realizing a mixed anomaly described by $\sim a_1\wedge a_2\wedge a_3\wedge a_4$, where the $a_i$ denote background gauge fields for the global symmetries. 
Through explicit lattice gauging, we demonstrate the emergence of higher symmetry structures, including 2-group, non-invertible, and higher fusion categorical symmetries.  We also provide a field theoretical understanding of these results. 
Applying this framework to systems with Lieb–Schultz–Mattis anomalies, obtained by promoting part of the internal symmetries to translational symmetries, we demonstrate that modulated (dipole) symmetries and non-invertible translation symmetry arise as direct counterparts of those in systems with purely internal type-IV anomalies. 
Importantly, we uncover a qualitatively new feature absent in previously studied modulated symmetries and non-invertible translation: their realization can become intrinsically defect-dependent. In particular, the emergent symmetry structure changes depending on whether symmetry defects are present. 
This work establishes a concrete lattice realization of mixed anomalies and reveals a rich structure of emergent symmetries, thereby clarifying their role in constraining quantum phases of matter.

\end{abstract}

\newpage
\pagenumbering{arabic}

\tableofcontents
\section{Introduction}
Quantum anomalies constitute one of the most fundamental principles in modern physics, imposing stringent and universal constraints on quantum systems. In particular, ’t Hooft anomalies characterize the obstruction to gauging global symmetries and provide a powerful organizing principle for quantum phases of matter. Through the anomaly inflow mechanism, they play a central role in the classification of symmetry-protected topological (SPT) phases and constrain possible low-energy dynamics~\cite{Chen:2011pg,Kapustin:2014lwa,Turner:2011vfe,chen2014symmetry}.

Understanding how such anomalous symmetries are realized in lattice systems is therefore a fundamental problem. In recent years, significant progress has been made in the microscopic realization and diagnosis of
’t Hooft anomalies, especially for anomalies involving discrete symmetries. Among them, the ``type~III anomaly'',~\cite{propitius1995topological,PhysRevLett.114.031601} characterized by the topological action
\begin{equation}
S_{III}= \int_{M_3} A \wedge B \wedge C,\label{iii}
\end{equation}
where $A,B,C$ denote $1$-form background gauge fields corresponding to the global symmetries and $M_d$ is
a $d$-dimensional manifold, 
has been extensively studied. Systems with this anomaly exhibit rich structures. 
For instance, partial gauging leads to the emergence of 
exotic symmetries, such as 
non-invertible symmetries. 


Despite these developments, the systematic structure associated with higher-order anomalies, in particular the type-IV anomaly described by
\begin{equation}
S_{IV}= \int_{M_4} A \wedge B \wedge C \wedge D,
\label{abcd}
\end{equation}
has remained much less explored, especially from the perspective of explicit lattice gauging constructions and Lieb–Schultz–Mattis (LSM) realizations. Here, $A,B,C,D$ denote $1$-form background gauge fields for global symmetries. While lattice realizations of type IV group cohomology anomalies themselves have been discussed previously, such as in~\cite{Yoshida:2015cia,Tantivasadakarn:2021wdv,tantivasadakarn2023building,Kawagoe:2025ldx,Kobayashi:2024dqj,Kapustin:2025nju,4zfz-x9xh}, it has not been systematically clarified how gauging different subgroups generates a hierarchy of emergent symmetry structures — including 2-group, non-invertible, and higher fusion-categorical symmetries — at the microscopic lattice level. Furthermore, as we explain below, the corresponding crystalline/LSM realization, obtained by replacing part of the internal symmetry with lattice translation symmetry, and the resulting defect-dependent modulated symmetry structure, have remained largely unexplored.

\begin{table}[t]
\centering
\small
\renewcommand{\arraystretch}{1.2}

\begin{tabular}{
p{3.2cm}
p{1.8cm}
p{3.2cm}
p{6.0cm}
}
\hline\hline

Anomaly
& Gauged symmetry
& Defect / background
& Emergent symmetry 
\\
\hline

$A\wedge B\wedge C\wedge D$
& $A$
& $B$-symmetry defect
&  2-group symmetry
\\

$A\wedge B\wedge C\wedge D$
& $A,B$
& $C$-symmetry defect
& Defect-induced non-invertible symmetry
\\

$A\wedge B\wedge C\wedge D$
& $A,B,C$
& ---
& $2\text{-}\mathrm{Rep}(\Gamma)$ symmetry
\\

\hline

$A\wedge B\wedge e^x\wedge e^y$
& $A$
& Translation defect
& Modulated 2-group / dipole symmetry
\\

$A\wedge B\wedge e^x\wedge e^y$
& $A,B$
& Translation defect
& Non-invertible translation
\\

\hline

$A\wedge B\wedge C\wedge e^x$
& $A$
& Translation defect
& Crystalline 2-group symmetry
\\

$A\wedge B\wedge C\wedge e^x$
& $A$
& $B$-symmetry defect
& Modulated 1-form / dipole symmetry
\\

$A\wedge B\wedge C\wedge e^x$
& $A,B$
& Translation defect
& Defect-induced non-invertible symmetry
\\

$A\wedge B\wedge C\wedge e^x$
& $A,B,C$
& ---
& Non-invertible translation with higher-categorical fusion rule
\\

\hline\hline
\end{tabular}

\caption{
Emergent symmetries obtained by partially
gauging the type-IV anomaly and its crystalline LSM counterparts explored in this work.
Internal symmetry and translation defects can qualitatively modify
the resulting emergent symmetry structure.
}
\label{defect_hierarchy}
\end{table}

In this work, we systematically analyze emergent
symmetry structures that arise upon gauging its subgroups in a concrete lattice model with type IV anomaly.
We show that gauging different choices of subgroups generates a rich hierarchy of emergent symmetries, including 2-group symmetries~\cite{Cordova:2018cvg, Benini:2018reh, kapustin2017higher, Delcamp:2018wlb}, non-invertible symmetries, and higher fusion categorical symmetries~\cite{douglas2018fusion,Bartsch:2022mpm,Bartsch:2022ytj,Decoppet:2024htz, Bhardwaj:2022yxj,Bhardwaj:2022lsg, Delcamp:2023kew}.  

We further support this result through field-theoretical analysis, demonstrating that these structures are universal and independent of microscopic details. 
The microscopic perspective is important because many recent developments in higher groups, non-invertible symmetries, and higher fusion-categorical symmetries have been formulated primarily in field-theoretic or categorical language. 

We further investigate crystalline realizations of the type-IV anomaly in systems with LSM anomalies~\cite{Lieb1961,PhysRevLett.84.1535,PhysRevB.69.104431,Cheng:2015kce,metlitski2018intrinsic}. Replacing one or two internal symmetries with lattice translations yields two classes of mixed internal–translation anomalies, for which we construct explicit lattice models and gauge an internal $\mathbb{Z}_2$ symmetry. 
Microscopic realizations of group-cohomology anomalies, general lattice constructions of higher categorical symmetries~\cite{Choi:2024rjm,Inamura:2025cum,SciPostPhys.19.2.054}, and connections between LSM anomalies and modulated symmetries~\cite{Aksoy:2023hve,Seifnashri:2023dpa, Pace:2024acq,Pace:2025hpb,SciPostPhys.20.4.117} have been developed previously. 
What has remained less clear is how these structures are tied together microscopically by a common higher anomaly, and in particular how background symmetry defects can qualitatively modify the symmetry structure generated by partial gauging. In this work, we provide such a bridge starting from a single lattice realization of a type-IV anomaly. Gauging one, two, or three $\mathbb Z_2$ factors produces, respectively, a 2-group symmetry, a non-invertible symmetry structure, and a symmetry described by a 2-representation category, in agreement with a complementary field-theoretical analysis. Crucially, in the case of partial gauging by two symmetry factors, the residual symmetry is not intrinsically non-invertible in the defect-free theory. Rather, it is invertible in the absence of a background defect and becomes non-invertible only after an appropriate symmetry defect is inserted. Its fusion then closes onto condensation operators associated with the emergent 1-form symmetries. This defect-induced change of invertibility is a key distinction from constructions in which non-invertibility is an intrinsic symmetry property of the gauged theory.

We further show that the same mechanism has crystalline descendants. By replacing one or two internal symmetries with lattice translations, we identify the type-IV cocycle as the internal parent of a class of crystalline LSM anomalies. Partial gauging then produces modulated symmetries and non-invertible translation whose realization and algebra depend on translation-twisted backgrounds, system-size parity, and internal symmetry defects. 
%
%
 %
To the best of our knowledge, this 
constitutes the first explicit lattice realization in which a residual symmetry changes from 
invertible to non-invertible upon insertion of a background symmetry defect in a purely 
type-IV anomalous system, as well as a defect-dependent partial-gauging response in an LSM 
system descending from the crystalline counterpart of the same anomaly. Thus, the central novelty of our construction lies not in the occurrence of a $2\mathrm{Rep}$ symmetry alone, but in unifying the type-IV gauging hierarchy, and
with its crystalline/LSM descendants and uncovering the defect-dependent origin of the resulting modulated symmetries and non-invertible translation. 
%
%
%
%
The results of this paper are summarized in Table~\ref{defect_hierarchy}.
\par
The rest of this work is organized as follows. In Sec.~\ref{sec2}, we explore emergent symmetries obtained by gauging subgroups of four internal symmetries exhibiting the type IV anomaly.  Sec.~\ref{sec4} is devoted to corroborating our findings by employing field-theoretical analysis. 
In Sec.~\ref{sec3}, we introduce a few lattice models with the LSM anomaly in the form of type IV~\eqref{abcd} where some of the internal symmetries are replaced with the translational ones and reveal rich symmetry structures, in particular, modulated symmetries.
Finally, in Sec.~\ref{sec5}, we conclude and discuss several future directions. Technical issues are provided in appendices.   

\section{Emergent symmetries via gauging subgroups} \label{sec2}
In this section,
we systematically construct the gauging structure associated with the type~IV anomaly in a lattice model and analyze the resulting emergent symmetries. 
To this end, we study a $\mb^{A} \times \mb^{B} \times \mb^{C} \times \mb^{D} $ $0$-form symmetry that exhibits the type~IV anomaly, whose anomaly inflow action is described by
\begin{equation}
    (-1)^{\int_{M_4}A\cup B\cup C\cup D}.
\end{equation}
Here $A, B, C,D$ represent background gauge fields for $\mb^{A,B,C,D}$ symmetries, respectively. 
We analyze a lattice system with this anomalous symmetry by introducing
a triangular lattice, which is partitioned into three sublattices labeled by $\Lambda_a, \Lambda_b,\Lambda_c$, such that each triangle contains one site from each sublattice. This construction provides a minimal lattice realization of the type IV anomaly while allowing for an explicit implementation of gauging.
We assign Pauli operators~$X_i$ and~$ Z_i$ on each site. In this setting, $\mb^{A} \times \mb^{B} \times \mb^{C} \times \mb^{D} $ symmetry operators that exhibit the type~IV anomaly
are described by
\begin{equation}
    U_A=\prod_{a\in\Lambda_a}X_a, \quad U_B=\prod_{b\in\Lambda_b}X_b, \quad U_C=\prod_{c\in\Lambda_c}X_c, \quad U_D=\prod_{\langle abc\rangle}CCZ_{abc},\label{symmetries}
\end{equation}
where $\langle abc\rangle$ represents a triangle of the lattice and the operator 
$CCZ_{abc}$ is given by
\begin{eqnarray}
    CCZ_{abc}\vcentcolon=(-1)^{(1-Z_a)(1-Z_b)(1-Z_c)/8},
\end{eqnarray}
corresponding to the controlled-controlled-$Z$ (CCZ)~gate acting on the triangle. Under conjugation by $CCZ_{ijk}$, local operators are transformed as
\begin{align}
    X_i \rightarrow X_iCZ_{jk}, \quad Z_i \rightarrow Z_i.
\end{align}
For the subsequent purposes, it is convenient to define the controlled-$Z$~(CZ) gate operator as
\begin{align}
    CZ_{ab}\vcentcolon=(-1)^{(1-Z_a)(1-Z_b)/4
    }\label{7}
\end{align}
and similarly for $CZ_{bc}$, $CZ_{ca}$.\par
Physically, $U_D$ acts as an SPT entangler corresponding to $\mathbb{Z}_2^A\times\mathbb{Z}_2^B\times\mathbb{Z}_2^C$ SPT phase characterized by a nontrivial $3$-cocycle~\cite{Chen:2011pg, Yoshida:2015cia}. As a result, the symmetry operators exhibit a mixed~\tht~anomaly consistent with the type IV anomaly inflow action.
\par
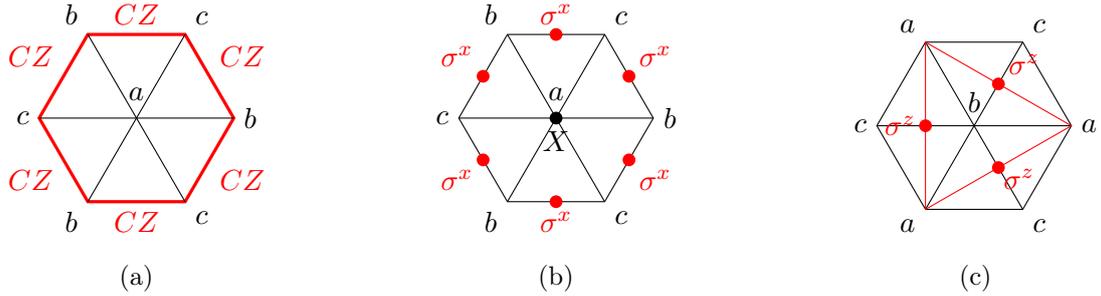
\begin{figure}[t]
\centering
\begin{subfigure}[t]{0.31\linewidth}\centering
\begin{tikzpicture}[scale=0.8]
\def\r{1.6}
\coordinate (v1) at (120:\r);
\coordinate (v2) at (60:\r);
\coordinate (v3) at (0:\r);
\coordinate (v4) at (-60:\r);
\coordinate (v5) at (-120:\r);
\coordinate (v6) at (180:\r);
\coordinate (c)  at (0,0);

\draw[red, line width=1.2pt]
  (v1)--(v2)--(v3)--(v4)--(v5)--(v6)--cycle;

\draw (v1)--(v4);
\draw (v2)--(v5);
\draw (v3)--(v6);

\node at ($(c)+(0,0.4)$) {$a$};

\node at (v1) [above left] {$b$};
\node at (v2) [above right] {$c$};
\node at (v3) [right] {$b$};
\node at (v4) [below right] {$c$};
\node at (v5) [below left] {$b$};
\node at (v6) [left] {$c$};

\node at ($(v1)!0.5!(v2)$) [above] {$\color{red}CZ$};
\node at ($(v2)!0.5!(v3)$) [above right] {$\color{red}CZ$};
\node at ($(v3)!0.5!(v4)$) [below right] {$\color{red}CZ$};
\node at ($(v4)!0.5!(v5)$) [below] {$\color{red}CZ$};
\node at ($(v5)!0.5!(v6)$) [below left] {$\color{red}CZ$};
\node at ($(v6)!0.5!(v1)$) [above left] {$\color{red}CZ$};
\end{tikzpicture}
\caption{}
\label{fig:type4a}
\end{subfigure}\hfill
\begin{subfigure}[t]{0.31\linewidth}\centering
\begin{tikzpicture}[scale=0.8]
\def\r{1.6}
\coordinate (v1) at (120:\r);
\coordinate (v2) at (60:\r);
\coordinate (v3) at (0:\r);
\coordinate (v4) at (-60:\r);
\coordinate (v5) at (-120:\r);
\coordinate (v6) at (180:\r);
\coordinate (c)  at (0,0);

\draw
  (v1)--(v2)--(v3)--(v4)--(v5)--(v6)--cycle;

\draw (v1)--(v4);
\draw (v2)--(v5);
\draw (v3)--(v6);

\node at ($(c)+(0,0.4)$) {$a$};
\fill (c) circle (3pt);
\node[below=1pt] at (c) {$X$};

\node at (v1) [above left] {$b$};
\node at (v2) [above right] {$c$};
\node at (v3) [right] {$b$};
\node at (v4) [below right] {$c$};
\node at (v5) [below left] {$b$};
\node at (v6) [left] {$c$};

\node at ($(v1)!0.5!(v2)$) [above] {$\color{red}\sigma^x$};
\node at ($(v2)!0.5!(v3)$) [above right] {$\color{red}\sigma^x$};
\node at ($(v3)!0.5!(v4)$) [below right] {$\color{red}\sigma^x$};
\node at ($(v4)!0.5!(v5)$) [below] {$\color{red}\sigma^x$};
\node at ($(v5)!0.5!(v6)$) [below left] {$\color{red}\sigma^x$};
\node at ($(v6)!0.5!(v1)$) [above left] {$\color{red}\sigma^x$};

\fill[red] ($(v1)!0.5!(v2)$) circle (3pt);
\fill[red] ($(v2)!0.5!(v3)$) circle (3pt);
\fill[red] ($(v3)!0.5!(v4)$) circle (3pt);
\fill[red] ($(v4)!0.5!(v5)$) circle (3pt);
\fill[red] ($(v5)!0.5!(v6)$) circle (3pt);
\fill[red] ($(v6)!0.5!(v1)$) circle (3pt);
\end{tikzpicture}
\caption{}
\label{fig:type4b}
\end{subfigure}\hfill
\begin{subfigure}[t]{0.31\linewidth}\centering
\begin{tikzpicture}[scale=0.8]
\def\r{1.6}
\coordinate (v1) at (120:\r);
\coordinate (v2) at (60:\r);
\coordinate (v3) at (0:\r);
\coordinate (v4) at (-60:\r);
\coordinate (v5) at (-120:\r);
\coordinate (v6) at (180:\r);
\coordinate (c)  at (0,0);

\draw
  (v1)--(v2)--(v3)--(v4)--(v5)--(v6)--cycle;

\draw (v1)--(v4);
\draw (v2)--(v5);
\draw (v3)--(v6);

\draw (v1)--(v3) [red];
\draw (v1)--(v5) [red];
\draw (v3)--(v5) [red];

\node at ($(c)+(0,0.4)$) {$b$};

\node at (v1) [above left] {$a$};
\node at (v2) [above right] {$c$};
\node at (v3) [right] {$a$};
\node at (v4) [below right] {$c$};
\node at (v5) [below left] {$a$};
\node at (v6) [left] {$c$};

\node at ($(v2)!0.5!(c)$) [above right] {$\color{red}\sigma^z$};
\node at ($(v4)!0.6!(c)$) [below right] {$\color{red}\sigma^z$};
\node at ($(v6)!0.5!(c)$) [left] {$\color{red}\sigma^z$};

\fill[red] ($(v2)!0.5!(c)$) circle (3pt);
\fill[red] ($(v4)!0.5!(c)$) circle (3pt);
\fill[red] ($(v6)!0.5!(c)$) circle (3pt);

\end{tikzpicture}
\caption{}
\label{fig:type4c}
\end{subfigure}

\caption{
(a) The product of CZ terms in the Hamiltonian~\eqref{model with type4}.
(b) Configuration of the Gauss law $G_{A,a}$, which is described in~\eqref{type4:gauss1}. 
(c) Flux term $B_A$ given in~\eqref{type4:flatness01}.
}

\label{fig:type4}
\end{figure}

An explicit example of the Hamiltonian that respects the symmetries~\eqref{symmetries} is given by
\begin{equation}
\begin{aligned}
    H_0=-\sum_{\langle a_1,a_2\rangle}Z_{a_1}Z_{a_2}- \sum_{a\in\Lambda_a}X_a\left(1 + \prod_{\langle bc\rangle \subset \langle abc\rangle}CZ_{bc}\right)  \\
    -\sum_{\langle b_1,b_2\rangle}Z_{b_1}Z_{b_2}- \sum_{b\in\Lambda_b}X_b\left(1 + \prod_{\langle ca\rangle \subset \langle abc\rangle}CZ_{ca}\right) \\
    -\sum_{\langle c_1,c_2\rangle}Z_{c_1}Z_{c_2}- \sum_{c\in\Lambda_c}X_c\left(1 + \prod_{\langle ab\rangle \subset \langle abc\rangle}CZ_{ab}\right).
    \label{model with type4}
\end{aligned}
\end{equation}
Here, there is a qubit degree of freedom (d.o.f) on each node of the $\Lambda_k\ (k=a,b,c)$. 
Also, the sum $\sum_{\langle k_1,k_2\rangle}$ runs over nearest-neighbor sites on the sublattice $\Lambda_k\ (k=a,b,c)$, and $\prod_{\langle ab\rangle \subset \langle abc\rangle}CZ_{bc}$ 
denotes the product of CZ gates along a closed loop formed by links connecting $b$ and $c$,  which surrounds a site $a$. See
 Fig.~\ref{fig:type4a}
 for its visual illustration. 
 The other products of the~CZ gates are analogously defined. 

In the following, we implement gauging of various subgroups of $\mb^{A} \times \mb^{B} \times \mb^{C} \times \mb^{D}$ and investigate the corresponding symmetry structures. For concreteness, we perform the gauging using an explicit lattice Hamiltonian \eqref{model with type4}. Importantly, as we will see in Sec.~\ref{sec4}, the resulting symmetry structure is universal and does not depend on microscopic details of the Hamiltonian.

\subsection{Gauging $\mathbb{Z}_2^A$ : 2-group symmetry} \label{sec.2.1}
We begin by gauging the minimal subgroup of $\mathbb{Z}_2^{A} \times \mathbb{Z}_2^{B} \times \mathbb{Z}_2^{C} \times \mathbb{Z}_2^{D}$ symmetry, 
$\mathbb{Z}_2^A$, 
generated by $U_A$, and investigate the resulting symmetry structure. To implement gauging, we introduce auxiliary qubits on links $\langle bc\rangle\in \Lambda_{bc}$ where $b\in \Lambda_b, c\in \Lambda_c$, whose Pauli operators are given by~$\sigma_{\langle bc\rangle}^x$ and $\sigma_{\langle bc\rangle}^z$. The gauging is achieved by enforcing the following  Gauss law:
\begin{equation}
    G_{A,a}:=X_a\prod_{\langle bc\rangle\subset \langle abc \rangle}\sigma_{\langle bc\rangle}^x = 1, \quad \forall a\in\Lambda_a \label{type4:gauss1}
\end{equation}
which is shown in Fig.~\ref{fig:type4b}, together with the following flatness condition
\begin{equation}
    B_{A}(\gamma_a):=\prod_{\langle bc\rangle\subset \gamma_a}\sigma_{\langle bc\rangle}^z =1, \quad \forall\gamma_a, \label{type4:flatness01}
\end{equation}
which  ensures that the gauge theory does not admit excess flux.
Here, $\gamma_a$ denotes a contractible loop on the sublattice $\Lambda_a$, and the product 
is defined for a trajectory of the links
$\langle bc\rangle$ intersected by~$\gamma_a$ (See Fig.~\ref{fig:type4c}.). 
To proceed, we modify the local terms in the Hamiltonian~\eqref{model with type4} so that they commute with the Gauss law \eqref{type4:gauss1}, corresponding to the minimal coupling. To wit,  
\begin{equation}
    Z_{a_1}Z_{a_2} \rightarrow Z_{a_1}\sigma_{\langle bc\rangle}^zZ_{a_2}. 
\end{equation}
Here, $\langle bc\rangle$ denotes a link intersected by the line connecting $a_1$ and $a_2$. To solve the Gauss law~\eqref{type4:gauss1}, we act a unitary transformation on local operators so that 
\begin{align}
   G_{A,a} \rightarrow X_a, \quad Z_{a_1}\sigma_{\langle bc\rangle}^zZ_{a_2} \rightarrow \sigma_{\langle bc\rangle}^z.
\end{align}
Under this transformation, the Gauss law~\eqref{type4:gauss1} becomes $X_a=1, \forall a\in \Lambda_a$.

After gauging, local and quadratic operators transform as 
\begin{align}
    X_a \rightarrow \prod_{\langle bc\rangle\subset \langle abc \rangle}\sigma_{\langle bc\rangle}^x, \quad Z_{a_1}Z_{a_2} \rightarrow \sigma_{\langle bc\rangle}^z. \label{gauging operation}
\end{align}
In addition, 
for a given $c$, the product of CZ gate operators\footnote{
Here, we employ an alternative form of the CZ gate~\eqref{7}
in the $Z$ basis: $CZ_{ab}|a,b\rangle=(-1)^{ab}|a,b\rangle$ with $a,b\in\{0,1\}$. 
}
\begin{eqnarray*}
    \prod_{\langle ab\rangle \subset \langle abc\rangle}CZ_{ab}&=&(-1)^{  a_1b_1+b_1a_2+a_2b_2+b_2a_3+a_3b_3+b_3a_1}\nonumber\\
    &=&(-1)^{(a_1+a_2)b_1 + (a_2+a_3)b_2 +(a_3+a_1)b_3},
\end{eqnarray*}
with $a_i$ and $b_i~(i=1,2,3)$ 
denotes sites surrounding $c$ ordered clockwise with increasing $i$. This
is transformed into
\begin{eqnarray}
    (-1)^{ \tilde{a}_{12}b_1 + \tilde{a}_{23}b_2 +\tilde{a}_{31}b_3}:= \prod_{\langle bc\rangle \subset \langle abc\rangle}CZ_{b\langle bc\rangle},
\end{eqnarray}
where $\tilde{a}_{ij}$ is a gauge variable on the link $\langle bc\rangle$ intersected by the line connecting $a_i$ and $a_j$, and~$CZ_{b\langle bc\rangle}$ is a CZ gate acting on site $b$ and link $\langle bc\rangle$. Graphically, this procedure is described by 
\begin{equation}
\begin{tikzpicture}[baseline=(current bounding box.center),scale=0.6]
\def\r{1.6}
\coordinate (v1) at (120:\r);
\coordinate (v2) at (60:\r);
\coordinate (v3) at (0:\r);
\coordinate (v4) at (-60:\r);
\coordinate (v5) at (-120:\r);
\coordinate (v6) at (180:\r);
\coordinate (c)  at (0,0);

\draw[red, line width=1.2pt]
  (v1)--(v2)--(v3)--(v4)--(v5)--(v6)--cycle;

\draw (v1)--(v4);
\draw (v2)--(v5);
\draw (v3)--(v6);

\node at ($(c)+(0,0.4)$) {$c$};

\node at (v1) [above left] {$a_1$};
\node at (v2) [above right] {$b_1$};
\node at (v3) [right] {$a_2$};
\node at (v4) [below right] {$b_2$};
\node at (v5) [below left] {$a_3$};
\node at (v6) [left] {$b_3$};

\node at ($(v1)!0.5!(v2)$) [above] {$\color{red}CZ$};
\node at ($(v2)!0.5!(v3)$) [above right] {$\color{red}CZ$};
\node at ($(v3)!0.5!(v4)$) [below right] {$\color{red}CZ$};
\node at ($(v4)!0.5!(v5)$) [below] {$\color{red}CZ$};
\node at ($(v5)!0.5!(v6)$) [below left] {$\color{red}CZ$};
\node at ($(v6)!0.5!(v1)$) [above left] {$\color{red}CZ$};
\end{tikzpicture}
\;\xrightarrow{\ \mathrm{gauging}\ }\;
\begin{tikzpicture}[baseline=(current bounding box.center),scale=0.6]
\def\r{1.6}
\coordinate (v1) at (120:\r);
\coordinate (v2) at (60:\r);
\coordinate (v3) at (0:\r);
\coordinate (v4) at (-60:\r);
\coordinate (v5) at (-120:\r);
\coordinate (v6) at (180:\r);
\coordinate (c)  at (0,0);
\coordinate (d1) at ($(v2)!0.5!(c)$);
\coordinate (d2) at ($(v4)!0.5!(c)$);
\coordinate (d3) at ($(v6)!0.5!(c)$);

\draw (v1)--(v2)--(v3)--(v4)--(v5)--(v6)--cycle;

\draw (v1)--(v4);
\draw (v2)--(v5);
\draw (v3)--(v6);

\draw[red, line width=1pt] (v2)--(d1);
\draw[red, line width=1pt] (v4)--(d2);
\draw[red, line width=1pt] (v6)--(d3);

\node at ($(c)+(0,0.4)$) {$c$};

\node at (v1) [above left] {$a_1$};
\node at (v2) [above right] {$b_1$};
\node at (v3) [right] {$a_2$};
\node at (v4) [below right] {$b_2$};
\node at (v5) [below left] {$a_3$};
\node at (v6) [left] {$b_3$};

\node at (d1) [right] {${\tilde{a}_{12}}$};
\node at (d2) [right] {${\tilde{a}_{23}}$};
\node at (d3) [below] {${\tilde{a}_{31}}$};

\node at ($(v2)!0.5!(v1)$) {$\color{red}CZ$};
\node at ($(v4)!0.5!(v5)$) {$\color{red}CZ$};
\node at ($(v6)!0.5!(v1)$) {$\color{red}CZ$};

\fill[red] (d1) circle (3pt);
\fill[red] (d2) circle (3pt);
\fill[red] (d3) circle (3pt);
\end{tikzpicture}.
\end{equation}
A similar computation applies to  $\prod_{\langle ca\rangle \subset \langle abc\rangle}CZ_{ca}$.

Overall, after gauging, we arrive at the following Hamiltonian:
\begin{equation}
\begin{aligned}
    H_1=&-\sum_{\langle bc\rangle \in \Lambda_{bc}}\sigma_{\langle bc\rangle}^z- \sum_{a\in\Lambda_a} \left[\prod_{\langle bc\rangle\subset \langle abc \rangle}\sigma_{\langle bc\rangle}^x \left(1 + \prod_{\langle bc\rangle \subset \langle abc\rangle}CZ_{bc}\right)\right]  \\
    &-\sum_{\langle b_1,b_2\rangle}Z_{b_1}Z_{b_2}- \sum_{b\in\Lambda_b}X_b\left(1 + \prod_{\langle bc\rangle \subset \langle abc\rangle}CZ_{c\langle bc\rangle}\right) \\
    &-\sum_{\langle c_1,c_2\rangle}Z_{c_1}Z_{c_2}- \sum_{c\in\Lambda_c}X_c\left(1 + \prod_{\langle bc\rangle \subset \langle abc\rangle}CZ_{b\langle bc\rangle}\right).
    \label{gauged model1 with type4}
\end{aligned}
\end{equation}
This model respects one $\hat{\mathbb{Z}}_2^{A}$ $1$-form and three $\mb$ $0$-form symmetries, generated by
\begin{align}
    \eta_{A}^{(1)}(\gamma_a)=\prod_{\langle bc\rangle\in\gamma_a}\sigma_{\langle bc\rangle}^z, \quad U_B=\prod_{b\in\Lambda_b}X_b, \quad U_C=\prod_{c\in\Lambda_c}X_c, \quad U_{D,1}=\prod_{\langle bc\rangle}CCZ_{bc\langle bc\rangle}, \label{2-group sym}
\end{align}
where $\gamma_a$ is noncontractible loop on the sublattice $\Lambda_a$, and $CCZ_{bc\langle bc\rangle}$ denotes the CCZ gate acting on two sites $b,c$ and one link $\langle bc\rangle$. 
Note that through the gauging procedure, the symmetry operator $U_D$ given in~\eqref{symmetries} is modified, giving rise to $U_{D,1}$.
Also, $1$-form symmetry is topological, i.e., the corresponding operators depend solely on the homology class due to 
the flatness condition~\eqref{type4:flatness01}. 

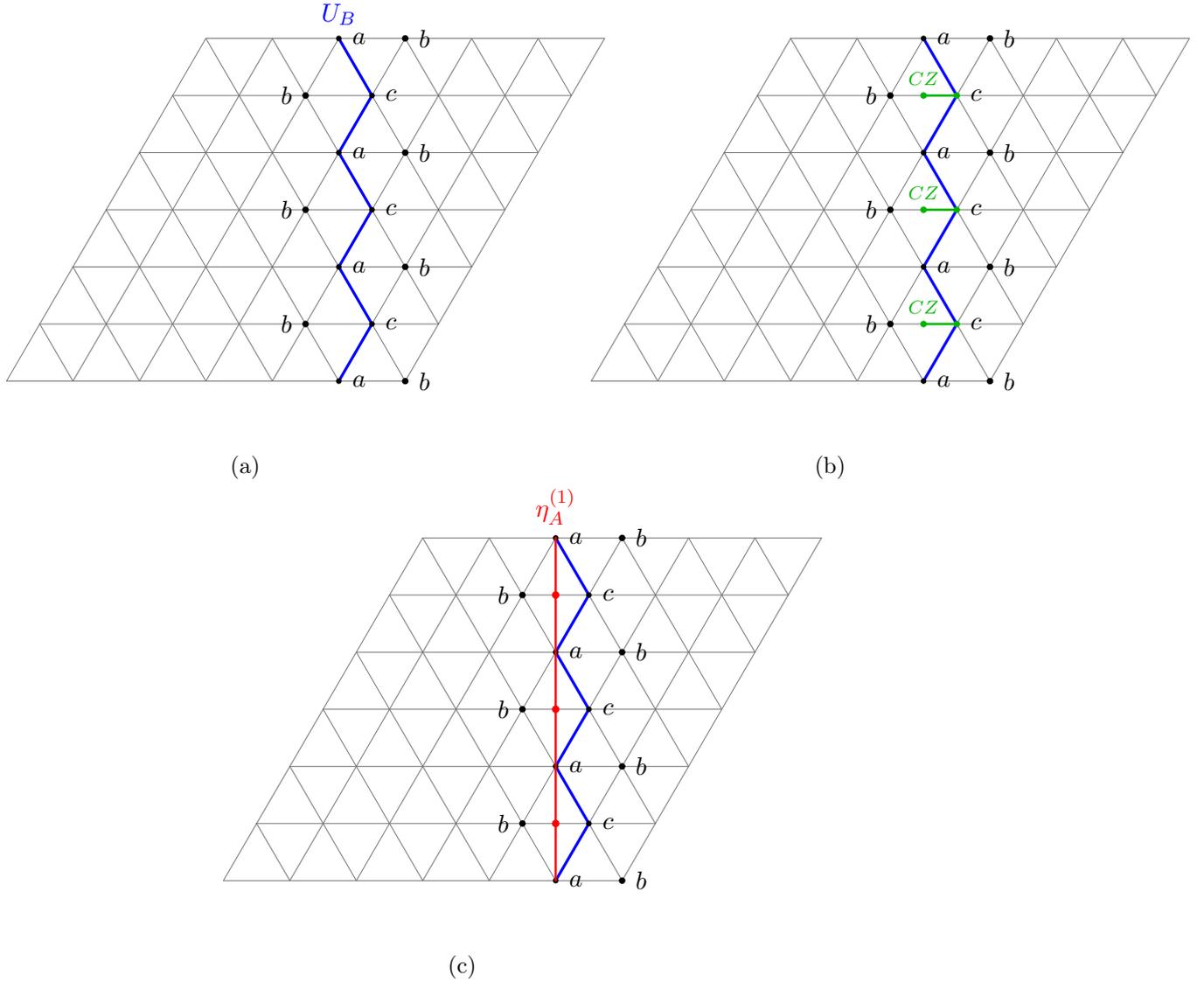
\begin{figure}[t]
\centering

\begin{subfigure}{0.45\textwidth}
\centering
\begin{tikzpicture}[scale=1, line cap=round, line join=round] 
\def\h{0.8660254} 
\def\Nx{6}        
\def\Ny{6}        

\foreach \j in {0,...,\Ny}{
  \foreach \i in {0,...,\Nx}{
    \pgfmathsetmacro{\x}{\i + 0.5*\j}
    \pgfmathsetmacro{\y}{\h*\j}

    \ifnum\i<\Nx
      \pgfmathsetmacro{\xB}{(\i+1) + 0.5*\j}
      \pgfmathsetmacro{\yB}{\h*\j}
      \draw[black!55, line width=0.35pt] (\x,\y) -- (\xB,\yB);
    \fi

    \ifnum\j<\Ny
      \pgfmathsetmacro{\xB}{\i + 0.5*(\j+1)}
      \pgfmathsetmacro{\yB}{\h*(\j+1)}
      \draw[black!55, line width=0.35pt] (\x,\y) -- (\xB,\yB);
    \fi

    \ifnum\j<\Ny
      \ifnum\i>0
        \pgfmathsetmacro{\xB}{(\i-1) + 0.5*(\j+1)}
        \pgfmathsetmacro{\yB}{\h*(\j+1)}
        \draw[black!55, line width=0.35pt] (\x,\y) -- (\xB,\yB);
      \fi
    \fi
  }
}

\coordinate (Q1) at ({5 + 0.5*0},{\h*0}); 
\coordinate (Q2) at ({5 + 0.5*1},{\h*1}); 
\coordinate (Q3) at ({4 + 0.5*2},{\h*2}); 
\coordinate (Q4) at ({4 + 0.5*3},{\h*3}); 
\coordinate (Q5) at ({3 + 0.5*4},{\h*4}); 
\coordinate (Q6) at ({3 + 0.5*5},{\h*5}); 
\coordinate (Q7) at ({2 + 0.5*6},{\h*6}); 

\coordinate (P1) at ({5 + 0.5*2},{\h*0}); 
\coordinate (P2) at ({4 + 0.5*1},{\h*1}); 
\coordinate (P3) at ({5 + 0.5*2},{\h*2}); 
\coordinate (P4) at ({4 + 0.5*1},{\h*3}); 
\coordinate (P5) at ({5 + 0.5*2},{\h*4}); 
\coordinate (P6) at ({4 + 0.5*1},{\h*5}); 
\coordinate (P7) at ({5 + 0.5*2},{\h*6}); 

\draw[blue, line width=1.2pt] (Q1)--(Q2)--(Q3)--(Q4)--(Q5)--(Q6)--(Q7);

\foreach \q in {Q1,Q2,Q3,Q4,Q5,Q6,Q7}{
  \fill[black] (\q) circle (1.2pt);
}

\foreach \p in {P1,P2,P3,P4,P5,P6,P7}{
  \fill[black] (\p) circle (1.4pt);
}

\node[right=2pt] at (Q1) {$a$};
\node[right=2pt] at (Q2) {$c$};
\node[right=2pt] at (Q3) {$a$};
\node[right=2pt] at (Q4) {$c$};
\node[right=2pt] at (Q5) {$a$};
\node[right=2pt] at (Q6) {$c$};
\node[right=2pt] at (Q7) {$a$};

\node[right=2pt] at (P1) {$b$};
\node[left=2pt] at (P2) {$b$};
\node[right=2pt] at (P3) {$b$};
\node[left=2pt] at (P4) {$b$};
\node[right=2pt] at (P5) {$b$};
\node[left=2pt] at (P6) {$b$};
\node[right=2pt] at (P7) {$b$};

\node[above=2pt] at (Q7) {\textcolor{blue}{$U_B$}};

\end{tikzpicture}
\caption{}
\label{Fig:2-groupa}
\end{subfigure}
\hfill
\begin{subfigure}{0.45\textwidth}
\centering
\begin{tikzpicture}[scale=1, line cap=round, line join=round] 
\def\h{0.8660254} 
\def\Nx{6}        
\def\Ny{6}        

\foreach \j in {0,...,\Ny}{
  \foreach \i in {0,...,\Nx}{
    \pgfmathsetmacro{\x}{\i + 0.5*\j}
    \pgfmathsetmacro{\y}{\h*\j}

    \ifnum\i<\Nx
      \pgfmathsetmacro{\xB}{(\i+1) + 0.5*\j}
      \pgfmathsetmacro{\yB}{\h*\j}
      \draw[black!55, line width=0.35pt] (\x,\y) -- (\xB,\yB);
    \fi

    \ifnum\j<\Ny
      \pgfmathsetmacro{\xB}{\i + 0.5*(\j+1)}
      \pgfmathsetmacro{\yB}{\h*(\j+1)}
      \draw[black!55, line width=0.35pt] (\x,\y) -- (\xB,\yB);
    \fi

    \ifnum\j<\Ny
      \ifnum\i>0
        \pgfmathsetmacro{\xB}{(\i-1) + 0.5*(\j+1)}
        \pgfmathsetmacro{\yB}{\h*(\j+1)}
        \draw[black!55, line width=0.35pt] (\x,\y) -- (\xB,\yB);
      \fi
    \fi
  }
}

\coordinate (Q1) at ({5 + 0.5*0},{\h*0}); 
\coordinate (Q2) at ({5 + 0.5*1},{\h*1}); 
\coordinate (Q3) at ({4 + 0.5*2},{\h*2}); 
\coordinate (Q4) at ({4 + 0.5*3},{\h*3}); 
\coordinate (Q5) at ({3 + 0.5*4},{\h*4}); 
\coordinate (Q6) at ({3 + 0.5*5},{\h*5}); 
\coordinate (Q7) at ({2 + 0.5*6},{\h*6}); 

\coordinate (P1) at ({5 + 0.5*2},{\h*0}); 
\coordinate (P2) at ({4 + 0.5*1},{\h*1}); 
\coordinate (P3) at ({5 + 0.5*2},{\h*2}); 
\coordinate (P4) at ({4 + 0.5*1},{\h*3}); 
\coordinate (P5) at ({5 + 0.5*2},{\h*4}); 
\coordinate (P6) at ({4 + 0.5*1},{\h*5}); 
\coordinate (P7) at ({5 + 0.5*2},{\h*6}); 

\coordinate (R2) at ({4 + 0.5*2},{\h*1}); 
\coordinate (R4) at ({4 + 0.5*2},{\h*3}); 
\coordinate (R6) at ({4 + 0.5*2},{\h*5}); 

\draw[blue, line width=1.2pt] (Q1)--(Q2)--(Q3)--(Q4)--(Q5)--(Q6)--(Q7);

\foreach \q in {Q1,Q3,Q5,Q7}{
  \fill[black] (\q) circle (1.2pt);
}

\foreach \q in {Q2,Q4,Q6}{
  \fill[green!70!black] (\q) circle (1.4pt);
}

\foreach \p in {P1,P2,P3,P4,P5,P6,P7}{
  \fill[black] (\p) circle (1.4pt);
}

\foreach \q in {R2,R4,R6}{
  \fill[green!70!black] (\q) circle (1.4pt);
}

\draw[green!70!black, line width=1.0pt] (R2)--(Q2);
\draw[green!70!black, line width=1.0pt] (R4)--(Q4);
\draw[green!70!black, line width=1.0pt] (R6)--(Q6);

\node[right=2pt] at (Q1) {$a$};
\node[right=2pt] at (Q2) {$c$};
\node[right=2pt] at (Q3) {$a$};
\node[right=2pt] at (Q4) {$c$};
\node[right=2pt] at (Q5) {$a$};
\node[right=2pt] at (Q6) {$c$};
\node[right=2pt] at (Q7) {$a$};

\node[right=2pt] at (P1) {$b$};
\node[left=2pt] at (P2) {$b$};
\node[right=2pt] at (P3) {$b$};
\node[left=2pt] at (P4) {$b$};
\node[right=2pt] at (P5) {$b$};
\node[left=2pt] at (P6) {$b$};
\node[right=2pt] at (P7) {$b$};

\node[above=0.8pt, font=\scriptsize] at (R2) {$\textcolor{green!70!black}{CZ}$};
\node[above=0.8pt, font=\scriptsize] at (R4) {$\textcolor{green!70!black}{CZ}$};
\node[above=0.8pt, font=\scriptsize] at (R6) {$\textcolor{green!70!black}{CZ}$};

\end{tikzpicture}
\caption{}
\label{Fig:2-groupb}
\end{subfigure}

\vspace{-0.2em}
\hspace{-0.15\textwidth}
\begin{subfigure}{0.45\textwidth}
\centering
\begin{tikzpicture}[scale=1, line cap=round, line join=round] 
\def\h{0.8660254} 
\def\Nx{6}        
\def\Ny{6}        

\foreach \j in {0,...,\Ny}{
  \foreach \i in {0,...,\Nx}{
    \pgfmathsetmacro{\x}{\i + 0.5*\j}
    \pgfmathsetmacro{\y}{\h*\j}

    \ifnum\i<\Nx
      \pgfmathsetmacro{\xB}{(\i+1) + 0.5*\j}
      \pgfmathsetmacro{\yB}{\h*\j}
      \draw[black!55, line width=0.35pt] (\x,\y) -- (\xB,\yB);
    \fi

    \ifnum\j<\Ny
      \pgfmathsetmacro{\xB}{\i + 0.5*(\j+1)}
      \pgfmathsetmacro{\yB}{\h*(\j+1)}
      \draw[black!55, line width=0.35pt] (\x,\y) -- (\xB,\yB);
    \fi

    \ifnum\j<\Ny
      \ifnum\i>0
        \pgfmathsetmacro{\xB}{(\i-1) + 0.5*(\j+1)}
        \pgfmathsetmacro{\yB}{\h*(\j+1)}
        \draw[black!55, line width=0.35pt] (\x,\y) -- (\xB,\yB);
      \fi
    \fi
  }
}

\coordinate (Q1) at ({5 + 0.5*0},{\h*0}); 
\coordinate (Q2) at ({5 + 0.5*1},{\h*1}); 
\coordinate (Q3) at ({4 + 0.5*2},{\h*2}); 
\coordinate (Q4) at ({4 + 0.5*3},{\h*3}); 
\coordinate (Q5) at ({3 + 0.5*4},{\h*4}); 
\coordinate (Q6) at ({3 + 0.5*5},{\h*5}); 
\coordinate (Q7) at ({2 + 0.5*6},{\h*6}); 

\coordinate (P1) at ({5 + 0.5*2},{\h*0}); 
\coordinate (P2) at ({4 + 0.5*1},{\h*1}); 
\coordinate (P3) at ({5 + 0.5*2},{\h*2}); 
\coordinate (P4) at ({4 + 0.5*1},{\h*3}); 
\coordinate (P5) at ({5 + 0.5*2},{\h*4}); 
\coordinate (P6) at ({4 + 0.5*1},{\h*5}); 
\coordinate (P7) at ({5 + 0.5*2},{\h*6}); 

\coordinate (R2) at ({4 + 0.5*2},{\h*1}); 
\coordinate (R4) at ({4 + 0.5*2},{\h*3}); 
\coordinate (R6) at ({4 + 0.5*2},{\h*5}); 

\draw[blue, line width=1.2pt] (Q1)--(Q2)--(Q3)--(Q4)--(Q5)--(Q6)--(Q7);

\foreach \q in {Q1,Q3,Q5,Q7}{
  \fill[black] (\q) circle (1.2pt);
}

\foreach \q in {Q2,Q4,Q6}{
  \fill[black] (\q) circle (1.2pt);
}

\foreach \p in {P1,P2,P3,P4,P5,P6,P7}{
  \fill[black] (\p) circle (1.4pt);
}

\foreach \q in {R2,R4,R6}{
  \fill[red] (\q) circle (1.6pt);
}

\draw[red, line width=0.9pt] (Q1)--(Q7);

\node[right=2pt] at (Q1) {$a$};
\node[right=2pt] at (Q2) {$c$};
\node[right=2pt] at (Q3) {$a$};
\node[right=2pt] at (Q4) {$c$};
\node[right=2pt] at (Q5) {$a$};
\node[right=2pt] at (Q6) {$c$};
\node[right=2pt] at (Q7) {$a$};

\node[right=2pt] at (P1) {$b$};
\node[left=2pt] at (P2) {$b$};
\node[right=2pt] at (P3) {$b$};
\node[left=2pt] at (P4) {$b$};
\node[right=2pt] at (P5) {$b$};
\node[left=2pt] at (P6) {$b$};
\node[right=2pt] at (P7) {$b$};

\node[above=2pt] at (Q7) {$\textcolor{red}{\eta_A^{(1)}}$};
\end{tikzpicture}
\caption{}
\label{Fig:2-groupc}
\end{subfigure}

\caption{
(a) A configuration with a $\mb^B$ defect (blue line) inserted
(b) Action of $U_{D,1}$ around the $\mb^B$ defect
(c) The projective algebra between $U_C$ and $U_D$ under the defect insertion, which gives rise to a $1$-form symmetry (red line) along the defect.
}
\label{Fig:2-group}
\end{figure}

We discuss the symmetry structure generated by these symmetry operators. Naively, these operators mutually commute\footnote{Without a symmetry defect, the commutativity, $  U_B U_{D_1} = U_{D_1} U_B $ and $  U_C U_{D_1} = U_{D_1} U_C $, can be verified using the flatness condition~\eqref{type4:flatness01}.}, yet
%
%
the situation drastically changes when we 
insert a $\mb^B$ symmetry defect. 
To see this more explicitly, 
we place the defect along the line segment connecting $a$ and $c$ as shown in Fig.~\ref{Fig:2-groupa}~\footnote{See, e.g.,~\cite{Kawagoe:2021gqi,Seifnashri:2023dpa,hofmeier2026lattice} for relevant discussions on insertion of the symmetry defects.}. 
The defect is implemented by imposing a twisted boundary condition along a branch cut. 
More precisely, 
every Ising coupling between neighboring $B$-sublattice 
spins whose bond crosses $\gamma_B$ acquires an additional minus sign:
\begin{equation}
    -Z_b^{\text{left}}Z_b^{\text{right}} \to +Z_b^{\text{left}}Z_b^{\text{right}},\label{twbd}
\end{equation}
Here, $Z_b^{\text{left}}$ and $Z_b^{\text{right}}$ label the two distinct neighboring sites located on opposite sides of the branch cut.
%
%
In the presence of the defect, we find the modified algebra:
\begin{equation}
    U_CU_{D,1} = \eta_A^{(1)}(\gamma)U_{D,1}U_C,\label{p}
\end{equation}
where 
$\eta_A^{(1)}(\gamma)$ represents $1$-form symmetry operator along the defect line $\gamma$~(Fig.~\ref{Fig:2-groupc}).
This shows that the symmetry algebra is realized 
\textit{projectively} in the defect background\footnote{Such higher group structures involving the controlled gates have also been discussed in the context of finite gauge theories~\cite{Barkeshli:2022wuz,Barkeshli:2022edm,Barkeshli:2023bta}. }. \par
\begin{figure} 
    \centering
    \begin{tikzpicture}[scale=1.7, line cap=round, line join=round]
\def\h{0.8660254} 

\coordinate (A) at (0,0);
\coordinate (B) at (0.5,\h);
\coordinate (C) at (1.5,\h);
\coordinate (D) at (1,0);

\draw[black, line width=0.8pt] (A)--(B)--(C)--(D)--cycle;

\draw[blue, line width=1.2pt] (B)--(D);

\fill[black] (A) circle (1.2pt);
\fill[black] (B) circle (1.2pt);
\fill[black] (C) circle (1.2pt);
\fill[black] (D) circle (1.2pt);

\node[below left=2pt]  at (A) {$b_1$};
\node[above left=2pt]  at (B) {$a$};
\node[above right=2pt] at (C) {$b_2$};
\node[below right=2pt] at (D) {$c$};

\node at ({(0+0.5+1)/3},{(0+\h+0)/3}) {$1$};
\node at ({(0.5+1.5+1)/3},{(\h+\h+0)/3}) {$2$};
\end{tikzpicture}
    \caption{Two triangles close to the defect (blue line connecting $a$ and $c$). }
    \label{fig:triangle}
\end{figure}
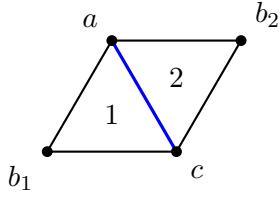

To verify this, we evaluate the operator algebra locally near the defect. 
Consider two adjacent triangles $\Delta_1$ and $\Delta_2$ separated by the defect (Fig.~\ref{fig:triangle}). 
In this setup, we evaluate $  U_{D,1}X_cU_{D,1}^{-1} $ on these triangles as 
\begin{equation}
\begin{aligned}   
     U_{D,1}X_cU_{D,1}^{-1} 
    &= CCZ_{b_1c\langle b_1c\rangle}CCZ_{b_2c\langle b_2c\rangle}X_cCCZ^{-1}_{b_2c\langle b_2c\rangle}CCZ^{-1}_{b_1c\langle b_1c\rangle} \\
    &=X_c\ (-1)^{(1-\sigma^z_{\langle b_1c\rangle})(1-Z_{b_1})/4}(-1)^{(1-\sigma^z_{\langle b_2c\rangle})(1-Z_{b_2})/4}.
\end{aligned}
\end{equation}
Using the twisted boundary condition \eqref{twbd} and $\sigma^z_{\langle b_1c\rangle}=\sigma^z_{\langle b_2c\rangle}$, one finds
\begin{equation}
    U_{D,1}X_cU_{D,1}^{-1} = X_c\sigma^z_{\langle b_1c\rangle}.
\end{equation}
This implies that around the defect, $U_{D,1}$ effectively acts as~(See also Fig.~\ref{Fig:2-groupb} for illustration.)
\begin{equation}
    U_{D,1}|_{\text{defect}} = \prod_{\langle bc\rangle}CZ_{c\langle bc\rangle},\label{21}
\end{equation}
leading to~\eqref{p}.
\par
To recap the argument in this subsection, we gauge one of the internal symmetries in  a system with the type~IV anomaly. In the presence of the symmetry defect, the model exhibits~2-group symmetry structure. Note that the above projective algebra originates from the nontrivial Postnikov class, described by the following third cohomology group:
\begin{equation}
    [\beta]\in H^3(\mathbb{Z}_2^3, \mathbb{Z}_2), \quad \beta\Big((b_1,c_1,d_1),(b_2,c_2,d_2),(b_3,c_3,d_3)\Big)=b_1c_2d_3.
\end{equation}
The result is consistent with the field-theoretical description, as we will discuss in detail in Sec.~\ref{sec4}.\par
In Appendix.~\ref{BAP}, we demonstrate gauging the non-onsite $\mathbb{Z}_2$ symmetry, $U_D$ resulting in the same emergent $2$-group symmetry.

\subsection{Gauging $\mathbb{Z}_2^A \times \mathbb{Z}_2^B$ : Non-invertible symmetry}\label{sec2.2}
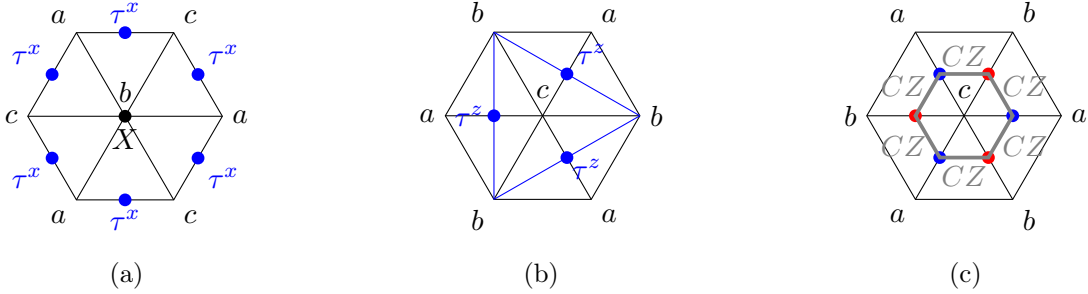
\begin{figure}[t]
\centering
\begin{subfigure}[t]{0.31\linewidth}\centering
\begin{tikzpicture}[scale=0.8]
\def\r{1.6}
\coordinate (v1) at (120:\r);
\coordinate (v2) at (60:\r);
\coordinate (v3) at (0:\r);
\coordinate (v4) at (-60:\r);
\coordinate (v5) at (-120:\r);
\coordinate (v6) at (180:\r);
\coordinate (c)  at (0,0);

\draw
  (v1)--(v2)--(v3)--(v4)--(v5)--(v6)--cycle;

\draw (v1)--(v4);
\draw (v2)--(v5);
\draw (v3)--(v6);

\node at ($(c)+(0,0.4)$) {$b$};
\fill (c) circle (3pt);
\node[below=1pt] at (c) {$X$};

\node at (v1) [above left] {$a$};
\node at (v2) [above right] {$c$};
\node at (v3) [right] {$a$};
\node at (v4) [below right] {$c$};
\node at (v5) [below left] {$a$};
\node at (v6) [left] {$c$};

\node at ($(v1)!0.5!(v2)$) [above] {$\color{blue}\tau^x$};
\node at ($(v2)!0.5!(v3)$) [above right] {$\color{blue}\tau^x$};
\node at ($(v3)!0.5!(v4)$) [below right] {$\color{blue}\tau^x$};
\node at ($(v4)!0.5!(v5)$) [below] {$\color{blue}\tau^x$};
\node at ($(v5)!0.5!(v6)$) [below left] {$\color{blue}\tau^x$};
\node at ($(v6)!0.5!(v1)$) [above left] {$\color{blue}\tau^x$};

\fill[blue] ($(v1)!0.5!(v2)$) circle (3pt);
\fill[blue] ($(v2)!0.5!(v3)$) circle (3pt);
\fill[blue] ($(v3)!0.5!(v4)$) circle (3pt);
\fill[blue] ($(v4)!0.5!(v5)$) circle (3pt);
\fill[blue] ($(v5)!0.5!(v6)$) circle (3pt);
\fill[blue] ($(v6)!0.5!(v1)$) circle (3pt);
\end{tikzpicture}
\caption{}
\label{Fig:B_type4a}
\end{subfigure}\hfill
\begin{subfigure}[t]{0.31\linewidth}\centering
\begin{tikzpicture}[scale=0.8]
\def\r{1.6}
\coordinate (v1) at (120:\r);
\coordinate (v2) at (60:\r);
\coordinate (v3) at (0:\r);
\coordinate (v4) at (-60:\r);
\coordinate (v5) at (-120:\r);
\coordinate (v6) at (180:\r);
\coordinate (c)  at (0,0);

\draw
  (v1)--(v2)--(v3)--(v4)--(v5)--(v6)--cycle;

\draw (v1)--(v4);
\draw (v2)--(v5);
\draw (v3)--(v6);

\draw (v1)--(v3) [blue];
\draw (v1)--(v5) [blue];
\draw (v3)--(v5) [blue];
\node at ($(c)+(0,0.4)$) {$c$};

\node at (v1) [above left] {$b$};
\node at (v2) [above right] {$a$};
\node at (v3) [right] {$b$};
\node at (v4) [below right] {$a$};
\node at (v5) [below left] {$b$};
\node at (v6) [left] {$a$};

\node at ($(v2)!0.5!(c)$) [above right] {$\color{blue}\tau^z$};
\node at ($(v4)!0.6!(c)$) [below right] {$\color{blue}\tau^z$};
\node at ($(v6)!0.5!(c)$) [left] {$\color{blue}\tau^z$};

\fill[blue] ($(v2)!0.5!(c)$) circle (3pt);
\fill[blue] ($(v4)!0.5!(c)$) circle (3pt);
\fill[blue] ($(v6)!0.5!(c)$) circle (3pt);

\end{tikzpicture}
\caption{}
\label{Fig:B_type4b}
\end{subfigure}
\hfill
\begin{subfigure}[t]{0.31\linewidth}\centering
\begin{tikzpicture}[scale=0.8]
\def\r{1.6}
\coordinate (v1) at (120:\r);
\coordinate (v2) at (60:\r);
\coordinate (v3) at (0:\r);
\coordinate (v4) at (-60:\r);
\coordinate (v5) at (-120:\r);
\coordinate (v6) at (180:\r);
\coordinate (c)  at (0,0);

\draw
  (v1)--(v2)--(v3)--(v4)--(v5)--(v6)--cycle;

\draw (v1)--(v4);
\draw (v2)--(v5);
\draw (v3)--(v6);

\node at ($(c)+(0,0.4)$) {$c$};

\node at (v1) [above left] {$a$};
\node at (v2) [above right] {$b$};
\node at (v3) [right] {$a$};
\node at (v4) [below right] {$b$};
\node at (v5) [below left] {$a$};
\node at (v6) [left] {$b$};

\node at ($(c)$) [above=15pt] {$\color{gray}CZ$};
\node at ($(c)+(1.0,0.5)$) {$\color{gray}CZ$};
\node at ($(c)+(-1.0,0.5)$) {$\color{gray}CZ$};
\node at ($(c)+(1.0,-0.5)$) {$\color{gray}CZ$};
\node at ($(c)+(-1.0,-0.5)$) {$\color{gray}CZ$};
\node at ($(c)$) [below=15pt] {$\color{gray}CZ$};

\fill[red] ($(v2)!0.5!(c)$) circle (3pt);
\fill[red] ($(v4)!0.5!(c)$) circle (3pt);
\fill[red] ($(v6)!0.5!(c)$) circle (3pt);
\fill[blue] ($(v1)!0.5!(c)$) circle (3pt);
\fill[blue] ($(v3)!0.5!(c)$) circle (3pt);
\fill[blue] ($(v5)!0.5!(c)$) circle (3pt);

\draw[gray, line width=1.4pt]
  ($(v1)!0.5!(c)$) --
  ($(v2)!0.5!(c)$) --
  ($(v3)!0.5!(c)$) --
  ($(v4)!0.5!(c)$) --
  ($(v5)!0.5!(c)$) --
  ($(v6)!0.5!(c)$) --
  cycle;
\end{tikzpicture}
\caption{}
\label{Fig:B_type4c}
\end{subfigure}

\caption{
(a) Gauss law $G_{B,b}$ given in~\eqref{type4:gauss2}.
(b) Flux term $B_B$ defined in~\eqref{type4:flatness2}.
(c) The product of~CZ terms in the Hamiltonian~\eqref{gauged model2 with type4}.
}
\label{Fig:B_type4}
\end{figure}

We next 
gauge $\mb^A$ and $\mb^B$
symmetries. Following the same procedure as the previous subsection, we introduce additional gauge d.o.f on links $\langle ac\rangle\in \Lambda_{ac}$ where $a\in \Lambda_a, c\in \Lambda_c$, and corresponding Pauli operators $\tau_{\langle ac\rangle}^x$ and $\tau_{\langle ac\rangle}^z$. We impose the following Gauss law constraints and the flatness condition (Fig.~\ref{Fig:B_type4a}): 
\begin{equation}
    G_{B,b}:=X_b\prod_{\langle ac\rangle\subset \langle abc \rangle}\tau_{\langle ac\rangle}^x = 1, \quad \forall b\in\Lambda_b, \label{type4:gauss2}
\end{equation}
\begin{equation}
    B_{B}(\gamma_b):=\prod_{\langle ac\rangle\subset \gamma_b}\tau_{\langle ac\rangle}^z =1, \quad \forall\gamma_b \label{type4:flatness2}.
\end{equation}
Here, $\gamma_b$ denotes a contractible loop on the sublattice $\Lambda_b$, and the product $\prod_{\langle ac\rangle\subset \gamma_b}$ 
is taken over all links
$\langle ac\rangle$ intersected by $\gamma$ (see Fig.~\ref{Fig:B_type4b}).
To proceed, we modify the local terms in the Hamiltonian~\eqref{gauged model1 with type4} so that they commute with the Gauss law \eqref{type4:gauss2}. After gauging, we arrive at the following gauged Hamiltonian:
\begin{equation}
\begin{aligned}
    H_2=&-\sum_{\langle bc\rangle \in \Lambda_{bc}}\sigma_{\langle bc\rangle}^z- \sum_{a\in\Lambda_a} \left[\prod_{\langle bc\rangle\subset \langle abc \rangle}\sigma_{\langle bc\rangle}^x \left(1 + \prod_{\langle bc\rangle \subset \langle abc\rangle}CZ_{bc}\prod_{\langle ac\rangle \subset \langle abc\rangle}CZ_{c\langle ac\rangle}\right)\right]  \\
    &-\sum_{\langle b_1,b_2\rangle}Z_{b_1}\tau_{\langle ac\rangle}^zZ_{b_2}- \sum_{b\in\Lambda_b} \left[\prod_{\langle ac\rangle\subset \langle abc \rangle}\tau_{\langle ac\rangle}^x  \big(1 + \prod_{\langle bc\rangle \subset \langle abc\rangle}CZ_{c\langle bc\rangle}\big)\right] \\
    &-\sum_{\langle c_1,c_2\rangle}Z_{c_1}Z_{c_2}- \sum_{c\in\Lambda_c}X_c\left(1 + \prod_{\langle bc\rangle \subset \langle abc\rangle}CZ_{b\langle bc\rangle}\prod_{\langle ac\rangle,\langle bc\rangle  \subset \langle abc\rangle}CZ_{\langle ac\rangle \langle bc\rangle}\right).
    \label{gauged model2 with type4}
\end{aligned}
\end{equation}
Here, $\prod_{\langle ac\rangle,\langle bc\rangle  \subset \langle abc\rangle}CZ_{\langle ac\rangle \langle bc\rangle}$ in last term corresponds to the product of  the CZ terms acting on two links $\langle ac\rangle, \langle bc\rangle$ as depicted in Fig.~\ref{Fig:B_type4c}. Note that unlike in Sec.~\ref{sec.2.1}, we do not solve the Gauss law here. Instead, the Gauss law \eqref{type4:gauss2} remains imposed. \par 
After gauging, the model~\eqref{gauged model2 with type4} has two $1$-form and two $0$-form $\mb$ symmetries whose symmetry operators are given by\footnote{Similar to $\eta_A^{(1)}$, $\eta_B^{(1)}$ is also topological due to the flatness conditions \eqref{type4:flatness2}.}
\begin{equation}
\begin{aligned}
    &\eta_{A}^{(1)}(\gamma_a)=\prod_{\langle bc\rangle\in\gamma_a}\sigma_{\langle bc\rangle}^z, \quad \eta_{B}^{(1)}(\gamma_b)=\prod_{\langle ac\rangle\in\gamma_b}\tau_{\langle ac\rangle}^z, \quad U_C=\prod_{c\in\Lambda_c}X_c, \\
    &U_{D,2}=B(\gamma_a, \gamma_b)\prod_{\langle bc\rangle}CCZ_{bc\langle bc\rangle}\prod_{c}CCZ_{\langle bc\rangle c\langle ac\rangle}, \label{non-invertible-1}
\end{aligned}
\end{equation}
where $\gamma_b$ represents a noncontractible loop on the sublattice $\Lambda_b$, and $CCZ_{\langle bc\rangle c\langle ac\rangle}$ does the CCZ  acting on one site $c$ and two links $\langle bc\rangle, \langle ac\rangle$. The CCZ  terms in the form of $U_{D,2}$ can be derived by modifying $U_{D,1}$ so that it commutes with the Gauss law \eqref{type4:gauss2}.
Also, we have introduced a phase factor~$B(\gamma_a,\gamma_b)$ which depends on the configuration of the~$1$-form loop operators, attached to the CCZ terms to ensure the commutativity between $U_{D,2}$ and other symmetry operators. More precisely, introducing states 
that label distinct homology classes of the $1$-form loops operators as  
$ \ket{(a_1, a_2), (b_1, b_2)} $ with the relation
\begin{eqnarray}
  \eta_A^{(1)} (\gamma_a)  \ket{(a_1, a_2), (b_1, b_2)} =
  \begin{cases}
        (-1)^{a_1} \ket{(a_1, a_2), (b_1, b_2)}~(\text{if }~\gamma_a~\text{goes around in the }~x\text{-direction}) \\
            (-1)^{a_2} \ket{(a_1, a_2), (b_1, b_2)}~(\text{if }~\gamma_a~\text{goes around in the }~y\text{-direction})
  \end{cases}
\end{eqnarray}
and similarly for the action of $\eta(\gamma_b)$, 
the phase factor $B(\gamma_a,\gamma_b)$ is defined by 
\begin{equation}
B(\gamma_a,\gamma_b) \ket{(a_1, a_2), (b_1, b_2)} 
= (-1)^{a_1 b_2 + a_2 b_1}
\ket{(a_1, a_2), (b_1, b_2)},
\qquad a_1,a_2,b_1,b_2 \in \{0,1\}.\label{28}
\end{equation}
%
%
%
With this phase factor, 
one can verify that all the symmetry operators~\eqref{non-invertible-1} commute. \par
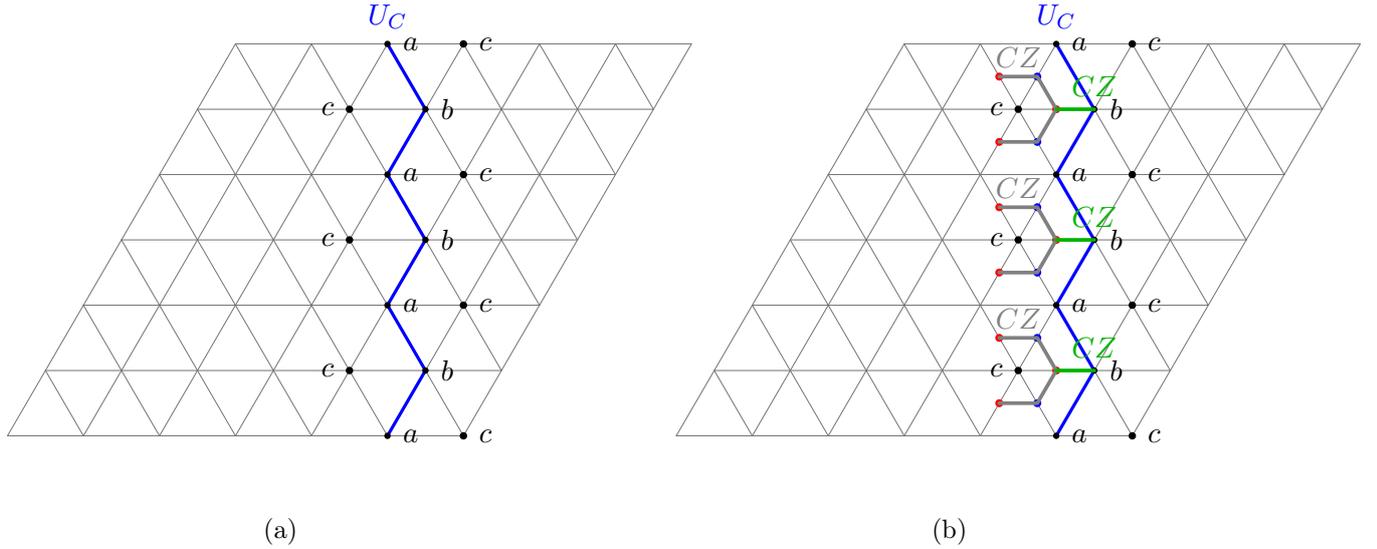
\begin{figure}
    \centering
    \begin{subfigure}{0.45\textwidth}
\centering
\begin{tikzpicture}[scale=1, line cap=round, line join=round] 
\def\h{0.8660254} 
\def\Nx{6}        
\def\Ny{6}        

\foreach \j in {0,...,\Ny}{
  \foreach \i in {0,...,\Nx}{
    \pgfmathsetmacro{\x}{\i + 0.5*\j}
    \pgfmathsetmacro{\y}{\h*\j}

    \ifnum\i<\Nx
      \pgfmathsetmacro{\xB}{(\i+1) + 0.5*\j}
      \pgfmathsetmacro{\yB}{\h*\j}
      \draw[black!55, line width=0.35pt] (\x,\y) -- (\xB,\yB);
    \fi

    \ifnum\j<\Ny
      \pgfmathsetmacro{\xB}{\i + 0.5*(\j+1)}
      \pgfmathsetmacro{\yB}{\h*(\j+1)}
      \draw[black!55, line width=0.35pt] (\x,\y) -- (\xB,\yB);
    \fi

    \ifnum\j<\Ny
      \ifnum\i>0
        \pgfmathsetmacro{\xB}{(\i-1) + 0.5*(\j+1)}
        \pgfmathsetmacro{\yB}{\h*(\j+1)}
        \draw[black!55, line width=0.35pt] (\x,\y) -- (\xB,\yB);
      \fi
    \fi
  }
}

\coordinate (Q1) at ({5 + 0.5*0},{\h*0}); 
\coordinate (Q2) at ({5 + 0.5*1},{\h*1}); 
\coordinate (Q3) at ({4 + 0.5*2},{\h*2}); 
\coordinate (Q4) at ({4 + 0.5*3},{\h*3}); 
\coordinate (Q5) at ({3 + 0.5*4},{\h*4}); 
\coordinate (Q6) at ({3 + 0.5*5},{\h*5}); 
\coordinate (Q7) at ({2 + 0.5*6},{\h*6}); 

\coordinate (P1) at ({5 + 0.5*2},{\h*0}); 
\coordinate (P2) at ({4 + 0.5*1},{\h*1}); 
\coordinate (P3) at ({5 + 0.5*2},{\h*2}); 
\coordinate (P4) at ({4 + 0.5*1},{\h*3}); 
\coordinate (P5) at ({5 + 0.5*2},{\h*4}); 
\coordinate (P6) at ({4 + 0.5*1},{\h*5}); 
\coordinate (P7) at ({5 + 0.5*2},{\h*6}); 

\draw[blue, line width=1.2pt] (Q1)--(Q2)--(Q3)--(Q4)--(Q5)--(Q6)--(Q7);

\foreach \q in {Q1,Q2,Q3,Q4,Q5,Q6,Q7}{
  \fill[black] (\q) circle (1.2pt);
}

\foreach \p in {P1,P2,P3,P4,P5,P6,P7}{
  \fill[black] (\p) circle (1.4pt);
}

\node[right=2pt] at (Q1) {$a$};
\node[right=2pt] at (Q2) {$b$};
\node[right=2pt] at (Q3) {$a$};
\node[right=2pt] at (Q4) {$b$};
\node[right=2pt] at (Q5) {$a$};
\node[right=2pt] at (Q6) {$b$};
\node[right=2pt] at (Q7) {$a$};

\node[right=2pt] at (P1) {$c$};
\node[left=2pt] at (P2) {$c$};
\node[right=2pt] at (P3) {$c$};
\node[left=2pt] at (P4) {$c$};
\node[right=2pt] at (P5) {$c$};
\node[left=2pt] at (P6) {$c$};
\node[right=2pt] at (P7) {$c$};

\node[above=2pt] at (Q7) {\textcolor{blue}{$U_C$}};
\end{tikzpicture}
\caption{}
\label{Fig:non-invertible1a}
\end{subfigure}
\hfill
\begin{subfigure}{0.45\textwidth}
\centering
\begin{tikzpicture}[scale=1, line cap=round, line join=round] 
\def\h{0.8660254} 
\def\Nx{6}        
\def\Ny{6}        

\foreach \j in {0,...,\Ny}{
  \foreach \i in {0,...,\Nx}{
    \pgfmathsetmacro{\x}{\i + 0.5*\j}
    \pgfmathsetmacro{\y}{\h*\j}

    \ifnum\i<\Nx
      \pgfmathsetmacro{\xB}{(\i+1) + 0.5*\j}
      \pgfmathsetmacro{\yB}{\h*\j}
      \draw[black!55, line width=0.35pt] (\x,\y) -- (\xB,\yB);
    \fi

    \ifnum\j<\Ny
      \pgfmathsetmacro{\xB}{\i + 0.5*(\j+1)}
      \pgfmathsetmacro{\yB}{\h*(\j+1)}
      \draw[black!55, line width=0.35pt] (\x,\y) -- (\xB,\yB);
    \fi

    \ifnum\j<\Ny
      \ifnum\i>0
        \pgfmathsetmacro{\xB}{(\i-1) + 0.5*(\j+1)}
        \pgfmathsetmacro{\yB}{\h*(\j+1)}
        \draw[black!55, line width=0.35pt] (\x,\y) -- (\xB,\yB);
      \fi
    \fi
  }
}

\coordinate (Q1) at ({5 + 0.5*0},{\h*0}); 
\coordinate (Q2) at ({5 + 0.5*1},{\h*1}); 
\coordinate (Q3) at ({4 + 0.5*2},{\h*2}); 
\coordinate (Q4) at ({4 + 0.5*3},{\h*3}); 
\coordinate (Q5) at ({3 + 0.5*4},{\h*4}); 
\coordinate (Q6) at ({3 + 0.5*5},{\h*5}); 
\coordinate (Q7) at ({2 + 0.5*6},{\h*6}); 

\coordinate (P1) at ({5 + 0.5*2},{\h*0}); 
\coordinate (P2) at ({4 + 0.5*1},{\h*1}); 
\coordinate (P3) at ({5 + 0.5*2},{\h*2}); 
\coordinate (P4) at ({4 + 0.5*1},{\h*3}); 
\coordinate (P5) at ({5 + 0.5*2},{\h*4}); 
\coordinate (P6) at ({4 + 0.5*1},{\h*5}); 
\coordinate (P7) at ({5 + 0.5*2},{\h*6}); 

\coordinate (S1) at ({5 - 0.5*2},{\h*6});
\coordinate (S2) at ({5 - 0.5*2},{\h*4});
\coordinate (S3) at ({5 - 0.5*2},{\h*2}); 
\coordinate (S4) at ({5 - 0.5*2},{\h*0}); 

\coordinate (T1) at ({5 - 0.5*2},{\h*0});
\coordinate (T3) at ({4 - 0.5*0},{\h*2});
\coordinate (T5) at ({3 + 0.5*2},{\h*4});

\draw[blue, line width=1.2pt] (Q1)--(Q2)--(Q3)--(Q4)--(Q5)--(Q6)--(Q7);

\foreach \q in {Q1,Q2,Q3,Q4,Q5,Q6,Q7}{
  \fill[black] (\q) circle (1.2pt);
}

\foreach \p in {P1,P2,P3,P4,P5,P6,P7}{
  \fill[black] (\p) circle (1.4pt);
}

\node[right=2pt] at (Q1) {$a$};
\node[right=2pt] at (Q2) {$b$};
\node[right=2pt] at (Q3) {$a$};
\node[right=2pt] at (Q4) {$b$};
\node[right=2pt] at (Q5) {$a$};
\node[right=2pt] at (Q6) {$b$};
\node[right=2pt] at (Q7) {$a$};

\node[right=2pt] at (P1) {$c$};
\node[left=2pt] at (P2) {$c$};
\node[right=2pt] at (P3) {$c$};
\node[left=2pt] at (P4) {$c$};
\node[right=2pt] at (P5) {$c$};
\node[left=2pt] at (P6) {$c$};
\node[right=2pt] at (P7) {$c$};

\node at ($(P6)$) [above=12pt] {$\textcolor{gray}{CZ}$};
\node at ($(P4)$) [above=12pt] {$\textcolor{gray}{CZ}$};
\node at ($(P2)$) [above=12pt] {$\textcolor{gray}{CZ}$};

\node[above=2pt] at (Q7) {\textcolor{blue}{$U_C$}};

\fill[blue] ($(P6)!0.5!(Q7)$) circle (1.5pt); 
\fill[red] ($(P6)!0.5!(Q6)$) circle (1.5pt); 
\fill[blue] ($(P6)!0.5!(Q5)$) circle (1.5pt); 
\fill[blue] ($(P4)!0.5!(Q5)$) circle (1.5pt); 
\fill[red] ($(P4)!0.5!(Q4)$) circle (1.5pt); 
\fill[blue] ($(P4)!0.5!(Q3)$) circle (1.5pt); 
\fill[blue] ($(P2)!0.5!(Q3)$) circle (1.5pt); 
\fill[red] ($(P2)!0.5!(Q2)$) circle (1.5pt); 
\fill[blue] ($(P2)!0.5!(Q1)$) circle (1.5pt); 

\fill[red] ($(S1)!0.5!(P6)$) circle (1.5pt); 
\fill[red] ($(S2)!0.5!(P4)$) circle (1.5pt); 
\fill[red] ($(S3)!0.5!(P2)$) circle (1.5pt); 

\fill[red] ($(T1)!0.5!(P2)$) circle (1.5pt); 
\fill[red] ($(T3)!0.5!(P4)$) circle (1.5pt); 
\fill[red] ($(T5)!0.5!(P6)$) circle (1.5pt); 

\draw[gray, line width=1.4pt]
  ($(T1)!0.5!(P2)$) --
  ($(P2)!0.5!(Q1)$) --
   ($(P2)!0.5!(Q2)$)--
   ($(P2)!0.5!(Q3)$)--
   ($(S3)!0.5!(P2)$) ;
 \draw[gray, line width=1.4pt] 
   ($(T3)!0.5!(P4)$)--
   ($(P4)!0.5!(Q3)$)--
   ($(P4)!0.5!(Q4)$) --
   ($(P4)!0.5!(Q5)$) --
   ($(S2)!0.5!(P4)$) ;
\draw[gray, line width=1.4pt]
    ($(T5)!0.5!(P6)$)--
    ($(P6)!0.5!(Q5)$)--
   ($(P6)!0.5!(Q6)$)--
    ($(P6)!0.5!(Q7)$)--
    ($(S1)!0.5!(P6)$);
    
\draw[green!70!black, line width=1.4pt] (R2)--(Q2);
\draw[green!70!black, line width=1.4pt] (R4)--(Q4);
\draw[green!70!black, line width=1.4pt] (R6)--(Q6);

\node[above=1pt] at (Q2) {$\textcolor{green!70!black}{CZ}$};
\node[above=1pt] at (Q4) {$\textcolor{green!70!black}{CZ}$};
\node[above=1pt] at (Q6) {$\textcolor{green!70!black}{CZ}$};
\end{tikzpicture}
\caption{}
\label{Fig:non-invertible1b}
\end{subfigure}
\caption{(a) A configuration with a $\mb^C$ defect (blue line) inserted
(b)~Action of $U_{D,2}$ around the~$\mb^C$ defect}
\end{figure}

%
The situation drastically changes in the presence of a symmetry defect, exhibiting nontrivial algebraic relations between the operators. To see this, 
we insert a $\mb^C$ symmetry defect along the line segment connecting $a$ and $b$ as shown in Fig.~\ref{Fig:non-invertible1a}
and see how the operator $U_{D,2}$ acts on other symmetry operators. Similar to the previous subsection, 
the defect is implemented by imposing a twisted boundary condition along a branch cut:
\begin{equation}
    -Z_c^{\text{left}}Z_c^{\text{right}} \to +Z_c^{\text{left}}Z_c^{\text{right}},\label{twbdd}
\end{equation}
namely, every Ising coupling between neighboring $C$-sublattice, denoted by $ Z_c^{\text{left}}$ and $ Z_c^{\text{right}}$, 
spins whose bond crosses the defect acquires an additional minus sign.
 %
%
 Focusing on the sector $\ket{(a_1, a_2), (b_1, b_2)}=\ket{(0,0), (0, 0)}$\footnote{The case of other sectors is similarly discussed.},  
 and following the similar argument around~\eqref{21}, 
 the operator~$U_{D,2}$ acts around the defect as
\begin{equation}
    U_{D,2}|_{\text{defect}} = \prod_{\langle bc \rangle}CZ_{b\langle bc \rangle}\prod_{c}CZ_{\langle bc \rangle\langle ac \rangle}.
\end{equation}
We investigate how $  U_{D,2}|_{\text{defect}}$ acts on the following product of the Pauli X operators around a hexagon in the vicinity of the defect:
\begin{eqnarray*}
    \prod_{\langle bc\rangle\subset \langle abc \rangle}\sigma_{\langle bc\rangle}^x,\quad \prod_{\langle ac\rangle\subset \langle abc \rangle}\tau_{\langle ac\rangle}^x
\end{eqnarray*}
After some algebra, one can show that 
%
%
%
\begin{equation}
\begin{tikzpicture}[baseline=(current bounding box.center),scale=0.6]
\def\r{1.6}
\coordinate (v1) at (120:\r);
\coordinate (v2) at (60:\r);
\coordinate (v3) at (0:\r);
\coordinate (v4) at (-60:\r);
\coordinate (v5) at (-120:\r);
\coordinate (v6) at (180:\r);
\coordinate (c)  at (0,0);

\draw
  (v1)--(v2)--(v3)--(v4)--(v5)--(v6)--cycle;

\draw (v1)--(v4);
\draw (v2)--(v5);
\draw (v3)--(v6);

\draw [blue, line width = 1.2pt] (v2)--(c)--(v4);

\node at ($(c)+(0,0.4)$) {$a$};

\node at (v1) [above left] {$c$};
\node at (v2) [above right] {$b$};
\node at (v3) [right] {$c$};
\node at (v4) [below right] {$b$};
\node at (v5) [below left] {$c$};
\node at (v6) [left] {$b$};

\node at ($(v1)!0.5!(v2)$) [above] {$\color{red}\sigma^x$};
\node at ($(v2)!0.5!(v3)$) [above right] {$\color{red}\sigma^x$};
\node at ($(v3)!0.5!(v4)$) [below right] {$\color{red}\sigma^x$};
\node at ($(v4)!0.5!(v5)$) [below] {$\color{red}\sigma^x$};
\node at ($(v5)!0.5!(v6)$) [below left] {$\color{red}\sigma^x$};
\node at ($(v6)!0.5!(v1)$) [above left] {$\color{red}\sigma^x$};

\fill[red] ($(v1)!0.5!(v2)$) circle (3pt);
\fill[red] ($(v2)!0.5!(v3)$) circle (3pt);
\fill[red] ($(v3)!0.5!(v4)$) circle (3pt);
\fill[red] ($(v4)!0.5!(v5)$) circle (3pt);
\fill[red] ($(v5)!0.5!(v6)$) circle (3pt);
\fill[red] ($(v6)!0.5!(v1)$) circle (3pt);

\end{tikzpicture}
\;\xrightarrow{\ U_{D,2}|_{\text{defect}}\ }\;
\begin{tikzpicture}[baseline=(current bounding box.center),scale=0.6]
\def\r{1.6}
\coordinate (v1) at (120:\r);
\coordinate (v2) at (60:\r);
\coordinate (v3) at (0:\r);
\coordinate (v4) at (-60:\r);
\coordinate (v5) at (-120:\r);
\coordinate (v6) at (180:\r);
\coordinate (c)  at (0,0);
\coordinate (d1) at ($(v3)!0.5!(c)$);

\draw (v1) -- ($(v1)+(-1.5,0)$);
\draw (v5) -- ($(v5)+(-1.5,0)$);
\fill[blue] ($(v1)+(-0.75,0)$) circle (3pt);
\fill[blue] ($(v5)+(-0.75,0)$) circle (3pt);
\node at ($(v1)+(-0.75,0)$) [above] {$\color{blue}{\tau^z}$};
\node at ($(v5)+(-0.75,0)$) [below] {$\color{blue}{\tau^z}$};

\draw (v1)--(v2)--(v3)--(v4)--(v5)--(v6)--cycle;
\draw [blue, line width = 1.2pt] (v2)--(c)--(v4);

\draw (v1)--(v4);
\draw (v2)--(v5);
\draw (v3)--(v6);

\node at ($(c)+(0,0.4)$) {$a$};

\node at (v1) [above] {$c$};
\node at (v2) [above right] {$b$};
\node at (v3) [right] {$c$};
\node at (v4) [below right] {$b$};
\node at (v5) [below] {$c$};
\node at (v6) [left] {$b$};

\node at ($(v1)!0.5!(v2)$) [above] {$\color{red}\sigma^x$};
\node at ($(v2)!0.5!(v3)$) [above right] {$\color{red}\sigma^x$};
\node at ($(v3)!0.5!(v4)$) [below right] {$\color{red}\sigma^x$};
\node at ($(v4)!0.5!(v5)$) [below] {$\color{red}\sigma^x$};
\node at ($(v5)!0.5!(v6)$) [left] {$\color{red}\sigma^x$};
\node at ($(v6)!0.5!(v1)$) [left] {$\color{red}\sigma^x$};

\fill[red] ($(v1)!0.5!(v2)$) circle (3pt);
\fill[red] ($(v2)!0.5!(v3)$) circle (3pt);
\fill[red] ($(v3)!0.5!(v4)$) circle (3pt);
\fill[red] ($(v4)!0.5!(v5)$) circle (3pt);
\fill[red] ($(v5)!0.5!(v6)$) circle (3pt);
\fill[red] ($(v6)!0.5!(v1)$) circle (3pt);

\fill[blue] (d1) circle (3pt);
\node at (d1) [above] {$\color{blue}{\tau^z}$};

\end{tikzpicture}, \label{sec2.2:action1}
\end{equation}
\begin{equation}
\begin{tikzpicture}[baseline=(current bounding box.center),scale=0.6]
\def\r{1.6}
\coordinate (v1) at (120:\r);
\coordinate (v2) at (60:\r);
\coordinate (v3) at (0:\r);
\coordinate (v4) at (-60:\r);
\coordinate (v5) at (-120:\r);
\coordinate (v6) at (180:\r);
\coordinate (c)  at (0,0);

\draw
  (v1)--(v2)--(v3)--(v4)--(v5)--(v6)--cycle;

\draw (v1)--(v4);
\draw (v2)--(v5);
\draw (v3)--(v6);

\draw [blue, line width = 1.2pt] (v1)--(c)--(v5);

\node at ($(c)+(0,0.4)$) {$b$};

\node at (v1) [above left] {$a$};
\node at (v2) [above right] {$c$};
\node at (v3) [right] {$a$};
\node at (v4) [below right] {$c$};
\node at (v5) [below left] {$a$};
\node at (v6) [left] {$c$};

\node at ($(v1)!0.5!(v2)$) [above] {$\color{blue}\tau^x$};
\node at ($(v2)!0.5!(v3)$) [above right] {$\color{blue}\tau^x$};
\node at ($(v3)!0.5!(v4)$) [below right] {$\color{blue}\tau^x$};
\node at ($(v4)!0.5!(v5)$) [below] {$\color{blue}\tau^x$};
\node at ($(v5)!0.5!(v6)$) [below left] {$\color{blue}\tau^x$};
\node at ($(v6)!0.5!(v1)$) [above left] {$\color{blue}\tau^x$};

\fill[blue] ($(v1)!0.5!(v2)$) circle (3pt);
\fill[blue] ($(v2)!0.5!(v3)$) circle (3pt);
\fill[blue] ($(v3)!0.5!(v4)$) circle (3pt);
\fill[blue] ($(v4)!0.5!(v5)$) circle (3pt);
\fill[blue] ($(v5)!0.5!(v6)$) circle (3pt);
\fill[blue] ($(v6)!0.5!(v1)$) circle (3pt);

\end{tikzpicture}
\;\xrightarrow{\ U_{D,2}|_{\text{defect}}\ }\;
\begin{tikzpicture}[baseline=(current bounding box.center),scale=0.6]
\def\r{1.6}
\coordinate (v1) at (120:\r);
\coordinate (v2) at (60:\r);
\coordinate (v3) at (0:\r);
\coordinate (v4) at (-60:\r);
\coordinate (v5) at (-120:\r);
\coordinate (v6) at (180:\r);
\coordinate (c)  at (0,0);

\draw
  (v1)--(v2)--(v3)--(v4)--(v5)--(v6)--cycle;

\draw (v1)--(v4);
\draw (v2)--(v5);
\draw (v3)--(v6);

\draw [blue, line width = 1.2pt] (v1)--(c)--(v5);

\node at ($(c)+(0,0.4)$) {$b$};

\node at (v1) [above left] {$a$};
\node at (v2) [above right] {$c$};
\node at (v3) [right] {$a$};
\node at (v4) [below right] {$c$};
\node at (v5) [below left] {$a$};
\node at (v6) [left] {$c$};

\node at ($(v1)!0.5!(v2)$) [above] {$\color{blue}\tau^x$};
\node at ($(v2)!0.5!(v3)$) [above right] {$\color{blue}\tau^x$};
\node at ($(v3)!0.5!(v4)$) [below right] {$\color{blue}\tau^x$};
\node at ($(v4)!0.5!(v5)$) [below] {$\color{blue}\tau^x$};
\node at ($(v5)!0.5!(v6)$) [below left] {$\color{blue}\tau^x$};
\node at ($(v6)!0.5!(v1)$) [above left] {$\color{blue}\tau^x$};

\fill[blue] ($(v1)!0.5!(v2)$) circle (3pt);
\fill[blue] ($(v2)!0.5!(v3)$) circle (3pt);
\fill[blue] ($(v3)!0.5!(v4)$) circle (3pt);
\fill[blue] ($(v4)!0.5!(v5)$) circle (3pt);
\fill[blue] ($(v5)!0.5!(v6)$) circle (3pt);
\fill[blue] ($(v6)!0.5!(v1)$) circle (3pt);

\fill[red] ($(v6)!0.5!(c)$) circle (3pt);
\node at ($(v6)!0.5!(c)$) [above] {$\color{red}\sigma^z$};

\end{tikzpicture} .\label{sec2.2:action2}
\end{equation}
Here, the blue line indicates the defect line. 
Note that, under the action of $U_{D,2}|_{\text{defect}}$, the  product of $\sigma^x$ or $\tau^x$ around a hexagon picks up additional $\tau^Z$ and $\sigma^Z$ factors in the vicinity of the defect. Away from the defect,~$U_{D,2}|_{\text{defect}}$ also acts nontrivially on these products, but this action is canceled by the variation of the adjacent plaquette operators. Hence, multiplying the operator $\prod_{\langle bc\rangle\subset \langle abc \rangle}\sigma_{\langle bc\rangle}^x$ and $\prod_{\langle ac\rangle\subset \langle abc \rangle}\tau_{\langle ac\rangle}^x$ on entire lattice, we have
\begin{eqnarray}
     \prod_{a\in\Lambda_a}\left(\prod_{\langle bc\rangle\subset \langle abc \rangle}\sigma_{\langle bc\rangle}^x\right) \;\xrightarrow{\ U_{D,2}|_{\text{defect}}\ }\;
      \prod_{a\in\Lambda_a}\left(\prod_{\langle bc\rangle\subset \langle abc \rangle}\sigma_{\langle bc\rangle}^x\right) \times \eta^{(1)}_B(\gamma_b^{\text{defect}})\nonumber\\
      \prod_{b\in\Lambda_b}\left(\prod_{\langle ac\rangle\subset \langle abc \rangle}\tau_{\langle ac\rangle}^x\right)\;\xrightarrow{\ U_{D,2}|_{\text{defect}}\ }\; \prod_{b\in\Lambda_b}\left(\prod_{\langle ac\rangle\subset \langle abc \rangle}\tau_{\langle ac\rangle}^x\right)\times \eta^{(1)}_A(\gamma_a^{\text{defect}})
\end{eqnarray}
where $\eta^{(1)}_A(\gamma_a^{\text{defect}})$ and $\eta^{(1)}_B(\gamma_b^{\text{defect}})$ denote the $1$-form symmetry operators supported along the defect.
However, 
the product of the plaquette operators on entire lattice equals identity, namely, 
\begin{equation*}
    \prod_{a\in\Lambda_a}\left(\prod_{\langle bc\rangle\subset \langle abc \rangle}\sigma_{\langle bc\rangle}^x\right) =1, \quad \prod_{b\in\Lambda_b}\left(\prod_{\langle ac\rangle\subset \langle abc \rangle}\tau_{\langle ac\rangle}^x\right)=1. 
\end{equation*}
 Hence, in order for $  U_{D,2}|_{\text{defect}} $ to be well-defined,
it is legitimate to modify it as
%
\begin{equation}
    \widetilde{U}_{D,2}|_{\text{defect}} := \frac{1}{2}C_A(\gamma_a^{\text{defect}})C_B(\gamma_b^{\text{defect}})
    \times U_{D,2}|_{\text{defect}}. \label{sec2.2:non-invetible}
\end{equation}
Here, $C_A(\gamma_a^{\text{defect}})$ and $C_B(\gamma_b^{\text{defect}})$ are referred to as  \emph{condensation operators}~\cite{Roumpedakis:2022aik, Choi:2024rjm} associated with $1$-form symmetry operators $\eta_A^{(1)}$ and $\eta_B^{(1)}$, supported along the defect. These are defined as 
\begin{eqnarray}
    C_A(\gamma_a^{\text{defect}}) \vcentcolon= 1+\eta_A^{(1)}(\gamma_a^{\text{defect}}), \quad C_B(\gamma_b^{\text{defect}}) \vcentcolon= 1+\eta_B^{(1)}(\gamma_b^{\text{defect}}).
\end{eqnarray}
From this representation \eqref{sec2.2:non-invetible}, we obtain the following fusion rules:
%
\begin{equation}
    \begin{aligned}
        &\widetilde{U}_{D,2}|_{\text{defect}} \times \widetilde{U}_{D,2}|_{\text{defect}} = C_A(\gamma_a^{\text{defect}})C_B(\gamma_b^{\text{defect}}), \\
        &\widetilde{U}_{D,2}|_{\text{defect}} \times \eta_{\alpha}^{(1)} = \eta_{\alpha}^{(1)} \times \widetilde{U}_{D,2}|_{\text{defect}} =\widetilde{U}_{D,2}|_{\text{defect}}\ (\alpha=A,B),\label{36}
    \end{aligned}
\end{equation}
indicating $\widetilde{U}_{D,2}|_{\text{defect}}$ is non-invertible. 
\par
To summarize the argument, 
in the absence of the $0$-form symmetry defect, symmetry operators form a direct product structure consisting of two $\mb$ $0$-form symmetries and two $\mb$~$1$-form symmetries \eqref{non-invertible-1}. By contrast, in the presence of a symmetry defect associated with one of the~$0$-form symmetries, the other $0$-form symmetry becomes non-invertible along the defect. 
It is worth addressing that along the defect, the fusion rule of the non-invertible symmetry operator resembles the one found in  the so-called Kennedy–Tasaki (KT) transformation, a mapping between a symmetry breaking phase to an SPT phase protected by $\mathbb{Z}_2 \times \mathbb{Z}_2$ symmetry~\cite{Kennedy:1992ifl, Kennedy:1992tke, Li:2023ani, Oshikawa:1992smv}. Such a connection will become more transparent from a field-theoretical perspective, as we will discuss in the next section~(Sec.~\ref{sec4}).
%
%
%


\subsection{Gauging $\mathbb{Z}_2^A \times \mathbb{Z}_2^B \times \mathbb{Z}_2^C$ : 2-Representation category}\label{sec2.3}
Finally, we gauge three internal symmetries to explore emergent symmetries.   
The gauging is implemented by imposing the following Gauss law constraint and flatness condition jointly with~\eqref{type4:gauss1}~\eqref{type4:flatness01}~\eqref{type4:gauss2}~\eqref{type4:flatness2}:
\begin{equation}
    G_{C,c}:=X_c\prod_{\langle ab\rangle\subset \langle abc \rangle}\mu_{\langle ab\rangle}^x = 1, \quad \forall c\in\Lambda_c, \label{type4:gauss3}
\end{equation}
\begin{equation}
    B_{C}(\gamma_c):=\prod_{\langle ab\rangle\subset \gamma_c}\mu_{\langle ab\rangle}^z =1, \quad \forall\gamma_c \label{type4:flatness3}.
\end{equation}
Here, $\gamma_c$ denotes a contractible loop on the sublattice $\Lambda_c$, and the product $\prod_{\langle ab\rangle\subset \gamma_c}$ is taken over all links $\langle ab\rangle$ intersected by $\gamma_c$. The gauge-invariant Hamiltonian is then given by
\begin{equation}
\begin{aligned}
    H_3=-\sum_{\langle bc\rangle \in \Lambda_{bc}}\sigma_{\langle bc\rangle}^z- \sum_{a\in\Lambda_a} h_a  
    -\sum_{\langle b_1,b_2\rangle}Z_{b_1}\tau_{\langle ac\rangle}^zZ_{b_2}- \sum_{b\in\Lambda_b} h_b 
    -\sum_{\langle c_1,c_2\rangle}Z_{c_1}\mu_{\langle ab\rangle}^zZ_{c_2}- \sum_{c\in\Lambda_c} h_c,
    \label{gauged model3 with type4}
\end{aligned}
\end{equation}
where
\begin{equation}
    \begin{aligned}
        h_a &= \prod_{\langle bc\rangle\subset \langle abc \rangle}\sigma_{\langle bc\rangle}^x \Big(1 + \prod_{\langle bc\rangle \subset \langle abc\rangle}CZ_{bc}\prod_{\langle ac\rangle \subset \langle abc\rangle}CZ_{c\langle ac\rangle}\prod_{\langle ab\rangle \subset \langle abc\rangle}CZ_{b\langle ab\rangle}\prod_{\langle ab\rangle, \langle ac\rangle \subset \langle abc\rangle}CZ_{\langle ab\rangle\langle ac\rangle}\Big) \\
         h_b &= \prod_{\langle ac\rangle\subset \langle abc \rangle}\tau_{\langle ac\rangle}^x  \Big(1 + \prod_{\langle bc\rangle \subset \langle abc\rangle}CZ_{c\langle bc\rangle}\prod_{\langle ab\rangle,\langle bc\rangle  \subset \langle abc\rangle}CZ_{\langle ab\rangle \langle bc\rangle}\Big) \\
         h_c &= \prod_{\langle ab\rangle\subset \langle abc \rangle}\mu_{\langle ab\rangle}^x\Big(1 + \prod_{\langle bc\rangle \subset \langle abc\rangle}CZ_{b\langle bc\rangle}\prod_{\langle ac\rangle,\langle bc\rangle  \subset \langle abc\rangle}CZ_{\langle ac\rangle \langle bc\rangle}\Big).
    \end{aligned}
\end{equation}
\par
\begin{figure}
    \centering
    \begin{tikzpicture}[baseline=(current bounding box.center),scale=0.8]
\def\r{1.6}
\coordinate (v1) at (120:\r);
\coordinate (v2) at (60:\r);
\coordinate (v3) at (0:\r);
\coordinate (v4) at (-60:\r);
\coordinate (v5) at (-120:\r);
\coordinate (v6) at (180:\r);
\coordinate (c)  at (0,0);

\draw
  (v1)--(v2)--(v3)--(v4)--(v5)--(v6)--cycle;

\draw (v1)--(v4);
\draw (v2)--(v5);
\draw (v3)--(v6);

\node at ($(v1)!0.5!(v2)$) [above] {$\color{green!80!black}\alpha^x$};
\node at ($(v2)!0.5!(v3)$) [above right] {$\color{green!80!black}\alpha^x$};
\node at ($(v3)!0.5!(v4)$) [below right] {$\color{green!80!black}\alpha^x$};
\node at ($(v4)!0.5!(v5)$) [below] {$\color{green!80!black}\alpha^x$};
\node at ($(v5)!0.5!(v6)$) [below left] {$\color{green!80!black}\alpha^x$};
\node at ($(v6)!0.5!(v1)$) [above left] {$\color{green!80!black}\alpha^x$};

\fill [green!80!black] ($(v1)!0.5!(v2)$) circle (3pt);
\fill [green!80!black] ($(v2)!0.5!(v3)$) circle (3pt);
\fill [green!80!black] ($(v3)!0.5!(v4)$) circle (3pt);
\fill [green!80!black] ($(v4)!0.5!(v5)$) circle (3pt);
\fill [green!80!black] ($(v5)!0.5!(v6)$) circle (3pt);
\fill [green!80!black] ($(v6)!0.5!(v1)$) circle (3pt);

\end{tikzpicture}
\;$\xrightarrow{\quad \Bar{S}TS \quad }$\;
\begin{tikzpicture}[baseline=(current bounding box.center),scale=0.8]
\def\r{1.6}
\coordinate (v1) at (120:\r);
\coordinate (v2) at (60:\r);
\coordinate (v3) at (0:\r);
\coordinate (v4) at (-60:\r);
\coordinate (v5) at (-120:\r);
\coordinate (v6) at (180:\r);
\coordinate (c)  at (0,0);

\draw
  (v1)--(v2)--(v3)--(v4)--(v5)--(v6)--cycle;

\draw (v1)--(v4);
\draw (v2)--(v5);
\draw (v3)--(v6);

\node at ($(c)$) [above=15pt] {$\color{gray}CZ$};
\node at ($(c)+(1.0,0.5)$) {$\color{gray}CZ$};
\node at ($(c)+(-1.0,0.5)$) {$\color{gray}CZ$};
\node at ($(c)+(1.0,-0.5)$) {$\color{gray}CZ$};
\node at ($(c)+(-1.0,-0.5)$) {$\color{gray}CZ$};
\node at ($(c)$) [below=15pt] {$\color{gray}CZ$};

\fill[red] ($(v2)!0.5!(c)$) circle (3pt);
\fill[red] ($(v4)!0.5!(c)$) circle (3pt);
\fill[red] ($(v6)!0.5!(c)$) circle (3pt);
\fill[blue] ($(v1)!0.5!(c)$) circle (3pt);
\fill[blue] ($(v3)!0.5!(c)$) circle (3pt);
\fill[blue] ($(v5)!0.5!(c)$) circle (3pt);

\draw[gray, line width=1.4pt]
  ($(v1)!0.5!(c)$) --
  ($(v2)!0.5!(c)$) --
  ($(v3)!0.5!(c)$) --
  ($(v4)!0.5!(c)$) --
  ($(v5)!0.5!(c)$) --
  ($(v6)!0.5!(c)$) --
  cycle;
\end{tikzpicture}
    \caption{The action of $\Bar{S}TS$ on local operators, where $\alpha=\sigma, \tau,\mu$ labels the three types of link variables.}
    \label{Fig: type4 STS}
\end{figure}
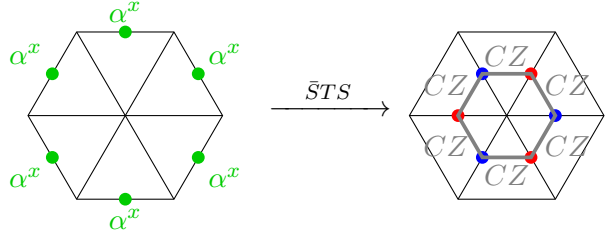

 The gauged Hamiltonian~\eqref{gauged model3 with type4} admits 
 three $1$-form symmetries, which are dual symmetries obtained by gauging three $\mathbb{Z}_2$ $0$-form symmetries. These are
generated by
 \begin{equation}
\begin{aligned}
    \eta_{A}^{(1)}(\gamma_a)=\prod_{\langle bc\rangle\in\gamma_a}\sigma_{\langle bc\rangle}^z, \quad \eta_{B}^{(1)}(\gamma_b)=\prod_{\langle ac\rangle\in\gamma_b}\tau_{\langle ac\rangle}^z, \quad 
    \eta_{C}^{(1)}(\gamma_c)=\prod_{\langle ab\rangle\in\gamma_c}\mu_{\langle ab\rangle}^z.\label{1-form sym1}
\end{aligned}
\end{equation}
Here, we have introduced $\gamma_i$ as a noncontractible loop on the sublattice $\Lambda_i$ $(i=a,b,c)$. 
 Regarding the $0$-form symmetry symmetry operator that is not gauged, $\mb^D$, 
 following the similar argument around~\eqref{non-invertible-1}$\sim$\eqref{28}, naively, it can be described by
the following form consisting of product of~CCZ gates and the phase factor:
 \begin{equation}
     U_{D,3} = B(\gamma_a, \gamma_b)\prod_{\langle bc\rangle}CCZ_{bc\langle bc\rangle}\prod_{c}CCZ_{\langle bc\rangle c\langle ac\rangle}\prod_{b}CCZ_{\langle bc\rangle b\langle ab\rangle}.
 \end{equation}
 Here, $B(\gamma_a,\gamma_b)$ is defined in~\eqref{28}. 
 However, this operator does not commute with the Gauss laws~\eqref{type4:gauss2} and \eqref{type4:gauss3}, and therefore is \emph{not} gauge-invariant. To circumvent this issue, 
 following the analogous argument outlined in~\cite{Oishi:2026sow}, 
 we modify it so that it is dressed with an additional operator. To wit, 
 \begin{equation}
     U_{D,3} \longrightarrow \mathcal{D}:=U_{D,3}\mathcal{D}_{\Bar{S}TS}.
 \end{equation}
 Here, $\mathcal{D}_{\Bar{S}TS}$ is an operator implementing topological manipulations $\Bar{S}TS$, where $S, T$ and $\Bar{S}$ corresponds to gauging three $\mb$ $1$-form symmetries \eqref{1-form sym1}, stacking of a  $\mb^A\times\mb^B\times\mb^C$ SPT phase $(-1)^{\int A\cup B\cup C}$, and gauging $\mb^A\times\mb^B\times\mb^C$ ~$0$-form symmetries symmetry, respectively. Deferring the details to Appendix~\ref{appendixA}, the action of $\mathcal{D}_{\Bar{S}TS}$ on local operators reads 
 \begin{equation}
 \begin{aligned}
     &\sigma_{\langle bc\rangle}^z \rightarrow  \sigma_{\langle bc\rangle}^z , \quad  \tau_{\langle ac\rangle}^z \rightarrow \tau_{\langle ac\rangle}^z , \quad \mu_{\langle ab\rangle}^z \rightarrow \mu_{\langle ab\rangle}^z, \\ 
     &\prod_{\langle bc\rangle\subset \langle abc \rangle}\sigma_{\langle bc\rangle}^x \rightarrow \prod_{\langle bc\rangle\subset \langle abc \rangle}\sigma_{\langle bc\rangle}^x\prod_{\langle ab\rangle,\langle ac\rangle\subset \langle abc \rangle}CZ_{\langle ab\rangle,\langle ac\rangle} \\ 
     &\prod_{\langle ac\rangle\subset \langle abc \rangle}\tau_{\langle ac\rangle}^x \rightarrow \prod_{\langle ac\rangle\subset \langle abc \rangle}\tau_{\langle ac\rangle}^x\prod_{\langle ab\rangle,\langle bc\rangle\subset \langle abc \rangle}CZ_{\langle ab\rangle,\langle bc\rangle} \\
     &\prod_{\langle ab\rangle\subset \langle abc \rangle}\mu_{\langle ab\rangle}^x \rightarrow \prod_{\langle ab\rangle\subset \langle abc \rangle}\mu_{\langle ab\rangle}^x\prod_{\langle ac\rangle,\langle bc\rangle\subset \langle abc \rangle}CZ_{\langle ac\rangle,\langle bc\rangle}.\label{44}
 \end{aligned}
 \end{equation} \label{action of STS}
See Fig.~\ref{Fig: type4 STS} for a graphical illustration of the last three lines of~\eqref{44}. One can verify that the modified symmetry operator $\mathcal{D}$ commutes with the Hamiltonian and the Gauss laws, ensuring the gauge invariance. 
\par
After the modification, the operator $\mathcal{D}$ becomes non-invertible, owing to the fact that 
$\mathcal{D}_{\Bar{S}TS}$ involves a gauging operation. Indeed, one finds the following fusion rule
\begin{equation}
    \mathcal{D}_{\Bar{S}TS} \times \mathcal{D}_{\Bar{S}TS} = \prod_{\alpha=A,B,C}\left[\frac{1}{2}\left(1+\eta_{\alpha}^{(1), x}\right)\left(1+\eta_{\alpha}^{(1), y}\right)\right],
\end{equation}
where $\eta_\alpha^{(1),x}$ and $\eta_\alpha^{(1),y}$ represents$1$-form symmetry operator along the $x$-direction and $y$-direction. respectively.
A detailed derivation of this fusion rule is discussed in Sec.~\ref{sec4}.
Incorporating all the emergent symmetries, we obtain the following fusion rules:
\begin{equation}
    \begin{aligned}
        &\mathcal{D} \times \mathcal{D} = C_1, \\
        &\mathcal{D} \times \eta_\alpha^{(1),i} = \eta_\alpha^{(1),i} \times \mathcal{D} = \mathcal{D} \quad (\alpha=A,B,C\quad i=x,y),\label{47}
    \end{aligned}
\end{equation}
where $C_1$ corresponds to the condensation defects \cite{Roumpedakis:2022aik, Choi:2024rjm}, associated with three $1$-form symmetries~\eqref{1-form sym1}:
\begin{equation}
    C_1 :=\prod_{\alpha=A,B,C}\frac{1}{2}\left(1+\eta_{\alpha}^{(1), x}\right)\left(1+\eta_{\alpha}^{(1), y}\right).
\end{equation}
This non-invertible symmetry is described by 
a special fusion category, \emph{higher fusion category}, a systematic framework to classify non-invertible symmetries in more than one spatial dimension~\cite{douglas2018fusion,Bartsch:2022mpm,Bartsch:2022ytj,Decoppet:2024htz, Bhardwaj:2022yxj, Bhardwaj:2022lsg}. In particular, 
the non-invertible symmetry that we find is described by
the \emph{fusion 2-category} 2-Rep$(\Gamma)$~\cite{Bartsch:2022mpm,Bartsch:2022ytj,Bhardwaj:2022lsg}. It is a 2-category of 2-representations of the 2-group~$\Gamma$ discussed in Sec.~\ref{sec.2.1}. \par
Note that emergent symmetry found here is dual to the one obtained by gauging one internal symmetry, which is 2-group (Sec.~\ref{sec.2.1}). 
    This relation can be understood by choosing a complementary gauging of the original type-IV anomalous theory $\mathcal T$. First, gauging $\mathbb Z_2^D$ produces a theory $\mathcal T_D$ with a (central) 2-group symmetry $\Gamma$. Its 0-form component is $G=\mathbb Z_2^A\times\mathbb Z_2^B\times\mathbb Z_2^C$, and its 1-form component is the dual symmetry $\widehat{\mathbb Z}_2^D$. The action of $G$ on the 1-form component is trivial, while the Postnikov class is nontrivial: $\beta=A\cup B\cup C.$ Accordingly, the background 2-form field~$D_2$ obeys $\delta D_2=A\cup B\cup C$. Full gauging of $\Gamma$ can be performed sequentially. Gauging its 1-form component first undoes the original $\mathbb Z_2^D$ gauging and recovers $\mathcal T$. In this step, the Postnikov class is converted back into the mixed type-IV anomaly between the restored $\mathbb Z_2^D$ symmetry and~$G$. Gauging the remaining 0-form component $G$ then gives precisely the theory constructed in this subsection. Therefore, with compatible gauging conventions and no additional discrete torsion, we have
\begin{equation*}
\mathcal T_D/\Gamma
\simeq
\mathcal T/
(\mathbb Z_2^A\times\mathbb Z_2^B\times\mathbb Z_2^C).
\end{equation*}
This provides the conceptual connection to the dual fusion 2-category $2\mathrm{Rep}(\Gamma)$: our lattice construction reaches the fully gauged theory through an equivalent sequence of ordinary subgroup gaugings. The identification therefore relies on this gauging relation and the general framework of higher-group gauging, rather than on the displayed fusion rule alone. Our explicit lattice calculations construct symmetry operators and establish their fusion relations.

%
 \section{Field-theoretical interpretation} \label{sec4}
In this section, we provide a field-theoretical derivation of our results, showing
 how gauging various subgroups gives rise to 2-group symmetry, non-invertible symmetry, and 2-fusion categorical symmetry
as discussed in Sec.~\ref{sec2}. 
Before delving into the details, let us first recall the setup.  
We consider a $(2+1)$-dimensional theory $\mathcal{T}$ with $\mathbb{Z}_2^A \times \mathbb{Z}_2^B \times \mathbb{Z}_2^C \times \mathbb{Z}_2^D$ $0$-form symmetry that exhibits a type~IV anomaly whose inflow action is given by
\begin{equation}
    (-1)^{\int_{M_4} A\cup B\cup C\cup D}, \label{type4 anomaly}
\end{equation}
where $A,B,C,D$ denote the background gauge fields for $\mathbb{Z}_2$. 
The presence of the anomaly implies nontrivial constraints on the symmetry structure after gauging.

\subsection{Gauging $\mb^A$ : 2-group symmetry}\label{3.1}
Following the argument presented in~\cite{Tachikawa:2017gyf}, we show that gauging the $\mb^A$ subgroup leads to the~2-group symmetry discussed in Sec.~\ref{sec.2.1}. We define a theory after gauging $\mb^A$ subgroup as $\mathcal{T}_1$, and write its partition function  as
\begin{equation}
    Z_{\mathcal{T}_1}[\hat{A}^{(2)}, B,C,D] = \# \sum_{a\in H^1(M_3, \mb)}Z_{\mathcal{T}}[a,B,C,D](-1)^{\int_{M_3}a\cup \hat{A}^{(2)}},
\end{equation}
where $\#$ denotes a normalization factor depending on the topology of spacetime manifold, which is not important in the subsequent discussion. Also, $\hat{A}^{(2)}$ represents a background 2-form gauge field for the dual $1$-form symmetry. \par
Due to the mixed anomaly of the original theory~\eqref{type4 anomaly}, gauge invariance requires that the background gauge fields satisfy 
\begin{equation}
    \delta\hat{A}^{(2)} = B\cup C\cup D.
\end{equation}
This relation indicates that the background gauge fields for the $0$-form and $1$-form symmetries are not independent, but instead form a 2-group structure\footnote{For the details of the  higher group structure, see e.g.,~\cite{girelli2008topological,PhysRevB.95.155118,Cordova:2018cvg, Benini:2018reh, kapustin2017higher}.}. 
Mathematically, this is captured by the following 2-group $\Gamma$:
\begin{equation}
    \Gamma :=(\hat{\mathbb{Z}}_2^{A,(1)}, \mb^B \times \mb^C\times\mb^D, \rho=\text{id}, [\beta]),
\end{equation}
where $\rho$ denotes the action of $0$-form symmetry on $1$-form symmetry, which is trivial in this case, and $[\beta]$ is a nontrivial Postnikov class given by
\begin{equation}
    [\beta]\in H^3(\mathbb{Z}_2^3, \mathbb{Z}_2), \quad \beta\Big((b_1,c_1,d_1),(b_2,c_2,d_2),(b_3,c_3,d_3)\Big)=b_1c_2d_3.
\end{equation}

\subsection{Gauging $\mb^A \times \mb^B$ : Non-invertible symmetry}\label{sec4.2}
We next examine the symmetry structure after gauging $\mb^A\times\mb^B$ subgroup. Defining the theory after gauging the symmetry by $\mathcal{T}_2$, we write the partition function as 
\begin{equation}
    Z_{\mathcal{T}_2}[\hat{A}^{(2)},\hat{B}^{(2)} ,C,D] = \# \sum_{a, b\in H^1(M_3, \mb)}Z_{\mathcal{T}}[a,b,C,D](-1)^{\int_{M_3}a\cup \hat{A}^{(2)} + b\cup\hat{B}^{(2)}},
\end{equation}
where
$\hat{A}^{(2)}, \hat{B}^{(2)}$ represents background 2-form gauge fields corresponding to dual $1$-form symmetries. 
Since $\mb^C$ and $\mb^D$ are not gauged, they can naively be treated as ordinary global symmetries. However, owing to the anomaly, in the presence of a symmetry defect associated with one of them, the other no longer acts as an ordinary symmetry. \par
%
%
To see this more explicitly, we insert a symmetry defect corresponding to $\mb^C$ along $M_2$, and investigate how  
$\mb^D$ symmetry acts on the gauged 
partition function\footnote{One can analogously discuss the case where we insert a $\mb^D$ defect along $M_2$, and investigate how  
$\mb^C$ symmetry behaves. }. Introducing $g$ and
$\mathrm{PD}(M_2)$ as
a generator of $\mb^D$ symmetry and  the Poincar\'e dual of a submanifold $M_2$, respectively, the gauged partition function is transformed as (See also Fig.~\ref{Fig:sym action w/ defect})
\footnote{The Poincar\'e duality gives an isomorphism between homology class and cohomology class of an oriented manifold, i.e., $H_k(X)\simeq H^{n-k}(X)$, where $X$ denotes $n$-dimensional oriented manifold. Physically, this implies a correspondence between the support of the symmetry defect (homology class) and the background gauge field (cohomology class) associated with the symmetry. The explicit correspondence is given by
\begin{eqnarray*}
    \int_{M^{(k)}} \omega^{(k)} = \int_{X} \omega^{(k)}\cup \mathrm{PD}(M^{(k)}), \quad M^{(k)}\in H_k(X), \quad \mathrm{PD}(M^{(k)})\in H^{n-k}(X).
\end{eqnarray*}
}
\begin{equation}
\begin{aligned}
    &Z_{g\mathcal{T}_2}[\hat{A}^{(2)},\hat{B}^{(2)}, \mathrm{PD}(M_2)] \\ 
    &= \# \sum_{a,b\in H^1(M_3,\mb)}Z_{\mathcal{T}}[a,b,\mathrm{PD}(M_2)](-1)^{\int_{M_3}a\cup \hat{A}^{(2)}+b\cup \hat{B}^{(2)} +a\cup b\cup \mathrm{PD}(M_2)}, \\
    &=\# \sum_{a,b\in H^1(M_3,\mb)}Z_{\mathcal{T}}[a,b,\mathrm{PD}(M_2)](-1)^{\int_{M_3}a\cup \hat{A}^{(2)}+b\cup \hat{B}^{(2)} +\int_{M_2}a\cup b}.\label{pt2}
\end{aligned}
\end{equation}
 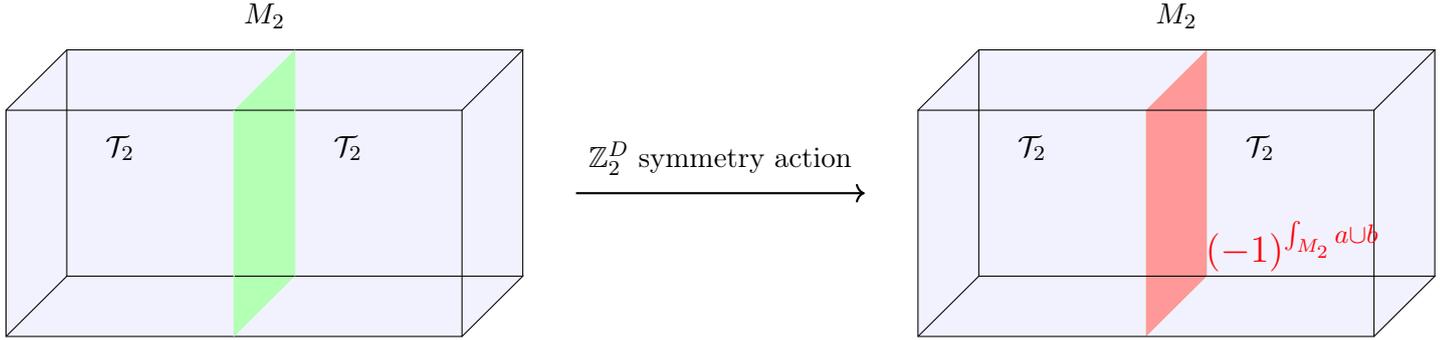
\begin{figure}[t]
    \centering
    \begin{tikzpicture}[scale=1]
\useasboundingbox (+1.0,0) rectangle (18.8,4.6);
\begin{scope}

\fill[blue!5] (0,0) rectangle (3,3);
\fill[blue!5] (3,0) rectangle (6,3);

\fill[blue!5] (0,3) -- (3,3) -- (3.8,3.8) -- (0.8,3.8) -- cycle;
\fill[blue!5] (3,3) -- (6,3) -- (6.8,3.8) -- (3.8,3.8) -- cycle;

\fill[blue!5] (0,0) -- (0,3) -- (0.8,3.8) -- (0.8,0.8) -- cycle;
\fill[blue!5] (6,0) -- (6,3) -- (6.8,3.8) -- (6.8,0.8) -- cycle;

\fill[blue!5] (0,0) -- (3,0) -- (3.8,0.8) -- (0.8,0.8) -- cycle;
\fill[blue!5] (3,0) -- (6,0) -- (6.8,0.8) -- (3.8,0.8) -- cycle;

\fill[green!30] (3,0) -- (3,3) -- (3.8,3.8) -- (3.8,0.8) -- cycle;

\draw (0,0) rectangle (6,3);
\draw (0,3) -- (0.8,3.8) -- (6.8,3.8) -- (6,3);
\draw (0,0) -- (0.8,0.8) -- (6.8,0.8) -- (6,0);

\draw (0.8,0.8) -- (0.8,3.8);
\draw (6.8,0.8) -- (6.8,3.8);

\draw [green!30](3,0) -- (3,3);
\draw  [green!30](3.8,0.8) -- (3.8,3.8);

\node at (3.4,4.25) {$M_2$};
\node at (1.5,2.5) {$\mathcal{T}_2$};
\node at (4.5,2.5) {$\mathcal{T}_2$};

\end{scope}

\draw[->,thick] (7.5,1.9) -- (11.3,1.9)
node[midway,above=3pt] {$\mb^D$ symmetry action};

\begin{scope}[shift={(12,0)}]

\fill[blue!5] (0,0) rectangle (3,3);
\fill[blue!5] (3,0) rectangle (6,3);

\fill[blue!5] (0,3) -- (3,3) -- (3.8,3.8) -- (0.8,3.8) -- cycle;
\fill[blue!5] (3,3) -- (6,3) -- (6.8,3.8) -- (3.8,3.8) -- cycle;

\fill[blue!5] (0,0) -- (0,3) -- (0.8,3.8) -- (0.8,0.8) -- cycle;
\fill[blue!5] (6,0) -- (6,3) -- (6.8,3.8) -- (6.8,0.8) -- cycle;

\fill[blue!5] (0,0) -- (3,0) -- (3.8,0.8) -- (0.8,0.8) -- cycle;
\fill[blue!5] (3,0) -- (6,0) -- (6.8,0.8) -- (3.8,0.8) -- cycle;

\fill[green!30] (3,0) -- (3,3) -- (3.8,3.8) -- (3.8,0.8) -- cycle;
\fill[red!40] (3,0) -- (3,3) -- (3.8,3.8) -- (3.8,0.8) -- cycle;

\draw (0,0) rectangle (6,3);
\draw (0,3) -- (0.8,3.8) -- (6.8,3.8) -- (6,3);
\draw (0,0) -- (0.8,0.8) -- (6.8,0.8) -- (6,0);

\draw (0.8,0.8) -- (0.8,3.8);
\draw (6.8,0.8) -- (6.8,3.8);


\node at (3.4,4.25) {$M_2$};
\node[text=red] at (4.93,1.2) {\Large $(-1)^{\int_{M_2}a\cup b}$};

\node at (1.5,2.5) {$\mathcal{T}_2$};
\node at (4.5,2.5) {$\mathcal{T}_2$};

\end{scope}

\end{tikzpicture}
    \caption{Action of the $\mb^D$ symmetry in the presence of a $C$ defect (green) supported on $M_2$. 
The symmetry action produces the phase $(-1)^{\int_{M_2} a \cup b}$ on the defect.}
    \label{Fig:sym action w/ defect}
\end{figure}
As seen from~\eqref{pt2}, the $\mb^D$ symmetry is no longer a symmetry of $\mathcal{T}_2$ in the presence of the defect. That is, 
\begin{equation*}
    Z_{g\mathcal{T}_2}[\hat{A}^{(2)},\hat{B}^{(2)}, \mathrm{PD}(M_2)] \neq Z_{\mathcal{T}_2}[\hat{A}^{(2)},\hat{B}^{(2)}, \mathrm{PD}(M_2)].
\end{equation*}
%
In order to promote $g$ to a genuine symmetry associated with $\mb^D$, we 
modify $g$ so that it is dressed with manipulations,
%
\begin{eqnarray}
    g\to TSTg.\label{modify}
\end{eqnarray}
%
Here, the manipulations $T$ and $S$ can be understood as topological manipulations supported on~$M_2$. In particular, the manipulation $T$ corresponds to stacking $\mathbb{Z}_2\times\mb$ SPT phase on $M_2$, 
\begin{equation}
(-1)^{\int_{M_2}\hat{A}^{(1)}\cup \hat{B}^{(1)}},
\end{equation}
while $S$ executes the gauging of the corresponding symmetries on $M_2$. Although $\hat{A}^{(2)}$, $\hat{B}^{(2)}$ are originally $1$-form symmetries in the bulk, when restricted to $M_2$, they effectively reduce to ordinary $0$-form symmetries with their gauge fields denoted by $\hat{A}^{(1)}, \hat{B}^{(1)}$. Such a consideration is reminiscent of the one studied in the context of higher gauging, i.e., gauging a symmetry on a higher co-dimensional manifold~\cite{Roumpedakis:2022aik}.
\par
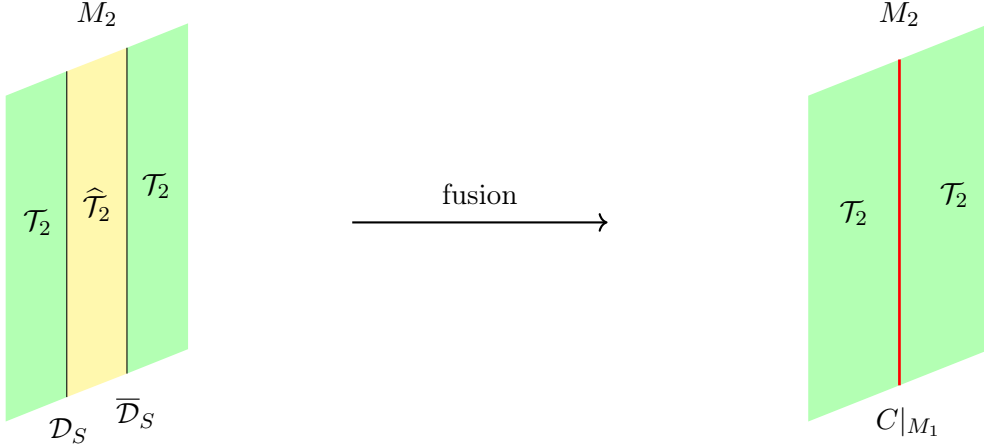
\begin{figure}
    \centering
    \begin{tikzpicture}[scale=1.2]

\begin{scope}

\fill[green!30] (2.4,-0.3) -- (2.4,3.3) -- (3.07,3.57) -- (3.07,-0.03) -- cycle;
\fill[yellow!40] (3.07,-0.03) -- (3.07,3.57) -- (3.73,3.83) -- (3.73,0.23) -- cycle;
\fill[green!30] (3.73,0.23) -- (3.73,3.83) -- (4.4,4.1) -- (4.4,0.5) -- cycle;

\draw (3.07,-0.03) -- (3.07,3.57);
\draw (3.73,0.23) -- (3.73,3.83);

\node at (2.75,1.9) {$\mathcal{T}_2$};
\node at (3.4,2.1) {$\widehat{\mathcal{T}}_2$};
\node at (4.05,2.3) {$\mathcal{T}_2$};

\node at (3.09,-0.4) {$\mathcal{D}_S$};
\node at (3.83,-0.2) {$\overline{\mathcal{D}}_S$};

\node at (3.4,4.2) {$M_2$};

\end{scope}

\draw[->,thick] (6.2,1.9) -- (9.0,1.9)
node[midway,above=3pt] {fusion};
\begin{scope}[shift={(8.8,0)}]

\fill[green!30] (2.4,-0.3) -- (2.4,3.3) -- (3.4,3.7) -- (3.4,0.1) -- cycle;
\fill[green!30] (3.4,0.1) -- (3.4,3.7) -- (4.4,4.1) -- (4.4,0.5) -- cycle;

\draw [red, line width=1pt] (3.4,0.1) -- (3.4,3.7);

\node at (2.9,2.0) {$\mathcal{T}_2$};
\node at (4.0,2.2) {$\mathcal{T}_2$};
\node at (3.5,-0.3) {$C|_{M_1}$};

\node at (3.4,4.2) {$M_2$};

\end{scope}

\end{tikzpicture}
    \caption{Fusion rule of the half-space gauging defects on the defect. 
Here $\widehat{\mathcal{T}}_2$ denotes the theory obtained by gauging the two $0$-form symmetries.}
    \label{Fig:higher gauging}
\end{figure}

With the modification~\eqref{modify},
one can verify that the partition function of $\mathcal{T}_2$ is invariant under the transformation $gTST$, i.e.,
\begin{equation}
    Z_{gTST\mathcal{T}_2}[\hat{A}^{(2)},\hat{B}^{(2)}, \mathrm{PD}(M_2)] = Z_{\mathcal{T}_2}[\hat{A}^{(2)},\hat{B}^{(2)}, \mathrm{PD}(M_2)]. \label{invariance_1}
\end{equation}
A detailed derivation is provided in Appendix~\ref{invariance}.
The topological manipulation $TST$ realizes the Kennedy-Tasaki transformation, mapping a $\mathbb{Z}_2 \times \mathbb{Z}_2$ symmetry breaking phase to an SPT phase along the defect \cite{Kennedy:1992ifl, Kennedy:1992tke, Li:2023ani, Oshikawa:1992smv}. Dressing such a manipulation with a symmetry operator is
consistent with the lattice description (Sec.~\ref{sec2.2}). Since this transformation involves a gauging operation, the resulting symmetry becomes non-invertible. Indeed, one finds non-invertible fusion rule: \footnote{Here $(\mathcal{D}_{TST})^\dagger = \overline{\mathcal{D}}_{TST}$.}
\begin{equation}
    g\mathcal{D}_{TST} \times  (g\mathcal{D}_{TST})^{\dagger} = C|_{M_1},
\end{equation}
where $C|_{M_1}$ is the condensation defect for two $0$-form symmetries, supported on $M_1\subset M_2$. Here, we have used the fact that $T^2=1$ and 
\begin{equation}
    \mathcal{D}_S \times \overline{\mathcal{D}}_S = C|_{M_1}.
\end{equation}
See also Fig.~\ref{Fig:higher gauging} for an illustration.
%

\begin{figure} [t]
    \centering
    \includegraphics[width=0.5\linewidth]{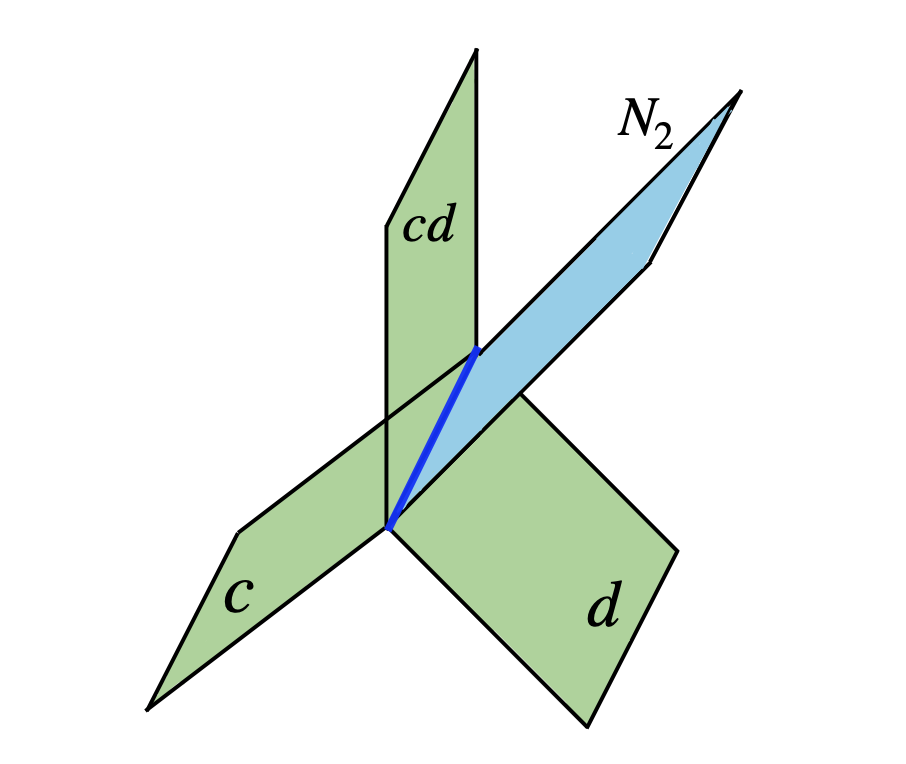}
    \caption{Junction (blue line) of the two $0$-form symmetry defects labeled by $c$ and $d$. An SPT phase defined on $N_2$ is attached to the junction.}
    \label{Fig:junction}
\end{figure}

One can give an alternative interpretation of the emergence of this non-invertible symmetry. We consider the junction of the defects associated with the two $0$-form symmetries. Due to the mixed anomaly \eqref{type4 anomaly}, the theory $\mathcal{T}_2$ acquires an SPT phase 
\begin{equation}
    (-1)^{\int_{N_2}a\cup b}
\end{equation}
whose boundary is the junction $M_1$, i.e., $\partial N_2=M_1$ (Fig.~\ref{Fig:junction}).
By the anomaly inflow of this SPT phase, this junction is not gauge invariant. To render the junction gauge invariant, one needs to couple a topological quantum field theory~(TQFT) on the junction that cancels the anomaly inflow from the SPT phase. In the present case, since the TQFT must have at least two-dimensional d.o.f, a non-invertible defect appears at the junction\footnote{To realize a projective representation of $\mathbb{Z}_2 \times \mathbb{Z}_2$, the Hilbert space must be at least two-dimensional.}. This argument is analogous to the construction of non-invertible defects discussed in \cite{Kaidi:2021xfk, Choi:2022jqy, Cordova:2022ieu, Kaidi:2023maf, Maeda:2025rxc}.

\subsection{Gauging $\mb^A \times \mb^B\times\mb^C$ : 2-Representation category}\label{sec4.3}
Finally, we gauge $\mb^A\times\mb^B\times\mb^C$ subgroup. We write a theory after gauging this symmetry as~$\mathcal{T}_3$, and its partition function as\footnote{Here, the background gauge field for the $\mb^D$ symmetry is turned off, which does not influence on the validness of our discussion, i.e., the emergence of fusion $2$-categorical symmetry.}
\begin{equation}
    Z_{\mathcal{T}_3}[\hat{A}^{(2)},\hat{B}^{(2)}, \hat{C}^{(2)}] = \# \sum_{a,b,c\in H^1(M_3,\mb)}Z_{\mathcal{T}}[a,b,c](-1)^{\int_{M_3}a\cup \hat{A}^{(2)}+b\cup \hat{B}^{(2)} +c\cup \hat{C}^{(2)}},
\end{equation}
where 
$\hat{A}^{(2)}, \hat{B}^{(2)},\hat{C}^{(2)}$ represent the background 2-form gauge fields for dual $1$-form symmetries generated by the Wilson loops of $a,b,c$. We now examine the action of the remaining $\mathbb{Z}_2^D$ symmetry on this theory. Due to the mixed anomaly \eqref{type4 anomaly}, the $\mathbb{Z}_2^D$ symmetry acts on $\mathcal{T}_3$ as
\begin{equation}
    Z_{g\mathcal{T}_3}[\hat{A}^{(2)},\hat{B}^{(2)}, \hat{C}^{(2)}] = \# \sum_{a,b,c\in H^1(M_3,\mb)}Z_{\mathcal{T}}[a,b,c](-1)^{\int_{M_3}a\cup \hat{A}^{(2)}+b\cup \hat{B}^{(2)} +c\cup \hat{C}^{(2)} +a\cup b\cup c},\label{kk}
\end{equation}
where $g$ denotes a generator of $\mb^D$ symmetry. 

The relation~\eqref{kk} indicates that the $\mathbb{Z}_2^D$ symmetry is not a symmetry of the theory $\mathcal{T}_3$. Yet, the symmetry can be restored by modifying the symmetry generator $g$ so that it is dressed with a topological manipulation $\Bar{S}TS$. Here, $S, T$ and $\Bar{S}$ denote gauging of three $\mb$ $1$-form symmetries, stacking of a $\mb^A\times\mb^B\times\mb^C$ SPT phase $(-1)^{\int A\cup B\cup C}$, and gauging $\mb^A\times\mb^B\times\mb^C$ $0$-form
symmetry, respectively. More explicitly, these topological manipulations are defined as
\begin{align}
    &S : Z[\hat{A}^{(2)},\hat{B}^{(2)},\hat{C}^{(2)}] \rightarrow \# \sum_{\substack{\hat{a}^{(2)},\hat{b}^{(2)},\hat{c}^{(2)} \\ \label{S}
    \in H^2(M_3,\mb)}}Z[\hat{a}^{(2)},\hat{b}^{(2)},\hat{c}^{(2)}] (-1)^{\int_{M_3}\hat{a}^{(2)}\cup A + \hat{b}^{(2)} \cup B + \hat{c}^{(2)} \cup C}, \\ \label{T}
    &T : Z[A,B,C] \rightarrow Z[A,B,C](-1)^{\int_{M_3} A\cup B\cup C}, \\ 
    &\Bar{S} : Z[A,B,C] \rightarrow \# \sum_{\substack{a,b,c \\
    \in H^1(M_3,\mb)}}Z[a,b,c] (-1)^{\int_{M_3}a\cup \hat{A}^{(2)} + b \cup \hat{B}^{(2)} + c\cup \hat{C}^{(2)}}, \label{bar_S}
\end{align}
where $A,B,C$ are background gauge fields for 0-form symmetries, and 
$\hat{A}^{(2)},\hat{B}^{(2)},\hat{C}^{(2)}$ are background 2-form gauge fields for $1$-form symmetries. One can verify that the partition function of gauged theory $\mathcal{T}_3$ is invariant under the combined action of $\Bar{S}TS$ and $\mb^D$ symmetry. To wit, 
\begin{equation}
    Z_{g\Bar{S}TS\mathcal{T}_3}[\hat{A}^{(2)},\hat{B}^{(2)}, \hat{C}^{(2)}] = Z_{\mathcal{T}_3}[\hat{A}^{(2)},\hat{B}^{(2)}, \hat{C}^{(2)}]. \label{invariance_2}
\end{equation}
A detailed derivation is given in Appendix~\ref{invariance}.

\begin{figure}
    \centering
    \begin{tikzpicture}[scale=0.9]

\begin{scope}

\node at (-0.5,3.5) {(a)};

\begin{scope}

\fill[gray!15] (0,0) rectangle (2,3);
\fill[blue!10] (2,0) rectangle (4,3);
\fill[gray!15] (4,0) rectangle (6,3);

\draw[thick] (2,0) -- (2,3);
\draw[thick] (4,0) -- (4,3);

\node at (2,3.3) {$\mathcal{D}_S$};
\node at (4,3.3) {$\overline{\mathcal{D}}_S$};

\node at (1,2.5) {$\mathcal{T}$};
\node at (3,2.5) {$\mathcal{T}_3$};
\node at (5,2.5) {$\mathcal{T}$};

\end{scope}

\draw[->,thick] (6.8,1.5) -- (8.2,1.5);
\node at (7.5,2) {fusion};

\begin{scope}[shift={(9,0)}]

\fill[gray!15] (0,0) rectangle (3,3);
\fill[gray!15] (3,0) rectangle (6,3);

\draw[thick] (3,0) -- (3,3);

\node at (3,3.3) {$C_0$};

\node at (1.5,2.5) {$\mathcal{T}$};
\node at (4.5,2.5) {$\mathcal{T}$};

\end{scope}

\end{scope}

\begin{scope}[shift={(0,-4.5)}]

\node at (-0.5,3.5) {(b)};

\begin{scope}

\fill[blue!10] (0,0) rectangle (2,3);
\fill[gray!15] (2,0) rectangle (4,3);
\fill[blue!10] (4,0) rectangle (6,3);

\draw[thick] (2,0) -- (2,3);
\draw[thick] (4,0) -- (4,3);

\node at (2,3.3) {$\overline{\mathcal{D}}_S$};
\node at (4,3.3) {$\mathcal{D}_S$};

\node at (1,2.5) {$\mathcal{T}_3$};
\node at (3,2.5) {$\mathcal{T}$};
\node at (5,2.5) {$\mathcal{T}_3$};

\end{scope}

\draw[->,thick] (6.8,1.5) -- (8.2,1.5);
\node at (7.5,2) {fusion};

\begin{scope}[shift={(9,0)}]

\fill[blue!10] (0,0) rectangle (3,3);
\fill[blue!10] (3,0) rectangle (6,3);

\draw[thick] (3,0) -- (3,3);

\node at (3,3.3) {$C_1$};

\node at (1.5,2.5) {$\mathcal{T}_3$};
\node at (4.5,2.5) {$\mathcal{T}_3$};

\end{scope}

\end{scope}

\end{tikzpicture}
    \caption{
Fusion rules of the half-space gauging defects. Here, $\mathcal{T}$ denotes the original theory with type IV anomaly.
(a) $\mathcal{D}_S \times \overline{\mathcal{D}}_S = C_0$.\ 
(b) $\overline{\mathcal{D}}_S \times \mathcal{D}_S = C_1$.
}
    \label{Fig:half-space gauging}
\end{figure}
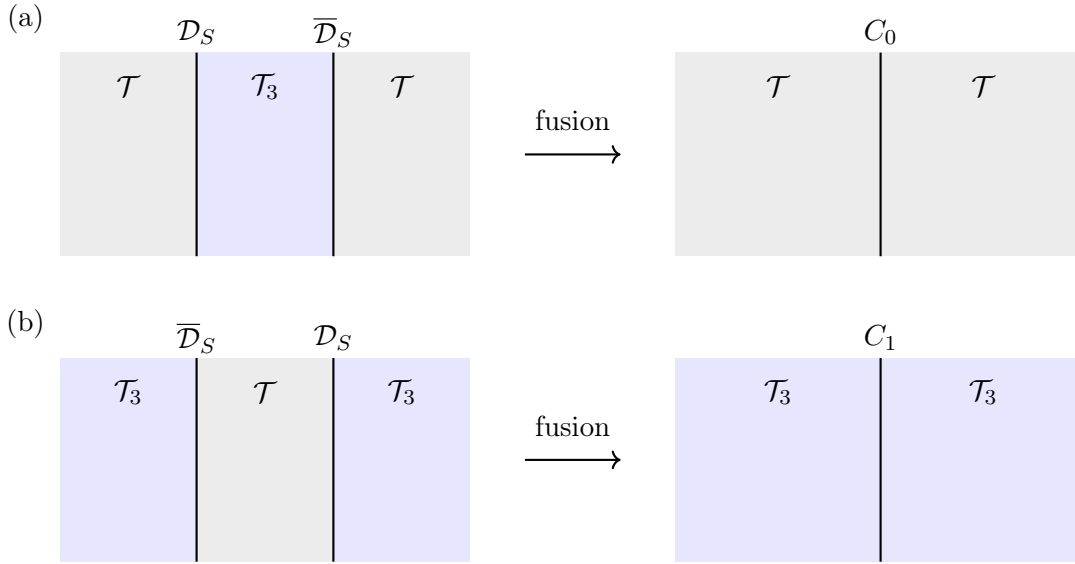

Finally, let us compute the fusion rule of this non-invertible symmetry. 
Generally, when we have a system that is invariant under a topological manipulation associated with gauging, such a manipulation can be implemented on a half-space of the entire system. This effectively allows us to introduce a domain wall located at the position where the manipulation terminates. The fusion of non-invertible operators can then be understood by 
introducing several domains and
fusing to such domain walls. We apply this consideration to the current situation.
Denoting the half-space gauging defects corresponding to~$S$, $\overline{S}$ by~$\mathcal{D}_S$, $\overline{\mathcal{D}}_{S}$, respectively, their fusion rules are given by
\begin{equation}
    \mathcal{D}_S \times \overline{\mathcal{D}}_{S} = C_0, \quad \overline{\mathcal{D}}_{S} \times \mathcal{D}_S = C_1, \label{sec4.3:fusion1}
\end{equation}
where $C_0$ and $C_1$ are condensation defects of three $0$-form symmetries and three $1$-form symmetries, respectively \cite{Roumpedakis:2022aik}~(Fig.~\ref{Fig:half-space gauging}){
\footnote{The fusion rules of half-space gauging defects were studied in
Ref.~\cite{Choi:2022zal}. For a lattice realization of these defects, see
Ref.~\cite{hofmeier2026lattice}.}}. Also, we have
\begin{equation}
    \mathcal{D}_S \times \eta_k^{(1)} = \mathcal{D}_S, \quad U_k \times \mathcal{D}_S = \mathcal{D}_S, \quad \overline{\mathcal{D}}_{S} \times U_k = \overline{\mathcal{D}}_{S}, \quad \eta_k^{(1)} \times \overline{\mathcal{D}}_{S}= \overline{\mathcal{D}}_{S}, \label{sec4.3:fusion2}
\end{equation}
where $U_k$ is the generator of the $0$-form symmetry and $\eta_k^{(1)}$ is the generator of the $1$-form symmetry, with $k\in\{A,B,C\}$. Using these fusion rules \eqref{sec4.3:fusion1}, \eqref{sec4.3:fusion2} and $T^2=1$, we find that
\begin{equation}
    g\mathcal{D}_{\Bar{S}TS} \times g\mathcal{D}_{\Bar{S}TS} = C_1,
\end{equation}
up to a normalization factor. 

In summary, we have provided a field-theoretical analysis on the emergent symmetries via gauging subgroups of a theory with the type IV anomaly.  The results show that the field-theoretical analysis is precisely consistent with all the symmetry structures obtained in the lattice construction (Table~\ref{defect_hierarchy}), demonstrating that they are universally dictated by the type-IV anomaly.
\section{Modulated symmetry and non-invertible translation from type~IV LSM anomalies}\label{sec3}
Having established the hierarchy of emergent symmetries obtained by partially gauging the purely internal type-IV anomaly in Secs.~\ref{sec2} and~\ref{sec4}, we now study its crystalline counterparts in translation-symmetric systems exhibiting LSM anomalies. Although the microscopic lattice models considered below differ from the triangular lattice model of Sec.~\ref{sec2}, their symmetry structures after gauging are governed by the same type-IV anomaly and are therefore not independent constructions. At the level of the anomaly inflow action, the two LSM anomalies studied in this section are obtained through the crystalline replacements
\begin{equation*}
A\wedge B\wedge C\wedge D
\ \longrightarrow\
A\wedge B\wedge e^{x}\wedge e^{y},
\qquad
A\wedge B\wedge C\wedge D
\ \longrightarrow\
A\wedge B\wedge C\wedge e^{x},
\end{equation*}
where $e^{x}$ and $e^{y}$ denote background fields for lattice translations. Upon gauging $\mathbb Z_{2}^{A}$, these replacements convert the defect-sensitive symmetry algebra of the internal model into an algebra involving lattice translations and the dual 1-form symmetry. The results derived in this section are therefore direct crystalline counterparts of the single-gauging analysis in Secs.~\ref{sec.2.1} and~\ref{3.1}. In particular, an odd linear system size realizes a nontrivial translation-twisted background and plays a role analogous to an internal-symmetry defect in probing the underlying type-IV anomaly.
\par

Previous studies have shown that gauging internal symmetries in LSM systems can generate \emph{modulated symmetries}, whose spatially varying actions are intertwined with lattice translations~\cite{Aksoy:2023hve,Seifnashri:2023dpa,Pace:2024acq,Pace:2025hpb,SciPostPhys.20.4.117}. Here, 
we identify its type-IV origin and explain why its algebra depends on the parity of the system size. In the second lattice model, we further demonstrate that inserting an internal-symmetry defect induces a modulated 1-form symmetry after gauging. Translation-twisted backgrounds and internal-symmetry defects therefore provide complementary realizations of the same anomaly-controlled symmetry structure. To the best of our knowledge, such a defect-induced modulated symmetry has not previously been derived in an explicit lattice model with a type-IV LSM anomaly. 

Furthermore, gauging larger subgroups in both models reveals richer symmetry structures, including non-invertible translations~\cite{Seifnashri:2023dpa,Pace:2024acq,Oishi:2026sow}. In both models, gauging two internal symmetries yields a dressed translation operator whose invertibility depends on the parity of the transverse system size. This dependence provides a crystalline counterpart of the defect-dependent non-invertibility discussed in Sec.~\ref{sec2.2}. Gauging all three internal symmetries in the second model further yields a non-invertible translation whose fusion rule is analogous to that of the $2$-representation category of a $2$-group symmetry, consistent with Sec.~\ref{sec2.3}.

\subsection{Two internal and two translational symmetries
}\label{4.1}
We begin with introducing a square lattice where each node accommodates a qubit. We also introduce Pauli operators on each node $v$ as $X_v$ and $Z_v$ where $v$ denotes the coordinate of the site~$v\vcentcolon=(\hx,\hy)\in\mathbb{Z}^2$. 
We define the following Hamiltonian:
\begin{equation}
    H_{2D,1}=-\sum_{\langle v,v^\prime\rangle}(X_vX_{v^\prime}+Z_vZ_{v^\prime}),\label{spin001}
\end{equation}
where $\langle v,v^\prime\rangle$ represents nearest-neighbor pairs. We impose periodic boundary condition with system size $L_x\times L_y$. 
The model possesses a global $\mathbb Z_2\times \mathbb Z_2$ $0$-form symmetry generated by
\begin{eqnarray}
    U^{(0)}_X=\prod_{\hx=1}^{L_x}\prod_{\hy=1}^{L_y}X_{v},\quad U^{(0)}_Z=\prod_{\hx=1}^{L_x}\prod_{\hy=1}^{L_y}Z_{v}.\label{global2d}
\end{eqnarray}
These symmetry operators satisfy the anomalous commutation relation
\begin{eqnarray}
    U^{(0)}_{Z}U^{(0)}_{X}=(-1)^{L_xL_y}U^{(0)}_{X}U^{(0)}_{Z},
\end{eqnarray}
which signals the presence of a LSM anomaly involving two internal symmetries and translational ones in the~$x$- and $y$-direction.

This anomaly can be captured by the following type IV inflow action: 
\begin{eqnarray}
    S=i\pi\int_{M_4}A^{(1)}\wedge B^{(1)}\wedge e^x\wedge e^y\label{exey}
\end{eqnarray}
with $A^{(1)}$ and $B^{(1)}$ being 
background gauge fields corresponding to two $\mathbb{Z}_2$
internal symmetries~\eqref{global2d}\footnote{Superscript of the gauge fields denote form, i.e., $A^{(p)}$ represents a $p$-form background gauge field. }
and $e^x=dx$, $e^y=dy$. Compared with~\eqref{abcd}, this LSM anomaly 
can be viewed as a type~IV anomaly in which two of the internal symmetries are replaced by lattice translation symmetries. 
Physically, this action admits an interpretation in terms of weak SPT phases~\cite{Cheng:2015kce,Pace:2025hpb,SciPostPhys.20.4.117,Oishi:2026sow}. Namely, $1+1$d $\mathbb{Z}_2\times\mathbb{Z}_2$ SPT chains extending in the $z$-direction, are stacked along the $x$- and $y$-directions. 
From this viewpoint, the LSM anomaly originates from a projective representation of $\mathbb{Z}_2\times\mathbb{Z}_2$ per unit cell and can be understood via anomaly inflow from these stacked SPT phases. 
This perspective will be crucial in what follows, where we show that gauging the internal symmetries leads to the emergence of modulated symmetries and non-invertible symmetries.
%
\subsubsection{Gauging one internal symmetry} \label{4.1.1}
Here, we gauge one of the global symmetries, $U_X^{(0)}$\footnote{Emergence of the modulated symmetry by gauging one internal symmetry in the model~\eqref{spin001} was discussed in Ref.~\cite{SciPostPhys.20.4.117}. However, we give a new insight to the result in view of type IV anomaly and symmetry defect, which is not elucidated in the previous study.}. To do so, we introduce extended Hilbert space on each link whose Pauli operator is denoted as $\Tilde{\tau}^{X/Z}_{v,v^\prime}$. Here, the subscript labels a link coordinate, connecting nodes $v$ and $v^\prime$. We interchangeably write the link coordinate as $l$, namely, depending on the context, we interchangeably write a operator
$\Tilde{\tau}^{X/Z}_{v,v^\prime}$
as $\Tilde{\tau}^{X/Z}_{l}$. The  
gauging is implemented by imposing the following Gauss law:
\begin{eqnarray}
 X_v\times\prod_{v\in l}\tilde{\tau}^X_l=1,~\forall v.\label{gauss 2d}
\end{eqnarray}
We minimally couple the spin terms to the gauge fields as
\begin{eqnarray}
    Z_vZ_{v^\prime}\to  Z_v\tilde{\tau}^Z_{v,v^\prime}Z_{v^\prime}
\end{eqnarray} 
in order for  them to commute with the Gauss law.
Further, we add the following flux operator of the gauge field
\begin{eqnarray}
    -g\sum_{p}\prod_{l\subset \partial p}\tilde{\tau}^Z_{l}
\end{eqnarray}
to the Hamiltonian to 
ensure that  the theory does not admit additional flux. Here, $p=(\hx+\frac{1}{2},\hy+\frac{1}{2})$ denotes coordinate of a plaquette. 
Rewriting operators as
\begin{eqnarray}
    \tau^{X}_{l}\vcentcolon=\Tilde{\tau}^{X}_{l},\quad 
\tau^Z_{v,v^\prime}\vcentcolon=Z_v\Tilde{\tau}^{Z}_{v,v^\prime}Z_{v^\prime},  
\end{eqnarray}
we obtain the following gauged Hamiltonian:
\begin{eqnarray}
   \Tilde{H}_{2D,1}=-\sum_{\langle v,v^\prime\rangle}A_v\times A_{v^\prime}-g\sum_{p} B_{p}-\sum_{l}\tau^Z_{l}\label{2toric}
\end{eqnarray}
with $g\to\infty$.
Here, 
\begin{eqnarray}
 A_v\vcentcolon=\prod_{v\in l}\tau^X_l,\quad B_p\vcentcolon=\prod_{l\subset \partial p}{\tau}^Z_{l}.
\end{eqnarray}
\begin{figure} [t]
    \centering
    \includegraphics[width=0.8\linewidth]{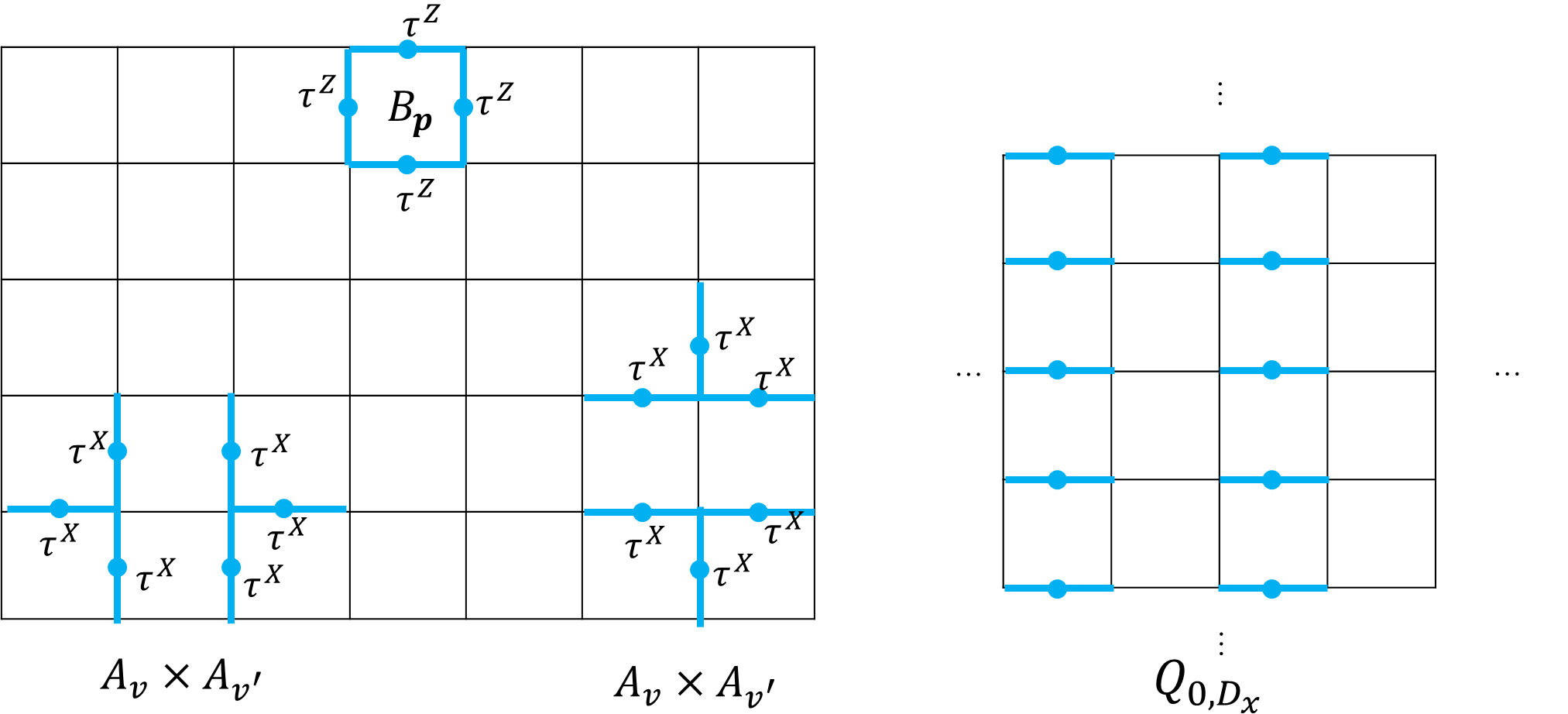}
    \caption{(Left) The first two terms that constitute the Hamiltonian~\eqref{2toric}. (Right) Configuration of the modulated $0$-form symmetry, $ Q_{0,D_x}$ given in~\eqref{0form}.  }
    \label{Fig:lsm}
\end{figure}
Here, $A_v$ is the product of $\tau^X_l$ over all links incident on $v$, whereas $B_p$ is the product of $\tau^Z_l$ over all links on the boundary of the plaquette $p$ (left of Fig.~\ref{Fig:lsm}). 
The first two terms in the model are reminiscent of the toric code, except that the electric terms is given by a product of adjacent star operators,~$A_v$,~$A_{v^\prime}$ rather than a single one. \par
The model~\eqref{2toric} respects $0$-form and $1$-form symmetries. 
The $0$-form symmetry is described by
\begin{eqnarray}
    Q_{0,D_x}\vcentcolon=\left[\prod_{\hx=1}^{L_x}\prod_{\hy=1}^{L_y}(\tau^Z_{l_x})^{\hx}\right]^{\alpha_x},\quad Q_{0,D_y}\vcentcolon=\left[\prod_{\hx=1}^{L_x}\prod_{\hy=1}^{L_y}(\tau^Z_{l_y})^{\hy}\right]^{\alpha_y},\label{0form}
\end{eqnarray}
where $l_{x}(l_y)$ denotes coordinate of a horizontal~(vertical) link, 
that is, $l_x=(\hx+\frac{1}{2},\hy)$, $l_y=(\hx,\hy+\frac{1}{2})$, 
and $\alpha_a\vcentcolon=\frac{2}{\gcd(L_a,2)}~(a=x,y)$.  One of the symmetries, $ Q_{0,D_x}$ is shown in the right of Fig.~\ref{Fig:lsm}. 
Symmetries~\eqref{0form} are $0$-form modulated symmetries (more precisely, dipole symmetries): 
the symmetry operators are supported on all horizontal or vertical links, with their exponents depending on the spatial coordinate.
%
Note 
that two global symmetries~\eqref{0form} are not independent; when both of $L_x$ and $L_y$ are even, by the flatness condition of the gauge fields, two symmetries are identical.  \par
There is also $1$-form symmetry in this model, described by the following noncontractible loops:
\begin{eqnarray}
    Q_{1,x}\vcentcolon=\prod_{\hx=1}^{L_x}\tau^Z_{l_x},\quad     Q_{1,y}\vcentcolon=\prod_{\hy=1}^{L_y}\tau^Z_{l_y}.\label{loop}
\end{eqnarray}
These loops are topological, that is, distinct loop operators solely depend on the homology class, owing to 
the flatness condition. Notably, the two types of the symmetries~\eqref{0form}~\eqref{loop}, are related via lattice translation:
\begin{eqnarray}
    T_xQ_{0,D_x}T_x^{-1}=Q_{0,D_x}(Q_{1,x})^{\alpha_xL_y},\quad   T_yQ_{0,D_x}T_y^{-1}=Q_{0,D_y}(Q_{1,y})^{\alpha_yL_x}\label{pp}
\end{eqnarray}
Other combination of symmetries and translations gives trivial commutation relation. Here, we have introduced a lattice translation operator $T_{x}~(T_y)$ in the $x~(y)$-direction which shifts a local operator in the positive $x$~($y$)-direction by a unit lattice constant.
The relation~\eqref{pp} is sometimes referred to as the \textit{dipole algebra}~\cite{Ebisu:2023idd,2024multipole} -- acting a lattice translation on a symmetry operator yields another symmetry, forming a hierarchical structure. 
The dipole algebra in~\eqref{pp} differs from previously studied cases, where lattice translation relates symmetry operators of the same form. In contrast, here the translation maps a $0$-form symmetry operator to a distinct~$1$-form  symmetry operator, reflecting a nontrivial extension structure. 

Moreover, the relation~\eqref{pp} is consistent with the 2-group structure discussed in Sec.~\ref{sec.2.1}. 
Recall that when gauging one of the internal symmetries in a system with the type IV anomaly, we obtain the 2-group described by the projective algebra~\eqref{p} in the presence of the symmetry defect. The relation~\eqref{pp} can also be understood by this argument once we replace some of the internal symmetries with the translational ones, following crystalline equivalence principle~\cite{PhysRevB.99.115116}.\par
To see this, we need to understand properly what is the translational defect in our context. 
Generally, a symmetry defect associated with an internal symmetry $g$, is defined as a codimension-one interface across which a local operator is transformed by the group action $U_g$. As we have seen in the previous argument, inserting such a defect amounts to imposing a twisted boundary condition. This construction only modifies the internal d.o.f and does not affect the underlying spatial geometry. Consequently, internal symmetry defects can be freely deformed in space (i.e., they are topological). 
By contrast, translation symmetry is not an internal symmetry but acts by shifting spatial coordinates. Hence, a translation defect cannot be implemented as a simple local twist of the internal d.o.f. Rather, it necessarily modifies the gluing of the lattice itself. More precisely, a translation defect can be interpreted as a geometric defect analogous to a lattice dislocation: when going around the defect, the system fails to return to the original position and instead acquires a net translation~\cite{Seifnashri:2023dpa}, which can be expressed as a modified periodic boundary condition
\begin{align}
    T_i^{L_i}=1\to T_i^{L_i}=T_i,
\end{align}
where we denote a lattice translation operator in the $i$-direction as $T_i$ and the length of the lattice in the direction as $L_i$. This 
effectively corresponds to removing a lattice site ($L_i\to L_i-1$). \par
With this in mind, we again turn to the relation~\eqref{pp}. The dipole algebra depends sensitively on the parity of the system size: it is trivial when $L_x$ and $L_y$ are even, while it becomes nontrivial when either of $L_x$ or $L_y$ is odd.
Therefore, the parity dependence of the dipole algebra can be understood as a direct consequence of translation symmetry defects. The nontrivial algebra arises precisely when the system effectively hosts such a defect, and thus provides a diagnostic of the underlying mixed anomaly. 
Here, the translation symmetries are effectively treated as~$\mathbb{Z}_2$ symmetries generated by lattice translations~\cite{metlitski2018intrinsic}; presence (absence) of translation defect corresponds to $L_i$ being odd~(even).
This establishes the dipole symmetry as an emergent structure dictated by anomaly matching, rather than an accidental feature of the lattice model. 

\subsubsection{Gauging two internal symmetries}
We next consider gauging two internal symmetries, $U_X^{(0)}$ and $U_Z^{(0)}$.
We focus on the case with $L_x$ even. To gauge the two symmetries, $U^{(0)}_X$ and $U^{(0)}_Z$, we 
introduce two spin d.o.fs on each link, $\tau$ and $\sigma$ corresponding to the gauge fields, 
and impose the following Gauss laws:
\begin{eqnarray}
     X_v\times\prod_{l\ni v}{\tau}^X_l\times(\sigma^Z_{v+e_x/2})^{\hx+1}=1,\quad    Z_v\times\prod_{l\ni v}{\sigma}^X_l\times(\tau^Z_{v-e_x/2})^{\hx}=1\label{129}
\end{eqnarray}
with $e_x=(1,0)$ (which should not be confused with the gauge field, $e^x$ corresponding to the translation symmetry in the $x$-direction). 
Note that we have introduced position dependent factors $(\sigma^Z_{v+e_x/2})^{\hx+1}$ and $(\tau^Z_{v-e_x/2})^{\hx}$ to ensure the commutativity between two Gauss laws.  
We minimally couple the spin terms to the gauge fields as
\begin{eqnarray}
    Z_vZ_{v^\prime}\to  Z_v{\tau}^Z_{v,v^\prime}Z_{v^\prime},\quad  X_vX_{v^\prime}\to  X_v{\sigma}^Z_{v,v^\prime}X_{v^\prime}
\end{eqnarray} 
in order for  them to commute with the Gauss laws. 
Hence, the gauged Hamiltonian reads
\begin{eqnarray}
    H_g=-\sum_{\langle v,v^\prime\rangle}(Z_v{\tau}^Z_{v,v^\prime}Z_{v^\prime}+X_v{\sigma}^Z_{v,v^\prime}X_{v^\prime}).\label{h_g}
\end{eqnarray}
We further impose the fluxless condition:
\begin{eqnarray}
      \prod_{l\subset \partial p}{\tau}^Z_{l}=1,\quad \prod_{l\subset \partial p}{\sigma}^Z_{l}=1~\forall p.\label{fluxtau}
\end{eqnarray}
We impose local flatness while retaining all noncontractible Wilson-loop sectors.
The gauged theory respects the dual $1$-form symmetry, $\mathbb{Z}_2^{(1)}\times \mathbb{Z}_2^{(1)}$, whose operators are described by
\begin{eqnarray}
Q_{1,x}^\tau\vcentcolon=\prod_{\hx=0}^{L_x-1}\tau^Z_{(\hx+1/2,0)},\quad     Q^\tau_{1,y}\vcentcolon=\prod_{\hy=0}^{L_y-1}\tau^Z_{(0,\hy+1/2)}\nonumber\\
Q_{1,x}^\sigma\vcentcolon=\prod_{\hx=0}^{L_x-1}\sigma^Z_{(\hx+1/2,0)},\quad     Q^\sigma_{1,y}\vcentcolon=\prod_{\hy=0}^{L_y-1}\sigma^Z_{(0,\hy+1/2)}.\label{132}
\end{eqnarray}
Note that these operators are topological,  i.e., independent operators depend solely on the homology class due to the fluxless condition~\eqref{fluxtau}.\par
In the gauged theory, the translation operator~$T_x$ is not well-defined due to the relation
\begin{eqnarray}
    T_x G^X_v \neq G^X_{v+e_x}T_x,\quad    T_x G^Z_v \neq G^Z_{v+e_x}T_x,\label{gggg}
\end{eqnarray}
where $G^X_v$ and $G^Z_v$ denotes the Gauss law term in the left hand side of~\eqref{129}.
The translation operator in the $y$-direction, $T_y$ is well-defined, as $ T_y G^{X/Z}_v T_y^{-1}=G^{X/Z}_{v+e_y}$.
To remedy the issue of~\eqref{gggg}, 
we modify the translation operator dressed by an operator $W$, $T_x\to T_x W$ where $W$ satisfies
\begin{eqnarray}
    W G^X_v =\sigma^Z_{v+e_x/2}G^X_vW,\quad WG^Z_v=\tau^Z_{v-e_x/2}G^Z_vW. \label{W}
\end{eqnarray}
Before delving into the details, we outline the construction of $W$.
We construct $W$ from two commuting unitary operators, $V_1$ and $V_2$, followed by a projection. The operator $V_1$ generates the link-$Z$ factors required by ~\eqref{W},
but leaves an additional noncontractible Wilson-loop. The operator $V_2$ cancels this factor away from the origin. The remaining obstruction at the origin is global and will be treated by the projection below. To that end, we introduce
\begin{eqnarray}
V_1&\vcentcolon=&\prod_{\hx,\hy}C\left(\eta^\tau_{(\hx,\hy)},\sigma^Z_{(\hx+1/2,\hy)}\right),\nonumber\\
V_2&\vcentcolon=&\prod_{\hy}C\left(\zeta^\sigma_{(0,\hy)},Q^\tau_{1,x}\right).
\end{eqnarray}
Here, $C(A,B)\vcentcolon=\frac{1}{2}(1+A+B-AB)$ is the  operator form of a CZ gate for $A$ and $B$ consisting of  Pauli~$Z$ operators, and 
$\eta^\tau_{(\hx,\hy)}$ represents string of gauge fields $\tau^Z$ connecting  an origin $(0,0)$ and a point $(\hx,\hy)$. Likewise, $\zeta^\sigma_{(0,\hy)}$ denotes string of gauge fields, $\sigma^Z$ from an origin $(0,0)$  to a point~$(0,\hy)$. More precisely, 
\begin{eqnarray}
    \eta^\tau_{(\hx,\hy)}&\vcentcolon=&\prod_{i=0}^{\hx-1}\tau^Z_{(i+1/2,0)}\times \prod_{j=0}^{\hy-1}\tau^Z_{(\hx,j+1/2)}\nonumber\\
    \zeta^\sigma_{(0,\hy)}&\vcentcolon=&\prod_{j=0}^{\hy-1}\sigma^Z_{(0,j+1/2)} 
\end{eqnarray}
with $\eta^\tau_{(0,0)}=\zeta^\sigma_{(0,0)}=I$.
The $1$-form operator $Q^\tau_{1,x}$ is given in~\eqref{132}.\par
The action of $V_1$ on $G_v^X$ and $G^Z_v$ follows from the endpoint algebra of the strings. The Gauss law terms, $G_v^X$, 
anticommutes with the string $\eta_s^\tau$ precisely when $v$ is one of its
end point. 
For~$G_v^Z$, the two incident horizontal $\sigma^Z$ operators contribute the product $\eta_v^\tau\eta_{v-e_x}^\tau$. By the flatness condition~\eqref{fluxtau}, this product reduces to the intervening $\tau^Z$ link, with an additional $Q^\tau_{1,x}$ when that link crosses the $y$-axis with $\hat x=0$, leading to
\begin{eqnarray}
    V_1 G^X_v V_1^{-1}&=& \sigma^Z_{v+e_x/2}G^X_v~(v\neq (0,0)),\nonumber\\
    V_1G_v^ZV_1^{-1}&=&\begin{cases}
        \tau^Z_{v-e_x/2}G_v^Z~(v=(\hx,\hy),\;\hx\neq 0)\\
          Q^\tau_{1,x}\tau^Z_{v-e_x/2}G_v^Z~(v=(0,\hy),\;\hy\neq 0).
    \end{cases}\label{137}
\end{eqnarray}
The operator $V_2$ supplies the boundary correction. At $v=(0,\hat y)\neq(0,0)$, $G_v^Z$ detects the endpoint of $\zeta^\sigma_{(0,\hat y)}$, so conjugation by $V_2$ attaches a factor $Q^\tau_{1,x}$ to $G_v^Z$. This cancels the additional factor produced by $V_1$.  Away from $\hat x=0$, $V_2$ leaves the Gauss operators unchanged. Explicitly,  
\begin{eqnarray}
    V_2G^X_v V_2^{-1}&=&G^X_v~(v\neq (0,0)),\nonumber\\
    V_2G_v^ZV_2^{-1}&=&\begin{cases}
        G_v^Z~(v=(\hx,\hy),\;\hx\neq 0)\\
          Q^\tau_{1,x}G_v^Z~(v=(0,\hy),\;\hy\neq 0).
          \end{cases}
\end{eqnarray}
Therefore, the operator $V\vcentcolon=V_1V_2$ plays the same role as $W$~\eqref{W} \emph{except} at the origin $v=(0,0)\vcentcolon=o$. Indeed, 
from
\begin{eqnarray*}
    V_1G_o^XV_1^{-1}=\left(\prod_{s\neq o}\sigma^Z_{s+e_x/2}\right)G_o^X,\quad  V_2G_o^XV_2^{-1}=G_o^X\\
    V_1 G_o^ZV_1^{-1}=Q^\tau_{1,x}\tau^Z_{o-e_x/2}G^Z_o,\quad  V_2 G_o^ZV_2^{-1}=(Q^\tau_{1,x})^{L_y-1}G^Z_o,
\end{eqnarray*}
it follows that 
\begin{eqnarray}
    VG^X_oV^{-1}=(Q^\sigma_{1,x})^{L_y}\sigma^Z_{o+e_x/2}G^X_o,\quad   VG^Z_oV^{-1}=(Q^\tau_{1,x})^{L_y}\tau^Z_{o-e_x/2}G^Z_o,
\end{eqnarray}
where we have used $\prod_s \sigma^Z_{s+e_x/2}=(Q^\sigma_{1,x})^{L_y}$. 
 When $L_y$ is even, $V$ implements the desired transformation, so we can take $W=V$. Therefore, the translation operator is modified as $\tilde{T}_x\vcentcolon=T_x W$ which satisfies
\begin{eqnarray*}
      \tilde{T}_x G^X_v= G^X_{v+e_x}\tilde{T}_x,\quad    \tilde{T}_x G^Z_v = G^Z_{v+e_x}\tilde{T}_x. 
\end{eqnarray*}
and $[\tilde{T}_x,H_g]=0$. Since $V$ is invertible, the modified translation operator is also invertible and satisfies the fusion rule 
\begin{align} 
    \tilde{T}_x \times \tilde{T}_{-x}=I, \quad \tilde{T}_{-x}:= \tilde{T}_x^{\dagger}.\label{158}
\end{align}
On the other hand, when $L_y$ is odd, $V$ does not implement the desired transformation. To rectify this, we multiply the following condensation operator
\begin{eqnarray*}
   \mathcal{P}\vcentcolon=
   (1+Q^\sigma_{1,x})\times (1+Q^\tau_{1,x}) 
\end{eqnarray*}
to $V$. The desired operator $W$ is then defined as 
  $W\vcentcolon=\frac{1}{2}\mathcal{P}V.$
With this $W$, the translation operator is modified as $\tilde{T}_x\vcentcolon=T_x W$ which satisfies
\begin{eqnarray*}
      \tilde{T}_x G^X_v= G^X_{v+e_x}\tilde{T}_x,\quad    \tilde{T}_x G^Z_v = G^Z_{v+e_x}\tilde{T}_x. 
\end{eqnarray*}
and $[\tilde{T}_x,H_g]=0$.
Furthermore, denoting the modified translation in the $-x$ direction as $\tilde{T}_{-x}\vcentcolon=\tilde{T}_x^{\dagger}=WT_x^{-1}$, we have 
\begin{eqnarray} 
    \tilde{T}_x\times \tilde{T}_{-x}=
      (1+Q^\sigma_{1,x})\times (1+Q^\tau_{1,x}).\label{159}
\end{eqnarray}
These fusion rules \eqref{158}, \eqref{159} indicate that depending on the $L_y$ even or odd, the translation operator becomes invertible or non-invertible. 
Following the interpretation of system size parity in Sec.~\ref{4.1}, this dependence reflects the nontrivial mod-two translation background in the $y$-direction. It parallels the defect-dependent non-invertible junction found in Sec.~\ref{sec2.2}: here, the corresponding non-invertibility appears in the action of the dressed $x$-translation.

\subsection{Three internal and one translational symmetries}\label{42new}
Let us turn to another lattice model that exhibits the LSM anomaly involving three internal and one translational symmetries. We introduce a square lattice with the periodic boundary condition. Assuming that system size in the $y$-direction is even, we introduce a unit cell which contains two consecutive  sites in the vertical direction, distinguished by the sublattice~$A$ and $B$~(Fig.~\ref{R1}). The Hamiltonian is defined by
\begin{eqnarray}
    H_{2D,2}=&-&\sum_v \left(X_{A,v}+Z_{B,v-e_y}X_{A,v}Z_{B,v}+X_{B,v}+Z_{A,v}X_{B,v}Z_{A,v+e_y}\right)\nonumber\\
    &-&\sum_v \left(Z_{A,v}Z_{A,v+e_x}+Z_{B,v}Z_{B,v+e_x}\right).\label{spin002}
\end{eqnarray}
Here, $v=(\hx,\hy)$ denotes spatial coordinate of a site and $e_x$ and $e_y$ does a unit vector in the $x$- and $y$-direction, respectively.
This Hamiltonian respects the following three $0$-form symmetries:
\begin{eqnarray}
    U_A=\prod_vX_{A,v},\quad U_B=\prod_{v}X_{B,v},\quad U_C=\prod_vCZ_{(A,v),(B,v)}\times CZ_{(B,v),(A,v+e_y)}. \label{lsmsym}
\end{eqnarray}
These three symmetries and translational symmetry in the $x$-direction exhibit the LSM anomaly. Indeed, when we decompose the lattice into vertical strip along the $x$-direction, accordingly,  by this decomposition, 
the symmetry operators~\eqref{lsmsym} act on each strip as 
\begin{eqnarray}
    U_A\to \prod_{\hy}X_{A,v}\Big|_{\hx=\hx_0},\quad  U_B\to \prod_{\hy}X_{B,v}\Big|_{\hx=\hx_0},\quad  U_C\to \prod_{\hy}CZ_{(A,v),(B,v)}\times CZ_{(B,v),(A,v+e_y)}\Big|_{\hx=\hx_0}.
\end{eqnarray}
The decomposed three symmetry operators exhibit the type III anomaly~\eqref{iii}\cite{Chen:2011pg,PhysRevB.86.115109, Seifnashri:2024dsd}. This implies that practically, the symmetry operators~\eqref{lsmsym} carry stack of such anomalies in the $x$-direction, allowing us to write an anomaly inflow action of these symmetries as
\begin{eqnarray}
     S=i\pi\int_{M_4}A^{(1)}\wedge B^{(1)}\wedge C^{(1)}\wedge e^x,
\end{eqnarray}
where $A^{(1)}$, $B^{(1)}$, $C^{(1)}$ correspond to gauge fields of the three $\mathbb{Z}_2$ $0$-form symmetries~\eqref{lsmsym}. 
The model exhibits the LSM anomaly involving three internal and one translational symmetries. In the following, we explore the dual symmetry structures that emerge upon gauging one, two, and three internal symmetries.

\subsubsection{Gauging one internal symmetry}
\begin{figure}
    \begin{center}
         \begin{subfigure}[h]{0.29\textwidth}
       \centering
  \includegraphics[width=0.7\textwidth]{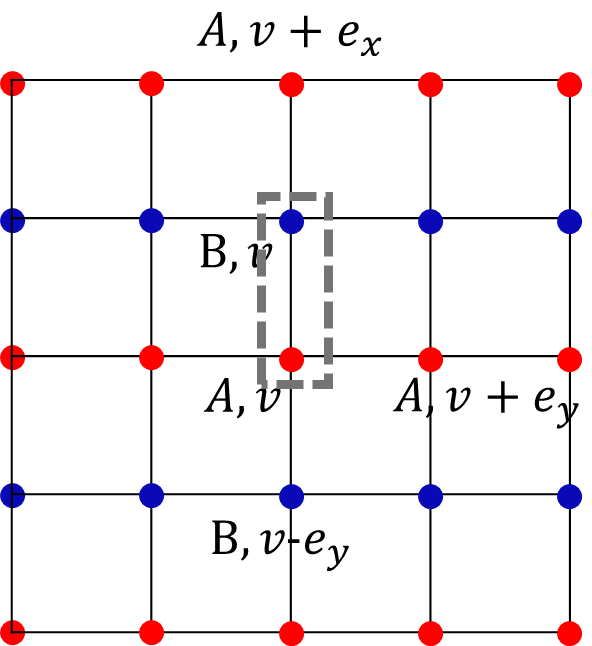}
         \caption{}\label{R1}
             \end{subfigure}
            \begin{subfigure}[h]{0.19\textwidth}
            \centering
  \includegraphics[width=1.2\textwidth]{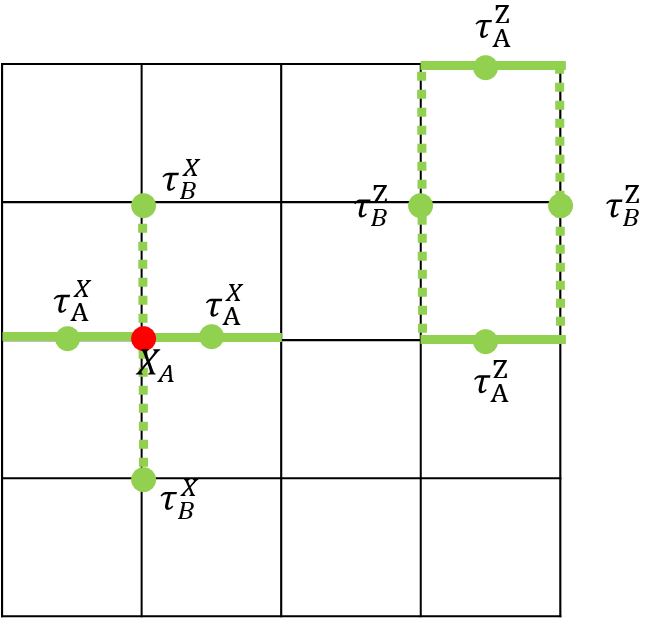}
         \caption{}\label{R2}
             \end{subfigure}

                      \begin{subfigure}[h]{0.2\textwidth}
            \centering
  \includegraphics[width=1.2\textwidth]{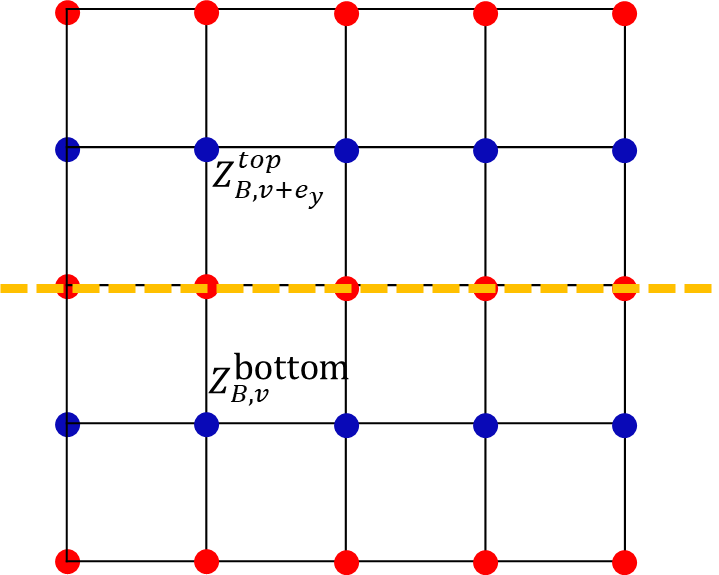}
         \caption{}\label{R3}
             \end{subfigure}
 \end{center}
 \caption{(a)~Configuration of the square lattice. We introduce a unite cell (gray dashed line) which contains two adjacent sites in the vertical direction, distinguished by $A$ (red) and $B$ (blue). (b)~Gauss law term and the plaquette operator that enters~\eqref{gas}
 and~\eqref{flat}, respectively. Coordinate dependence is omitted for simplicity. (c) Configuration of the symmetry defect regarding~$U_B$ (yellow dashed line) that goes along the~$x$-direction. 
 }
 \end{figure}

We gauge one of internal symmetries of this model, $U_A$, to discuss emergent symmetries. Introducing auxiliary qubits on sublattices $B$ and links between adjacent sites on $A$
whose Pauli operator is denoted as $\tau^{X/Z}_{B,v}$ and $\tau^{X/Z}_{A,l_x}$, the Gauss law reads
\begin{eqnarray}
    X_{A,v}\times\tau^X_{B,v}\tau^X_{B,v-e_y}\tau^X_{A,l_x}\tau^X_{A,l_x-e_x}=1.\label{gas}
\end{eqnarray}
Also, we impose the following flatness condition of the gauge fields:
\begin{eqnarray}
    \tau^Z_{B,v}\tau^Z_{B,v+e_x}\tau^Z_{A,l_x}\tau^Z_{A,l_x+e_y}=1.\label{flat}
\end{eqnarray}
See Fig.~\ref{R2}.
Gauging $\mathbb{Z}_2$ symmetry, $U_A$ induces a bijection map between the operators:
\begin{eqnarray}
    X_{A,v}\to \tau^X_{B,v}\tau^X_{B,v-e_y}\tau^X_{A,l_x}\tau^X_{A,l_x-e_x}(\vcentcolon=G^A_v),\quad Z_{A,v}Z_{A,v+e_y}\to \tau^Z_{B,v},\quad Z_{A,v}Z_{A,v+e_x}\to \tau^Z_{A,l_x}.\label{92}
\end{eqnarray}
Accordingly, the gauged Hamiltonian is given by
\begin{eqnarray}
    \widehat{H}_{2D,2}= -\sum_v \left(G^A_v+Z_{B,v-e_y}G^A_vZ_{B,v}+X_{B,v}+X_{B,v}\tau^Z_{B,v}\right)
    -\sum_v \left(\tau^Z_{A,l_x}+Z_{B,v}Z_{B,v+e_x}\right).\label{91}
\end{eqnarray}
\par
The gauged Hamiltonian~\eqref{91} has one $1$-form symmetry described by
\begin{eqnarray}
    \eta_x\vcentcolon=\prod_{l_x}\tau^Z_{A,l_x},\quad \eta_y\vcentcolon=\prod_{v}\tau^Z_{B,v},\label{ahat}
\end{eqnarray}
which is topological due to the flatness condition~\eqref{flat}. Also, the model~\eqref{91} possesses two $0$-form symmetries, corresponding to the original symmetries that are not gauged :
\begin{eqnarray}
    U_B=\prod_{v}X_{B,v},\quad \widehat{U}_C=\left(\prod_vCZ_{(A,v),(B,v)}\times CZ_{(B,v),(A,v+e_y)}\right)\times\prod_{v}  CZ_{(B,v),\tau_{B,v}}\label{bhat}
\end{eqnarray}
with 
\begin{eqnarray}
    CZ_{(B,v),\tau_{B,v}}\vcentcolon=e^{\frac{\pi i}{4}(1-Z_{B,v})(1-\tau^Z_{B,v})}.\label{cz}
\end{eqnarray}
Note that 
the symmetry operator $U_C$ given in~\eqref{lsmsym} is modified so that 
additional CZ terms~\eqref{cz} are multiplied to ensure that it commutes with the Gauss law~\eqref{gas}. \par
To study symmetry structures of~\eqref{ahat}\eqref{bhat}, we evaluate commutativity between $U_B$ and~$\widehat{U}_C$. After some algebra, we have 
\begin{eqnarray} \label{dipole_4.1_1}
    U_B\widehat{U}_C=\eta_y^{L_x}~\widehat{U}_CU_B
\end{eqnarray}
with $L_x$ being length of the lattice in the $x$-direction, indicating that depending on the parity of the system size, the two symmetries become projective. Similar to the LSM system in the previous case~(Sec.~\ref{4.1}), this is consistent with the previous consideration of symmetry structure
given in Sec.~\ref{sec.2.1}: two symmetries become projective, forming 2-group symmetry, in the presence of the translation symmetry defect, corresponding to~$L_x$ being odd. 
\par
    Furthermore, the modulated symmetry emerges in the presence of a defect associated with an internal symmetry. To see this, we set $L_x$ to be even (i.e., we turn off the translation symmetry defect), and accommodate a symmetry defect regarding~$U_B$ that goes along the $x$-direction at fixed $\hy=\hy_0$ as shown in Fig.~\ref{R3}. Following the similar argument presented in Sec.~\ref{sec.2.1}, such a defect is inserted by imposing a twisted boundary condition along a branch cut so that  
    \begin{eqnarray}
       -Z_{B,v+e_y}^{\text{top}} Z_{B,v}^{\text{bottom}}=\to +Z_{B,v+e_y}^{\text{top}}Z_{B,v}^{\text{bottom}}
    \end{eqnarray}
namely, Ising coupling between two  spin operators on sublattice $B$
located immediately adjacent to the defect, 
denoted by $   Z_{B,v}^{\text{bottom}}$ and $ Z_{B,v+e_y}^{\text{top}}$, 
acquires an additional minus sign. 
With this twisted boundary condition, we modify the symmetry operator $\widehat{U}_C$ so that is commutes with the twisted Hamiltonian:
%
%
\begin{eqnarray}
    \widehat{U}_C\Big|_{\text{defect}}=\prod_{\hx}\left(\tau^Z_{A,(\hx,\hy_0+1)}\right)^{\hx}\times \left(\prod_vCZ_{(A,v),(B,v)}\times CZ_{(B,v),(A,v+e_y)}\right)\times\prod_{v}  CZ_{(B,v),\tau_{B,v}}.
\end{eqnarray}
This implies that spatially modulated $1$-form symmetry is attached to the operator $\widehat{U}_C$ in the presence of the symmetry defect. Further, one can show that 
\begin{eqnarray} \label{dipole_4.1_2}
    T_x \widehat{U}_C\Big|_{\text{defect}}T_x^{-1}=\eta_{x}~\widehat{U}_C\Big|_{\text{defect}},
\end{eqnarray}
indicating $ \widehat{U}_C\Big|_{\text{defect}}$ and $\eta_x$ forms dipole algebra. As in Sec.~\ref{4.1.1}, these relations \eqref{dipole_4.1_1} and \eqref{dipole_4.1_2} are consistent with those in Sec.~\ref{sec.2.1}.


\subsubsection{Gauging two internal symmetries}
\begin{figure}
    \begin{center}
      \includegraphics[width=0.5\textwidth]{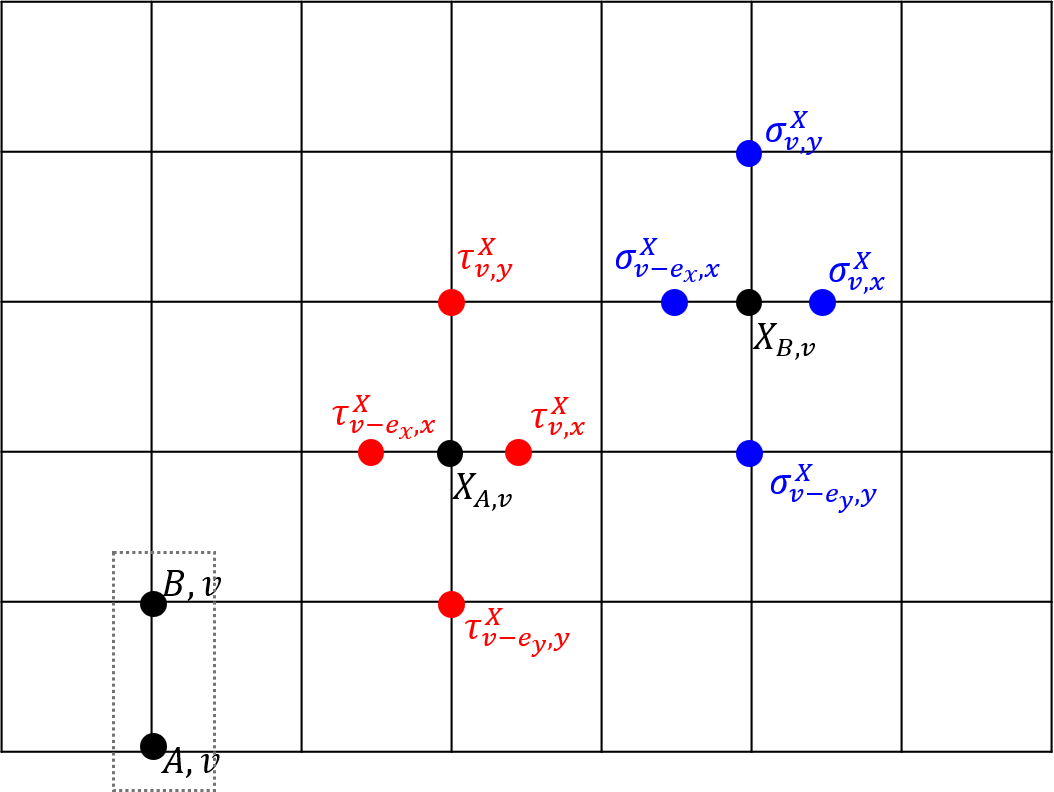}
         \caption{Gauss laws regarding gauging symmetries $U_A$ and $U_B$, described in~\eqref{gauss app}.}\label{R10}
          \end{center}
 \end{figure}
We explore gauging two internal symmetries, $U_A$ and $U_B$.
To gauge $U_A$ and $U_B$, we 
enlarge the Hilbert space by introducing two species of auxiliary qubits, denoted by $\tau$ and $\sigma$, associated with the $U_A$ and $U_B$ gauge fields, respectively. The variable~$\tau_{v,x}$ is assigned to the horizontal link connecting the neighboring $A$-sublattice sites at $v$ and $v+e_x$, while $\tau_{v,y}$ is placed at the intermediate position between the $A$-sublattice sites at $v$ and $v+e_y$, which coincides with the neighboring $B$-sublattice site (Although it is geometrically colocated with the neighboring $B$-sublattice site, it represents an independent auxiliary qubit.).  The $\sigma$ variables are introduced analogously with~$A$ and $B$ interchanged. The Gauss laws read (Fig.~\ref{R10})
\begin{eqnarray}
    X_{A,v}\times \tau^X_{v,x}\tau^X_{v-e_x,x}\tau^X_{v,y}\tau^X_{v-e_y,y}=1,\quad   X_{B,v}\times \sigma^X_{v,x}\sigma^X_{v-e_x,x}\sigma^X_{v,y}\sigma^X_{v-e_y,y}=1\quad\forall v\label{gauss app}
\end{eqnarray}
with fluxless condition
\begin{eqnarray}
     \tau^Z_{v,x}\tau^Z_{v+e_y,x}\tau^Z_{v,y}\tau^Z_{v+e_x,y}=1,\quad  \sigma^Z_{v,x}\sigma^Z_{v+e_y,x}\sigma^Z_{v,y}\sigma^Z_{v+e_x,y}=1\quad \forall v.\label{fluxless:app}
\end{eqnarray}
Using Gauss laws~\eqref{gauss app} and the standard gauging bijection, allowing us to identify gauge-invariant nearest neighboring $Z$-bilinears with the corresponding gauge link operators, e.g., 
~$Z_{A,v}Z_{A,v+e_y}\to \tau^Z_{v,y}$, 
$Z_{A,v}Z_{A,v+e_x}\to \tau^Z_{v,x}$, 
the gauged Hamiltonian is described by
\begin{eqnarray}
    H_{g,2}=-\sum_v\left[ \mathcal{A}_{v,\tau}(1+\sigma^Z_{v-e_y,y})+\mathcal{A}_{v,\sigma}(1+\tau^Z_{v,y}) + \tau^Z_{v,x}+\sigma^Z_{v,x} \right]\label{g2}
\end{eqnarray}
with fluxless condition~\eqref{fluxless:app}. Here, 
\begin{eqnarray*}
     \mathcal{A}_{v,\tau}\vcentcolon=\tau^X_{v,x}\tau^X_{v-e_x,x}\tau^X_{v,y}\tau^X_{v-e_y,y},\quad \mathcal{A}_{v,\sigma}\vcentcolon=\sigma^X_{v,x}\sigma^X_{v-e_x,x}\sigma^X_{v,y}\sigma^X_{v-e_y,y}.
\end{eqnarray*}
The gauged theory has two dual $1$-form symmetries, $\mathbb{Z}_2^{(1)}\times \mathbb{Z}_2^{(1)}$ whose corresponding topological operators are
\begin{eqnarray}
    Q_{A,\tau,x}=\prod _{\hx=0}^{L_x-1} \tau^Z_{(\hx,0),x},\quad   Q_{A,\tau,y}=\prod _{\hy=0}^{L_y-1} \tau^Z_{(0,\hy),y}\label{147}\\
     Q_{B,\sigma,x}=\prod _{\hx=0}^{L_x-1} \sigma^Z_{(\hx,0),x},\quad   Q_{B,\sigma,y}=\prod _{\hy=0}^{L_y-1} \sigma^Z_{(0,\hy),y}. \label{148}
\end{eqnarray}
Further, the gauged theory respects the symmetry, corresponding to the ungauged symmetry~$U_C$. To write its explicit form, we recall that before the gauging, the symmetry~$U_C$ acts on local Pauli~$X$ operators as
\begin{eqnarray}
    X_{A,v}\xrightarrow{\ {U_C}\ }Z_{B,v-e_y}X_{A,v}Z_{B,v},\quad   X_{B,v}\xrightarrow{\ {U_C}\ }Z_{A,v}X_{B,v}Z_{A,v+e_y}\label{Ucc}
\end{eqnarray}
while acting trivially on Pauli $Z$ operators. After gauging, such an action~\eqref{Ucc} is converted as 
\begin{eqnarray}
    \mathcal{A}_{v,\tau}\to \mathcal{A}_{v,\tau}\sigma^Z_{v-e_y,y},\quad \mathcal{A}_{v,\sigma}\to \mathcal{A}_{v,\sigma}\tau^Z_{v,y}.
\end{eqnarray}
Hence, we write a symmetry operator which is dual description of $U_c$ as
\begin{eqnarray}
    \widehat{U}_C=\prod_{\hx=0}^{L_x-1}\prod_{\hy=0}^{L_y-1} \left[CZ_{\tau_{v,x},\sigma_{v+e_x-e_y,y}}CZ_{\tau_{v,y},\sigma_{v,x}} \right]^{\lambda_x}.\label{uc1}
\end{eqnarray}
with $\lambda_x=\hx\mod 2$.
Here, for a moment, we assume $L_x$ is even. One can check that 
\begin{eqnarray}
    \widehat{U}_C\mathcal{A}_{v,\tau}\widehat{U}_C^\dagger=\left(\sigma^Z_{v+e_x-e_y,y}\sigma^Z_{v,x}\sigma^Z_{v-e_y,x}\right)^{\hx}\times (\sigma^Z_{v-e_y,y})^{\hx-1}\mathcal{A}_{v,\tau}=\mathcal{A}_{v,\tau}\sigma^Z_{v-e_y,y},\label{150}
\end{eqnarray}
where in the last equality, we have used the fluxless condition~\eqref{fluxless:app}. Analogous calculation leads to 
\begin{eqnarray}
    \widehat{U}_C\mathcal{A}_{v,\sigma}\widehat{U}^\dagger_C=\mathcal{A}_{v,\sigma}\tau^Z_{v,y}.\label{151}
\end{eqnarray}
From this relation, one can show that $\widehat{U}_C$ is the symmetry of the gauged Hamiltonian~\eqref{g2}. Obviously, this symmetry is invertible. 
\par
For $L_x$ odd, the operator $\widehat{U}_C$ no longer reproduces the gauged action of $U_C$ on all local operators at $\hx=0$, which can be seen from
\begin{eqnarray}
    \widehat{U}_C \mathcal{A}_{(0,\hy),\tau}\widehat{U}_C^\dagger=\mathcal{A}_{(0,\hy),\tau},\quad  \widehat{U}_C \mathcal{A}_{(0,\hy),\sigma}\widehat{U}_C^\dagger=\mathcal{A}_{(0,\hy),\sigma},
\end{eqnarray}
rather than~\eqref{150}\eqref{151}. 
To reproduce these transformations throughout the lattice, we supplement~$\widehat{U}_C$ with an operator supported along $\hx=0$. Let us define
the following operator:
\begin{eqnarray}
\mathcal{V}\vcentcolon=\prod_{\hy=0}^{L_y-1}C\left(\mathcal{S}_{\hy}^\tau,\sigma^Z_{(0,\hy-1),y}\right),
\end{eqnarray}
where $C(A.B)=\frac{1}{2}(1+A+B-AB)$ represents controlled phase gate-like operator for $A$ and $B$ consisting of Pauli $Z$ operators and   
\begin{eqnarray}
\mathcal{S}^\tau_{\hy}\vcentcolon=\prod_{j=0}^{\hy-1}\tau^Z_{(0,j),y},
\end{eqnarray}
which denotes string of gauge fields $\tau^Z$ from an origin to a point along the line with $\hx=0$. Also, $\mathcal{S}^\tau_0=1$ and $\hy-1$ can be understood as modulo $L_y$.\par
The operator $C\left(\mathcal{S}_{\hy}^\tau,\sigma^Z_{(0,\hy-1),y}\right)$ has odd number of overlaps with $\mathcal{A}_{(0,\hy),\tau}$ at the origin or endpoint of the string of the gauge fields. Therefore, for $v=(0,\hy)$, it follows that  
\begin{eqnarray}
\mathcal{V}\mathcal{A}_{(0,\hy),\tau}\mathcal{V}^\dagger=\begin{cases}
        \mathcal{A}_{(0,\hy),\tau}\sigma^Z_{(0,\hy-1),y}\quad (\hy\neq 0)\\
        \mathcal{A}_{(0,0),\tau}\sigma^Z_{(0,L_y-1),y}Q_{B,\sigma,y}\quad (\hy=0),
    \end{cases}
\end{eqnarray}
where $1$-form operator $Q_{B,\sigma,y}$ is given in~\eqref{148}.
Likewise, one finds
\begin{eqnarray}
    \mathcal{V}\mathcal{A}_{(0,\hy),\sigma}\mathcal{V}^\dagger=\begin{cases}
        \mathcal{A}_{(0,\hy),\sigma}\tau^Z_{(0,\hy),y}\quad (\hy\neq L_y-1)\\
        \mathcal{A}_{(0,L_y-1),\sigma}\tau^Z_{(0,L_y-1),y}Q_{A,\tau,y}\quad (\hy=L_y-1)
    \end{cases}
\end{eqnarray}
with $1$-form operator  $Q_{A,\tau,y}$ being defined in~\eqref{147}. Therefore, up to $1$-form loop operators, $\mathcal{V}$ is the desired operator to give the relation~\eqref{150}\eqref{151} along a line with $v=(0,\hy)$. We further multiply the following projector 
\begin{eqnarray}
   \Pi_1\vcentcolon=\frac{1}{4}(1+Q_{A,\tau,y})(1+Q_{B,\sigma,y})
\end{eqnarray}
and introduce a new operator as 
\begin{eqnarray}
    \widehat{V}_C\vcentcolon=2\widehat{U}_C\mathcal{V}\Pi_1,
\end{eqnarray}
which satisfies
\begin{eqnarray}
    \widehat{V}_C\mathcal{A}_{v,\tau}=\sigma^Z_{v-e_y,y}\mathcal{A}_{v,\tau}\widehat{V}_C,\quad  \widehat{V}_C\mathcal{A}_{v,\sigma}=\tau^Z_{v,y}\mathcal{A}_{v,\sigma}\widehat{V}_C
\end{eqnarray}
for any node $v$. From this property, $\widehat{V}_C$ is now the symmetry of the gauged theory~\eqref{g2}. This symmetry, however, is non-invertible as seen from 
the fusion rule:
\begin{eqnarray}
    \widehat{V}_C\times \widehat{V}_C=(1+Q_{A,\tau,y})(1+Q_{B,\sigma,y}). \label{162}
\end{eqnarray}
In summary, upon gauging the two internal symmetries, $U_A$ and $U_B$, the nature of the residual symmetry associated with $U_C$ depends qualitatively on the parity of $L_x$. For even $L_x$, it remains invertible, whereas for odd $L_x$, corresponding to the insertion of a nontrivial translation defect, it becomes non-invertible and obeys the fusion rule~\eqref{162}. This provides the crystalline counterpart of the defect-dependent non-invertibility found in the purely internal case of Sec.~\ref{sec2.2}.

\subsubsection{Gauging three internal symmetries}
\begin{figure}
    \begin{center}
      \includegraphics[width=0.5\textwidth]{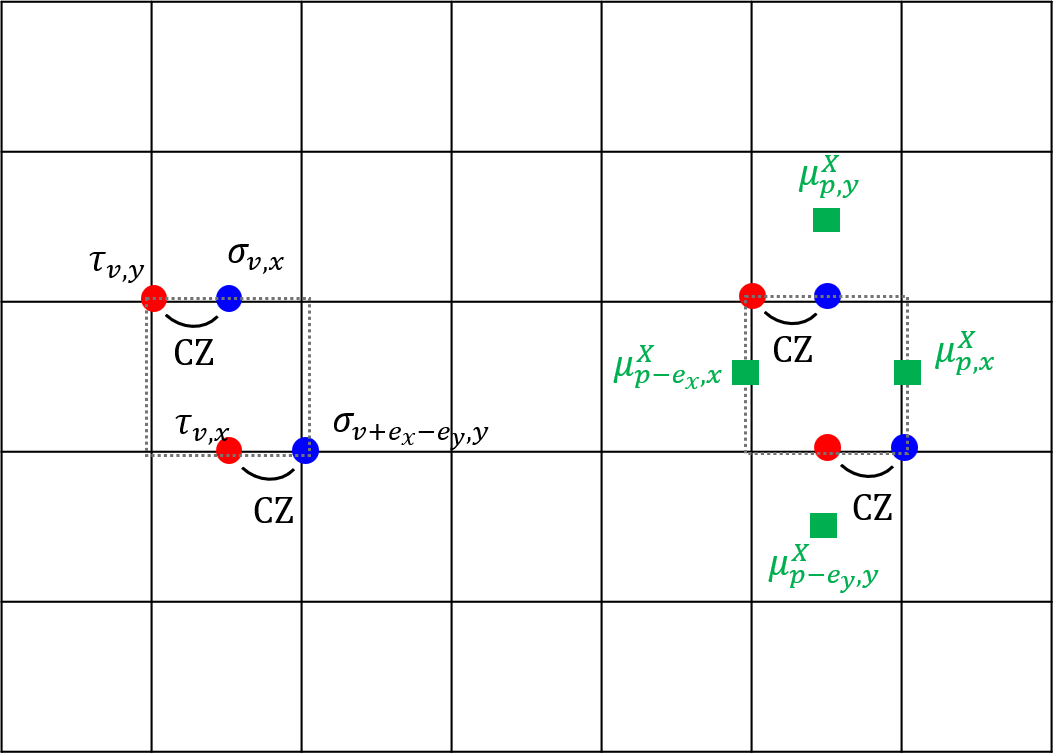}
         \caption{(Left) CZ operators that enter in the right hand side of~\eqref{uc1}. (Right)
         The Gauss Law when gauging  $\widehat{U}_C$.}\label{fgiC}
          \end{center}
 \end{figure}
Focusing on the case of even $L_x$, we further gauge the symmetry $\widehat{U}_C$~\eqref{uc1}, which amounts gauging three internal symmetries in the LSM system. To do so, we introduce additional qubit d.o.f.s on each vertical link emanating from $A$ sublattice $A$ to the adjacent $B$ sublattice in the $+y$ direction denoted as $\mu_{p,x}$, as well as on each plaquette below a horizontal link connecting adjacent $A$ sublattices, described by $\mu_{p,y}$. 
Here, $p$ denotes a coordinate of a plaquette, $p=(\hx+1/2,\hy+1/2)$. 
See also Fig.~\ref{fgiC}. 
The Gauss law is then given by
\begin{eqnarray}
\mathcal{G}_p^C=1\quad \forall p\label{gaussC}
\end{eqnarray}
with 
\begin{eqnarray}
  \mathcal{G}^C_p\vcentcolon=   \mu^X_{p,x}\mu^X_{p-e_x,x}\mu^X_{p,y}\mu^X_{p-e_y,y}\times\left[CZ_{\tau_{v,x},\sigma_{v+e_x-e_y,y}}CZ_{\tau_{v,y},\sigma_{v,x}} \right]^{\hx}.
\end{eqnarray}
It is straightforward to verify that  $\prod_p\mathcal{G}^C_p=\widehat{U}_C$ -- consistent with the fact that gauging is a procedure to promote a global symmetry to a local one. 
We further impose the fluxless condition as 
\begin{eqnarray}
\mu^Z_{p,x}\mu^Z_{p+e_y,x}\mu^Z_{p,y}\mu_{p+e_x,y}^Z=1~\forall p.    \label{fluxlesC}
\end{eqnarray}
After imposing the Gauss law~\eqref{gaussC}, terms in the Hamiltonian~\eqref{g2}, especially the first two terms, have to be modified to commute the Gauss law, corresponding to the minimal coupling. Indeed, we modify them as 
\begin{eqnarray}
    \mathcal{A}_{v,\tau}(1+\sigma^Z_{v-e_y,y})\to \mathcal{A}_{v,\tau}(1+\sigma^Z_{v-e_y,y}) \left[CZ_{\mu_{p-e_y,y},\sigma_{v-e_y,x}}\right]^{\hx},\nonumber\\
\mathcal{A}_{v,\sigma}(1+\tau^Z_{v,y})\to \mathcal{A}_{v,\sigma}(1+\tau^Z_{v,y})\left[CZ_{\mu_{p-e_x,y},\tau_{v-e_x+e_y,x}}\right]^{\hx-1} .   \label{tt}
\end{eqnarray}
One can verify that terms in the right hand side of~\eqref{tt} commute with~\eqref{gaussC}.
The gauged Hamiltonian is therefore given by
\begin{eqnarray}
    \widehat{H}_{g,2}=&-&\sum_v\Bigg(\mathcal{A}_{v,\tau}(1+\sigma^Z_{v-e_y,y}) \left[CZ_{\mu_{p-e_y,y},\sigma_{v-e_y,x}}\right]^{\hx}+\mathcal{A}_{v,\sigma}(1+\tau^Z_{v,y})\left[CZ_{\mu_{p-e_x,y},\tau_{v-e_x+e_y,x}}\right]^{\hx-1}\Biggr)\nonumber\\
   & -&\sum_v (\tau^Z_{v,x}+\sigma^Z_{v,x})\label{final}
\end{eqnarray}
with the fluxless conditions~\eqref{fluxless:app}~\eqref{fluxlesC}. 
In addition to the two $1$-form symmetries~\eqref{147}\eqref{148}, the gauged theory~\eqref{final} respects the following $1$-form symmetry
\begin{eqnarray}
       Q_{C,\mu,x}=\prod _{\hx=0}^{L_x-1} \mu^Z_{(\hx+1/2,1/2),x},\quad   Q_{C,\mu,y}=\prod _{\hy=0}^{L_y-1} \mu^Z_{(1/2,\hy+1/2),y}. \label{149}
\end{eqnarray}
Hence, in total, there are three $\mathbb{Z}_2^{(1)}$ symmetries after gauging $U_A$, $U_B$, and $U_C$.

In the gauged theory, 
since the Gauss law~\eqref{gaussC} contains position dependent term, 
the translation in the $x$-direction is not well-defined, that is, 
\begin{eqnarray}
    T_x \mathcal{G}^C_v \neq \mathcal{G}_{v+e_x}^CT_x.
\end{eqnarray}
To restore compatibility with the Gauss constraints, we dress the translation operator with
$\mathcal{V}_{\bar{S}TS}$, constructed through the sequence of topological manipulations described in
Appendix.~\ref{appendixA}. 
The construction consists of three steps, applied from right to left in $\bar{S}TS$.

 First, $S$ 
 gauges the three dual $1$-form symmetries, $\mathbb{Z}_2^{(1)}\times \mathbb{Z}_2^{(1)}\times \mathbb{Z}_2^{(1)}$. The resulting theory is described by three species of qubit d.o.f, with Pauli operators $(X_{A,v},Z_{A,v})$, $(X_{B,v},Z_{B,v})$, and $(X_{C,p},Z_{C,p})$. The first two species reside on the $A$ and $B$ sublattices, respectively, while the third one is introduced at the center of a plaquette. 
By this gauging, the gauge field star operators map to local Pauli $X$ operators as
\begin{eqnarray}
    \mathcal{A}_{v,\tau}\xrightarrow{\ {{S}}\ } X_{A,v},\quad  \mathcal{A}_{v,\sigma}\xrightarrow{\ {{S}}\ } X_{B,v},\quad \mu^X_{p,x}\mu^X_{p-e_x,x}\mu^X_{p,y}\mu^X_{p-e_y,y}\xrightarrow{\ {{S}}\ } X_{C,p}
\end{eqnarray}
Meanwhile, the link Pauli $Z$ operators map to nearest neighboring spin bilinears;
\begin{eqnarray}
    \tau^Z_{v,i}\xrightarrow{\ {{S}}\ }Z_{A,v}Z_{A,v+e_i},\quad \sigma^Z_{v,i}\xrightarrow{\ {{S}}\ }Z_{B,v}Z_{B,v+e_i},\quad \mu^Z_{p,i}\xrightarrow{\ {{S}}\ }Z_{C,p}Z_{C,p+e_i}\quad i=x,y.
\end{eqnarray}
Next, 
 $T$ acts on the resulting spin system by applying the unitary operator
$U_{CCZ}$ defined by
\begin{eqnarray}
    U_{CCZ}\vcentcolon=\prod_{\hx,\hy} C(Z_{C,p},Z_{A,v}Z_{A,v+e_y},Z_{B,v}Z_{B,v+e_x})\nonumber\\
    \times C(Z_{C,p},Z_{A,v}Z_{A,v+e_x},Z_{B,v+e_x}Z_{B,v+e_x-e_y}),
\end{eqnarray}
where 
\begin{eqnarray*}
    C(P,Q,R)\vcentcolon=1-\frac{1}{4}(1-P)(1-Q)(1-R)
\end{eqnarray*}
denotes generalized  CCZ gate with $P$, $Q$, and $R$ given by products of Pauli $Z$ operators. 
Finally, ~$\bar{S}$ gauges the three resulting $\mathbb{Z}_2$ $0$-form symmetries, returning to a description in terms of the gauge fields $\tau$, $\sigma$ and $\mu$. 
Together, these three steps define the dressing operator $\mathcal{V}_{\bar{S}TS}$.
Its action on the Gauss law reads
\begin{eqnarray}
    \mathcal{V}_{\bar{S}TS}\mathcal{G}^C_v=\mathcal{G}^C_v\left[CZ_{\tau_{v,x},\sigma_{v+e_x-e_y,y}}CZ_{\tau_{v,y},\sigma_{v,x}} \right] \mathcal{V}_{\bar{S}TS}. 
\end{eqnarray}
Hence, modify the translation operator as $\tilde{T}_x\vcentcolon= T_x\mathcal{V}_{\bar{S}TS}$, we have 
\begin{eqnarray}
     \tilde{T}_x \mathcal{G}^C_v =\mathcal{G}_{v+e_x}^C\tilde{T}_x. 
\end{eqnarray}
Further $\tilde{T}_x$ commutes with the gauged Hamiltonian~\eqref{final} as
\begin{eqnarray}
   \mathcal{V}_{\bar{S}TS}(\mathcal{A}_{v,\tau}(1+\sigma^Z_{v-e_y,y}) )&=&(\mathcal{A}_{v,\tau}(1+\sigma^Z_{v-e_y,y}) )CZ_{\mu_{p-e_y,y},\sigma_{v-e_y,x}} \mathcal{V}_{\bar{S}TS}\nonumber\\
      \mathcal{V}_{\bar{S}TS}(\mathcal{A}_{v,\sigma}(1+\tau^Z_{v,y}) )&=&(\mathcal{A}_{v,\sigma}(1+\tau^Z_{v,y}) )CZ_{\mu_{p-e_x,y},\tau_{v-e_x+e_y,x}}\mathcal{V}_{\bar{S}TS}.
\end{eqnarray}

The modified translation becomes non-invertible. Recalling the discussion around~\eqref{47} where multiplying $\mathcal{V}_{\bar{S}TS}$ twice gives condensation defects associated with three $1$-form symmetries, and defining dressed translation operator in the $-x$ direction as $\tilde{T}_{-x}\vcentcolon=\mathcal{V}_{\bar{S}TS}^\dagger T_{x}^\dagger$, one finds
\begin{eqnarray}
   \tilde{T}_x\times \tilde{T}_{-x}=\frac{1}{2^3}\prod_{i=x,y}\left[(1+Q_{A,\tau,i})(1+Q_{B,\sigma,i})(1+Q_{C,\mu,i})\right].\label{194}
\end{eqnarray}
Thus, gauging three symmetries yields a non-invertible translation whose fusion rule is analogous to that of the $2$-representation category of a $2$-group symmetry, consistent with Sec.~\ref{sec2.3}
\par 

In summary, we have constructed two lattice models whose LSM anomalies originate from the type IV anomaly. We have shown that lattice translations play a role analogous to that of background symmetry defects, so that the anomaly structure of the purely internal setting persists in a form intertwined with the lattice geometry. Gauging internal symmetries in these models yields emergent symmetry structures closely related to those discussed in Secs.~\ref{sec2} and~\ref{sec4}. These results connect LSM constraints, modulated symmetries, and emergent higher symmetries through the underlying type IV anomaly.
Dual symmetries explored in this section are listed in Table.~\ref{defect_hierarchy}.

\section{Conclusion}\label{sec5}
In this work, we have provided a concrete lattice realization of type IV~\tht~anomalies and systematically uncovered their emergent symmetries. By explicitly gauging subgroups of the anomalous symmetry, 
we have shown that different choices of gauged subgroups give rise to a hierarchy of generalized symmetries, including 2-group symmetries, non-invertible symmetries, and higher categorical symmetries, all of which admit a unified understanding within a single anomaly framework.
We provide a field-theoretical analysis on these results, emphasizing the universality of the emergent symmetry structures (Table~\ref{defect_hierarchy}). 
\par
As a key application, we have established a direct correspondence between type IV anomalies and systems with LSM anomalies by replacing part of internal symmetries with translational symmetries. In this setting, modulated symmetries and non-invertible translation emerge naturally upon gauging internal symmetries, providing a unified perspective on previously observed phenomena. Importantly, we have demonstrated that their emergence depends sensitively on the presence of symmetry defects and on global properties such as system size parity, revealing a qualitatively new aspect of modulated symmetries that is invisible in conventional formulations.
Our results thus uncover a common origin of a wide class of exotic symmetry structures---from higher-group to non-invertible and modulated symmetries---rooted in higher-order anomalies. This provides a concrete bridge between anomaly-based classification, lattice realizations, and LSM constraints, and thus suggests a unifying organizing principle for quantum many-body systems beyond conventional symmetry paradigms.

Several important directions remain for future investigations. While we focused on explicit lattice constructions, it would be desirable to develop a systematic classification of LSM constraints associated with higher (e.g., type IV and beyond) anomalies. In particular, understanding their relation to generalized cohomology classifications and spectral sequence approaches could clarify the distinction between weak and strong modulated SPT phases.\par

Another intriguing direction would be the connection to quantum information, in particular to symmetry-protected logical operations and transversal gates. Since our construction realizes higher and non-invertible symmetry structures on the lattice, it is natural to ask whether these symmetries can be leveraged to implement novel classes of fault-tolerant logical gates beyond conventional transversal operations. In particular, the interplay between modulated symmetries, higher-form symmetries, and non-invertible defects may provide a new mechanism for protected logical operations that are not captured by standard group-based symmetry frameworks. Exploring this possibility could shed light on generalized notions of fault tolerance and may lead to new paradigms for quantum error-correcting codes rooted in higher and categorical symmetries.
\section*{Acknowledgement}
We thank Takamasa Ando, Lucas Z. Brito,
Weiguang Cao, Bo Han, Masazumi Honda, Satoshi Iso,
Jun Maeda, Salvatore D. Pace,
Takuma Saito, Soichiro Shimamori, Nathanan Tantivasadakarn for helpful discussions. This work is in part supported by JST CREST (Grant No.~JPMJCR24I3), JST SPRING (Grant No.~JPMJSP2110).
\begin{appendix}

\section{Topological manipulation on the lattice} \label{appendixA}
In this appendix, we give a detailed derivation of~\eqref{44}, that is, 
the action of the topological manipulation~$\Bar{S}TS$ on local operators. To derive it, we first discuss the topological manipulations,~$\bar{S}$,~$T$, and $S$ one by one. 
\subsection{Gauging $1$-form symmetry}
We start by demonstrating how to gauge $\mb$ $1$-form symmetry on the triangular lattice, corresponding to~$\bar{S}$.
Consider the $1$-form symmetry generated by
\begin{equation}
    \eta_{A}^{(1)}(\gamma_a)=\prod_{\langle bc\rangle\in\gamma_a}\sigma_{\langle bc\rangle}^z,
\end{equation}
where $\gamma_a$ is a noncontractible loop on the sublattice $\Lambda_a$. The symmetric local terms are given by 
\begin{equation}
    \sigma_{\langle bc\rangle}^z, \quad \prod_{\langle bc\rangle\subset \langle abc \rangle}\sigma_{\langle bc\rangle}^x,
\end{equation}
where the second term is the plaquette term around the site $a$.
To gauge this symmetry, we introduce gauge variables $X_a, Z_a$ on the sites $a\in\Lambda_a$. The Gauss law constraint is given by
\begin{equation}
    G_{A,\langle bc\rangle}^{(1)}:=Z_{a_1}\sigma_{\langle bc\rangle}^zZ_{a_2} =1, \quad \forall\langle bc\rangle\in\Lambda_{bc},
\end{equation}
where $a_1$ and $a_2$ are the sites on $\Lambda_a$ adjacent to the link $\langle bc\rangle$. The symmetric local terms should be modified to be gauge-invariant as
\begin{equation}
     \sigma_{\langle bc\rangle}^z, \quad X_a\prod_{\langle bc\rangle\subset \langle abc \rangle}\sigma_{\langle bc\rangle}^x.
\end{equation}
Furthermore, to solve the Gauss law constraint, we implement the following unitary transformation,
\begin{equation}
    \sigma_{\langle bc\rangle}^z \rightarrow Z_{a_1}\sigma_{\langle bc\rangle}^zZ_{a_2}, \quad X_a\rightarrow X_a\prod_{\langle bc\rangle\subset \langle abc \rangle}\sigma_{\langle bc\rangle}^x.
\end{equation}
Under this transformation, the Gauss law constraint becomes $\sigma_{\langle bc\rangle}^z=1,\ \forall\langle bc\rangle\in\Lambda_{bc}$, allowing us to find the following gauging map:
\begin{equation}
    \sigma_{\langle bc\rangle}^z \rightarrow Z_{a_1}Z_{a_2},\quad \prod_{\langle bc\rangle\subset \langle abc \rangle}\sigma_{\langle bc\rangle}^x \rightarrow X_a, \label{Appendix:gauging map1}
\end{equation}
Similar map for the other labels $\tau, \mu$ can be obtained. 
Note that the theory after gauging three $1$-form symmetries has three dual $0$-form symmetries, generated by
\begin{equation}
    U_A=\prod_{a\in\Lambda_a} X_a, \quad U_B=\prod_{b\in\Lambda_b} X_b, \quad U_C=\prod_{c\in\Lambda_c} X_c. \label{Appendix:0-form sym}
\end{equation}
\subsection{Stacking of SPT phase}
We turn to  discuss the stacking of an SPT phase protected by three $\mb$ $0$-form symmetries~\eqref{Appendix:0-form sym}, corresponding to $T$. 
In particular, we focus on the SPT phase characterized by the $3$-cocycle~$(-1)^{\int A\cup B \cup C}$. The Hamiltonian realizing this SPT phase \cite{Chen:2011pg,Yoshida:2015cia} is given by
\begin{equation}
    H_{\text{SPT}} = -\sum_{a\in\Lambda_a} X_a\prod_{\langle bc\rangle \subset \langle abc\rangle}CZ_{bc} - \sum_{b\in\Lambda_b} X_b\prod_{\langle ac\rangle \subset \langle abc\rangle}CZ_{ac} - \sum_{c\in\Lambda_c} X_c\prod_{\langle ab\rangle \subset \langle abc\rangle}CZ_{ab},
\end{equation}
which can be obtained from the trivial SPT Hamiltonian
\begin{equation}
    H_0 = -\sum_{a\in\Lambda_a}X_a-\sum_{b\in\Lambda_b}X_b-\sum_{c\in\Lambda_c}X_c
\end{equation}
by a unitary transformation, called the SPT entangler. The SPT entangler is given by 
\begin{equation}
    V = \prod_{\langle abc\rangle}CCZ_{abc}. \label{Appendix:stacking SPT}
\end{equation}
Thus, stacking with this SPT phase $(T)$ is realized by the unitary operator $V$.

\subsection{Gauging $0$-form symmetry}
We review the gauging procedure of $0$-form symmetry generated by
\begin{equation}
    U_A=\prod_{a\in\Lambda_a}X_a,
\end{equation}
which corresponds to $S$.
Although this procedure is discussed in the main text (Sec.~\ref{sec.2.1}), we give a brief review of it for the sake of the completeness.\par
The symmetric local terms are given by the following two terms:
\begin{equation}
    X_a,\quad Z_{a_1}Z_{a_2}, \quad a,a_1,a_2\in\Lambda_a
\end{equation}
where $a_1$ and $a_2$ are the nearest-neighbor sites on the sublattice $\Lambda_a$. To gauge this symmetry, we introduce gauge variables $\sigma_{\langle bc \rangle}^x, \sigma_{\langle bc \rangle}^z$ on the links $\langle bc \rangle\in \Lambda_{bc}$, and impose the Gauss law constraint described by
\begin{equation}
    G_{A,a}:=X_a\prod_{\langle bc\rangle\subset \langle abc \rangle}\sigma_{\langle bc\rangle}^x = 1, \quad \forall a\in\Lambda_a.
\end{equation}
Also, we impose the flatness condition
\begin{equation}
    B_{A}(\gamma_a):=\prod_{\langle bc\rangle\subset \gamma_a}\sigma_{\langle bc\rangle}^z =1, \quad \forall\gamma_a, 
\end{equation}
to ensure that the gauged theory does not admit excess flux.
Here, $\gamma_a$ denotes a contractible loop on the sublattice $\Lambda_a$, and the product is taken over all links $\langle bc\rangle$ intersected by $\gamma_a$.
By minimally coupling to gauge variables, the gauge-invariant local terms take the form
\begin{equation}
    X_a,\quad Z_{a_1}\sigma_{\langle bc \rangle}^zZ_{a_2}, \quad a,a_1,a_2\in\Lambda_a,
\end{equation}
where $\langle bc\rangle$ denotes a link intersected by the line connecting $a_1$ and $a_2$. Furthermore, to solve the Gauss law constraint, we implement the unitary transformation so that 
\begin{equation}
    X_a \rightarrow X_a\prod_{\langle bc\rangle\subset \langle abc \rangle}\sigma_{\langle bc\rangle}^x, \quad \sigma_{\langle bc \rangle}^z \rightarrow Z_{a_1}\sigma_{\langle bc \rangle}^zZ_{a_2}.
\end{equation}
From this transformation, the Gauss law constraint becomes $X_a=1, \ \forall a\in\Lambda_a$. Thus, we find the gauging map
\begin{equation}
    X_a \rightarrow \prod_{\langle bc\rangle\subset \langle abc \rangle}\sigma_{\langle bc\rangle}^x, \quad Z_{a_1}Z_{a_2} \rightarrow \sigma_{\langle bc \rangle}^z, \label{Appendix:gauging map2}
\end{equation}
and similarly for the other labels $b\in\Lambda_b, c\in\Lambda_c$. The theory after gauging three $0$-form symmetries, has three dual $1$-form symmetries generated by 
 \begin{equation}
\begin{aligned}
    \eta_{A}^{(1)}(\gamma_a)=\prod_{\langle bc\rangle\in\gamma_a}\sigma_{\langle bc\rangle}^z, \quad \eta_{B}^{(1)}(\gamma_b)=\prod_{\langle ac\rangle\in\gamma_b}\tau_{\langle ac\rangle}^z, \quad 
    \eta_{C}^{(1)}(\gamma_c)=\prod_{\langle bc\rangle\in\gamma_c}\mu_{\langle ab\rangle}^z, \label{1-form sym}
\end{aligned}
\end{equation}
 where $\gamma_i$ is a noncontractible loop on the sublattice $\Lambda_i$ $(i=a,b,c)$.

 \subsection{Action of $\overline{S}TS$}
\begin{figure}
    \centering
    \begin{tikzpicture}[baseline=(current bounding box.center),scale=0.8]
\def\r{1.6}
\coordinate (v1) at (120:\r);
\coordinate (v2) at (60:\r);
\coordinate (v3) at (0:\r);
\coordinate (v4) at (-60:\r);
\coordinate (v5) at (-120:\r);
\coordinate (v6) at (180:\r);
\coordinate (c)  at (0,0);

\draw[red, line width=1.2pt]
  (v1)--(v2)--(v3)--(v4)--(v5)--(v6)--cycle;

\draw (v1)--(v4);
\draw (v2)--(v5);
\draw (v3)--(v6);

\node at ($(c)+(0,0.4)$) {$a$};

\node at (v1) [above left] {$b_1$};
\node at (v2) [above right] {$c_1$};
\node at (v3) [right] {$b_2$};
\node at (v4) [below right] {$c_2$};
\node at (v5) [below left] {$b_3$};
\node at (v6) [left] {$c_3$};

\node at ($(v1)!0.5!(v2)$) [above] {$\color{red}CZ$};
\node at ($(v2)!0.5!(v3)$) [above right] {$\color{red}CZ$};
\node at ($(v3)!0.5!(v4)$) [below right] {$\color{red}CZ$};
\node at ($(v4)!0.5!(v5)$) [below] {$\color{red}CZ$};
\node at ($(v5)!0.5!(v6)$) [below left] {$\color{red}CZ$};
\node at ($(v6)!0.5!(v1)$) [above left] {$\color{red}CZ$};
\end{tikzpicture}
\;$\xrightarrow{\quad \Bar{S} \quad }$\;
\begin{tikzpicture}[baseline=(current bounding box.center),scale=0.8]
\def\r{1.6}
\coordinate (v1) at (120:\r);
\coordinate (v2) at (60:\r);
\coordinate (v3) at (0:\r);
\coordinate (v4) at (-60:\r);
\coordinate (v5) at (-120:\r);
\coordinate (v6) at (180:\r);
\coordinate (c)  at (0,0);

\draw
  (v1)--(v2)--(v3)--(v4)--(v5)--(v6)--cycle;

\draw (v1)--(v4);
\draw (v2)--(v5);
\draw (v3)--(v6);
\node at ($(c)+(0,0.4)$) {$a$};

\node at (v1) [above left] {$b_1$};
\node at (v2) [above right] {$c_1$};
\node at (v3) [right] {$b_2$};
\node at (v4) [below right] {$c_2$};
\node at (v5) [below left] {$b_3$};
\node at (v6) [left] {$c_3$};

\node at ($(c)$) [above=15pt] {$\color{gray}CZ$};
\node at ($(c)+(1.0,0.5)$) {$\color{gray}CZ$};
\node at ($(c)+(-1.0,0.5)$) {$\color{gray}CZ$};
\node at ($(c)+(1.0,-0.5)$) {$\color{gray}CZ$};
\node at ($(c)+(-1.0,-0.5)$) {$\color{gray}CZ$};
\node at ($(c)$) [below=15pt] {$\color{gray}CZ$};

\fill[red] ($(v2)!0.5!(c)$) circle (3pt);
\fill[red] ($(v4)!0.5!(c)$) circle (3pt);
\fill[red] ($(v6)!0.5!(c)$) circle (3pt);
\fill[blue] ($(v1)!0.5!(c)$) circle (3pt);
\fill[blue] ($(v3)!0.5!(c)$) circle (3pt);
\fill[blue] ($(v5)!0.5!(c)$) circle (3pt);

\draw[gray, line width=1.4pt]
  ($(v1)!0.5!(c)$) --
  ($(v2)!0.5!(c)$) --
  ($(v3)!0.5!(c)$) --
  ($(v4)!0.5!(c)$) --
  ($(v5)!0.5!(c)$) --
  ($(v6)!0.5!(c)$) --
  cycle;
\end{tikzpicture}
    \caption{Action of $\Bar{S}$ on product of $CZ$ gates.}
    \label{Fig:appendixA}
\end{figure}
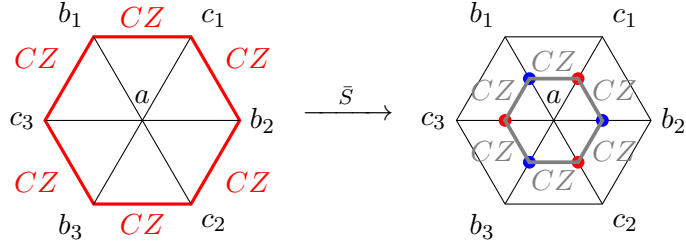

 Combing all the discussions in the previous subsections together, 
 we are now in a good place to derive the action of the topological manipulation $\Bar{S}TS$ on the local operators~\eqref{44}. Using the gauging maps \eqref{Appendix:gauging map1}, \eqref{Appendix:gauging map2} and the stacking of SPT phase \eqref{Appendix:stacking SPT}, we find
 \begin{align}
\sigma_{\langle bc\rangle}^z &\xrightarrow{\quad S\quad}Z_{a_1}Z_{a_2}  \xrightarrow{\quad T \quad} Z_{a_1}Z_{a_2} \xrightarrow{\quad \Bar{S}\quad} \sigma_{\langle bc\rangle}^z,
\end{align}
\begin{align}
     \prod_{\langle bc\rangle\subset \langle abc \rangle}\sigma_{\langle bc\rangle}^x &\xrightarrow{\quad S\quad} X_a \xrightarrow{\quad T \quad} X_a\prod_{\langle bc\rangle \subset \langle abc\rangle}CZ_{bc} \xrightarrow{\quad \Bar{S}\quad} 
     \prod_{\langle bc\rangle\subset \langle abc \rangle}\sigma_{\langle bc\rangle}^x\prod_{\langle ab\rangle,\langle ac\rangle\subset \langle abc \rangle}CZ_{\langle ab\rangle,\langle ac\rangle},
 \end{align}
 and similarly for the other labels $\tau, \mu$, leading to~\eqref{44}.
 The action of $\Bar{S}$ on $\prod_{\langle bc\rangle \subset \langle abc\rangle}CZ_{bc}$ can be obtained as follows. In the $Z$-basis, we have the following representation
\begin{align}
    \prod_{\langle bc\rangle \subset \langle abc\rangle}CZ_{bc} &= (-1)^{b_1c_1 + c_1b_2 + b_2c_2 + c_2b_3 + b_3c_3 + c_3b_1} ,
\end{align}
where $b_i, c_i\ (i=1,2,3)$ denote the spin variables taking values in $\{0,1\}$.
It can be rewritten as
\begin{align}
    &b_1c_1 + c_1b_2 + b_2c_2 + c_2b_3 + b_3c_3 + c_3b_1 \\ \nonumber
     =&(b_1+b_2)(c_1+c_2) + (c_1+c_2)(b_2+b_3) + (b_2+b_3)(c_2+c_3) \\ \nonumber
     &+(c_2+c_3)(b_3+b_1) + (b_3+b_1)(c_3 +c_1) + (c_3 +c_1)(b_1+b_2)\\ \nonumber
   \xrightarrow{\ \Bar{S}\ }\  &\tilde{b}_{12}\tilde{c}_{12} + \tilde{c}_{12}\tilde{b}_{23} + \tilde{b}_{23}\tilde{c}_{23} + \tilde{c}_{23}\tilde{b}_{31} + \tilde{b}_{31}\tilde{c}_{31} + \tilde{c}_{31}\tilde{b}_{12}, 
\end{align}
where $\tilde{b}_{ij}$ is a gauge variable on the link $\langle ac\rangle$, intersected by the line connecting $b_i$ and $b_j$, and similarly for $\tilde{c}_{ij}$.  This transformation is graphically described by Fig.~\ref{Fig:appendixA}.

\section{$2$-group symmetry by gauging non-onsite symmetry}\label{BAP}
In Sec.~\ref{sec.2.1}, we discussed a $2$-group symmetry obtained by gauging an onsite symmetry $U_A$. Here, we show that the same $2$-group symmetry can also be obtained by gauging a non-onsite symmetry~$U_D$. From a field-theoretic perspective, this is expected, as $A$, $B$, $C$, and $D$ enter the anomaly inflow action \eqref{type4 anomaly} on an equal footing. Nevertheless, it is instructive to see explicitly how the same $2$-group structure emerges from the microscopic gauging procedure. 
Since the symmetry structure induced by gauging does not depend on the microscopic details of the system, in the following we do not specify an explicit Hamiltonian and focus only on the symmetry operators.

We first introduce gauge variables $\sigma_{\ell}^x, \sigma_{\ell}^z$ on all links. The Gauss law constraint and the flatness condition are given by
\begin{align}
     G_{D, abc} := CCZ_{abc}\prod_{\ell\subset \langle abc \rangle}\sigma_{\ell}^x &=1, \quad \forall \langle abc\rangle, \\
     B_{D}(\gamma):= \prod_{\ell\subset\gamma}\sigma_{\ell}^z &=1, \quad \forall \gamma.
\end{align}
Here, $\ell\subset \langle abc \rangle$ denotes links contained in the triangle $\langle abc \rangle$, and $\gamma$ denotes a contractible loop on the dual lattice. The other three symmetry operators are not gauge invariant, so they must be modified to restore gauge invariance. Indeed, the gauge invariant symmetry operators are given by
\begin{align}
    \widetilde{U}_A = \prod_{a\in \Lambda_a} X_a\prod_{\langle bc\rangle}CCZ_{bc\langle bc\rangle}, \quad \widetilde{U}_B = \prod_{b\in \Lambda_b} X_b\prod_{\langle ac\rangle}CCZ_{ac\langle ac\rangle}, \quad \widetilde{U}_C = \prod_{c\in \Lambda_c} X_c\prod_{\langle ab\rangle}CCZ_{ab\langle ab\rangle}.
\end{align}
Owing to the flatness condition, one can show that these operators commute. Furthermore, the gauged theory possesses a dual $1$-form symmetry associated with the gauging procedure
\begin{align}
    \eta_D^{(1)}(\gamma) = \prod_{\ell\subset\gamma}\sigma_{\ell}^z,
\end{align}
where $\gamma$ denotes a non-contractible loop on the dual lattice. As in Sec.~\ref{sec.2.1}, the symmetry structure is simply a direct product in the absence of symmetry defect insertions. To reveal the $2$-group symmetry, we consider the insertion of a symmetry defect. In particular, we focus on a symmetry defect of $\widetilde{U}_A$. However, the non-onsite form of $\widetilde{U}_A$ makes the definition of the associated symmetry defect more subtle. To address this issue, we enlarge the Hilbert space by introducing ancilla qubits on the sublattice $\Lambda_b$, which allows the non-onsite symmetry to be represented as an onsite symmetry via a suitable unitary transformation~\cite{Zhang:2024mhl, Seifnashri:2025vhf, Brito:2026eqi}. We denote the Pauli operators action on the ancilla Hilbert space as $X', Z'$. Using ancilla qubits, we redefine the symmetry operator~$\widetilde{U}_A$ as
\begin{align}
    \widetilde{U}_{A, B'} := \widetilde{U}_{A}\widetilde{U}_{B'}, \quad \widetilde{U}_{B'}:=\prod_{b'\in\Lambda_b}X'_{b'} ,
\end{align}
 where $X'_{b'}$ denotes the pauli-$X$ operator on an ancilla qubit, and the prime denotes the label of an ancilla qubit.
 Note that this procedure does not change the symmetry structure. Moreover, we perform the unitary transformation with
\begin{align}
V
= \prod_{\langle bc\rangle} CCCZ_{b'bc\langle bc \rangle}
= \prod_{\langle bc\rangle}
e^{\frac{\pi i}{16}(1-Z'_{b'})(1-Z_b)(1-Z_c)(1-\sigma_{\langle bc \rangle}^z)}.
\end{align}
Here, for each triangle, $CCCZ_{b'bc\langle bc\rangle}$ denotes a four-qubit controlled-$Z$ gate acting on the ancilla qubit $b'$, the two qubits $b$ and $c$, and the gauge qubit living on the link $\langle bc\rangle$.
Under this unitary transformation, the symmetry operators and the Gauss law take the following form:
\begin{align}
\widetilde{U}_{A, B'} &\longrightarrow  \mathsf{U}_{A,B'} \coloneq U_A\widetilde{U}_{B'} =\prod_{a\in\Lambda_a}X_a\prod_{b'\in\Lambda_b}X'_{b'}, \\
    \widetilde{U}_B &\longrightarrow  \mathsf{U}_{B}\coloneq\prod_{b\in \Lambda_b} X_b\prod_{\langle ac\rangle}CCZ_{ac\langle ac\rangle}\prod_{\langle bc \rangle}CCZ_{b'c\langle bc\rangle}, \\
    \widetilde{U}_C &\longrightarrow \mathsf{U}_{C}\coloneq \prod_{c\in \Lambda_c} X_c\prod_{\langle ab\rangle}CCZ_{ab\langle ab\rangle}\prod_{\langle bc 
    \rangle}CCZ_{b'b\langle bc\rangle}. \\
    G_{D,abc} &\longrightarrow 
    \widetilde{G}_{D,abc} \coloneq CCZ_{abc}CCZ_{b'bc}\prod_{\ell\subset \langle abc \rangle}\sigma_{\ell}^x =1
\end{align}
With this reformulation, the construction of Sec.~\ref{sec.2.1} can be applied directly, allowing us to introduce the symmetry defect associated with the $\mathsf{U}_{A,B'}$ symmetry. Here, we choose the symmetry defect configuration shown in Fig.~\ref{app_B_fig_1}. 

\begin{figure}
  \centering
  \begin{subfigure}{0.48\textwidth}
    \centering
    \resizebox{\linewidth}{!}{\usetikzlibrary{backgrounds}
\begin{tikzpicture}[scale=1.4, line cap=round, line join=round]
\def\h{0.8660254}
\def\Nx{6}
\def\Ny{4}
\usetikzlibrary{backgrounds}
\foreach \j in {0,...,\Ny}{
  \foreach \i in {0,...,\Nx}{
    \pgfmathsetmacro{\x}{\i + 0.5*\j}
    \pgfmathsetmacro{\y}{\h*\j}

    \ifnum\i<\Nx
      \draw[black!55, line width=0.4pt]
        (\x,\y) -- ({\x+1},\y);
    \fi

    \ifnum\j<\Ny
      \draw[black!55, line width=0.4pt]
        (\x,\y) -- ({\x+0.5},{\y+\h});

      \ifnum\i>0
        \draw[black!55, line width=0.4pt]
          (\x,\y) -- ({\x-0.5},{\y+\h});
      \fi
    \fi
  }
}

\pgfmathtruncatemacro{\NxInner}{\Nx-1}

\foreach \j in {0,...,\Ny}{
  \foreach \i in {1,...,\NxInner}{
    \pgfmathsetmacro{\x}{\i + 0.5*\j}
    \pgfmathsetmacro{\y}{\h*\j}
    \pgfmathtruncatemacro{\sublattice}{mod(\i+2*\j,3)}

    \ifcase\sublattice
  \def\vertexlabel{b}
\or
  \def\vertexlabel{a}
\or
  \def\vertexlabel{c}
\fi

    \fill[black] (\x,\y) circle (1.5pt);
    \node[
      anchor=south west,
      xshift=3pt,
      yshift=2pt,
      font=\large,
      fill=white,
      inner sep=0.7pt
    ] at (\x,\y) {$\vertexlabel$};
  }
}

\node[above=12pt, red, font=\Large]
  at (4,{4*\h}) {$\widetilde{U}_{B'}$};

\node[above=12pt, blue, font=\Large]
  at (5,{4*\h}) {$U_A$};

\draw[blue, line width=1.5pt]
  (5,0)
  -- (4.5,{\h})
  -- (5,{2*\h})
  -- (4.5,{3*\h})
  -- (5,{4*\h});

\draw[red, line width=1.5pt]
  (4,0)
  -- (3.5,{\h})
  -- (4,{2*\h})
  -- (3.5,{3*\h})
  -- (4,{4*\h});
\end{tikzpicture}}
    \caption{}
    \label{app_B_fig_1}
  \end{subfigure}%
  \hspace{0.04\textwidth}%
  \begin{subfigure}{0.48\textwidth}
    \centering
    \resizebox{\linewidth}{!}{\usetikzlibrary{backgrounds}
\begin{tikzpicture}[scale=1.4, line cap=round, line join=round]
\def\h{0.8660254}
\def\Nx{6}
\def\Ny{4}
\usetikzlibrary{backgrounds}
\foreach \j in {0,...,\Ny}{
  \foreach \i in {0,...,\Nx}{
    \pgfmathsetmacro{\x}{\i + 0.5*\j}
    \pgfmathsetmacro{\y}{\h*\j}

    \ifnum\i<\Nx
      \draw[black!55, line width=0.4pt]
        (\x,\y) -- ({\x+1},\y);
    \fi

    \ifnum\j<\Ny
      \draw[black!55, line width=0.4pt]
        (\x,\y) -- ({\x+0.5},{\y+\h});

      \ifnum\i>0
        \draw[black!55, line width=0.4pt]
          (\x,\y) -- ({\x-0.5},{\y+\h});
      \fi
    \fi
  }
}

\pgfmathtruncatemacro{\NxInner}{\Nx-1}

\foreach \j in {0,...,\Ny}{
  \foreach \i in {1,...,\NxInner}{
    \pgfmathsetmacro{\x}{\i + 0.5*\j}
    \pgfmathsetmacro{\y}{\h*\j}
    \pgfmathtruncatemacro{\sublattice}{mod(\i+2*\j,3)}

    \ifcase\sublattice
  \def\vertexlabel{b}
\or
  \def\vertexlabel{a}
\or
  \def\vertexlabel{c}
\fi

    \fill[black] (\x,\y) circle (1.5pt);
    \node[
      anchor=south west,
      xshift=3pt,
      yshift=2pt,
      font=\large,
      fill=white,
      inner sep=0.7pt
    ] at (\x,\y) {$\vertexlabel$};
  }
}

\node[above=12pt, red, font=\Large]
  at (4,{4*\h}) {$\widetilde{U}_{B'}$};

\node[above=12pt, blue, font=\Large]
  at (5,{4*\h}) {$U_A$};

\begin{scope}[on background layer]
  \fill[gray!20]
    (4,0)
    -- (3.5,{\h})
    -- (4,{2*\h})
    -- (3.5,{3*\h})
    -- (4,{4*\h})
    -- (5,{4*\h})
    -- (4.5,{3*\h})
    -- (5,{2*\h})
    -- (4.5,{\h})
    -- (5,0)
    -- cycle;
\end{scope}

\draw[blue, line width=1.5pt]
  (5,0)
  -- (4.5,{\h})
  -- (5,{2*\h})
  -- (4.5,{3*\h})
  -- (5,{4*\h});

\draw[red, line width=1.5pt]
  (4,0)
  -- (3.5,{\h})
  -- (4,{2*\h})
  -- (3.5,{3*\h})
  -- (4,{4*\h});
\end{tikzpicture}}
    \caption{}
    \label{app_B_fig_2}
  \end{subfigure}

  \caption{%
    (a) A configuration with the $\mathsf{U}_{A,B'}$
    symmetry defect inserted. The red and blue lines represent
    the $\widetilde{U}_{B'}$ and $U_A$ symmetry defects, respectively.
    (b) The Gauss-law constraints are modified on the triangles
    in the gray region between the blue and red lines.
  }
  \label{fig:app_B}
\end{figure}

In this case, the insertion of a symmetry defect associated with $\mathsf{U}_{A,B'}$ modifies the Gauss law in the vicinity of the defect. More explicitly, the Gauss law on the triangles lying between $U_A$ and $\widetilde{U}_B$ symmetry defects is modified as\footnote{We introduce the symmetry defect by applying the symmetry transformation only to the region $R$ to the right of the defect line. Writing the restricted symmetry operator as $U_R$, we define the defect Hamiltonian $H_{\mathrm{defect}}:=U_RHU_R^\dagger$ and conjugate the Gauss-law constraints by the same operator. This implements a symmetry twist across the boundary of $R$.}
\begin{align}
\widetilde{G}_{D,abc}
\longrightarrow
CZ_{bc},\widetilde{G}_{D,abc}.
\end{align}
See Fig.~\ref{app_B_fig_2} for an illustration of this modification.
Accordingly, the symmetry operators must also be modified in the vicinity of the defect so as to retain gauge invariance. The gauge invariant symmetry operators read
\begin{align}
    \mathsf{U}_B &= \prod_{b\in \Lambda_b} X_b\prod_{\langle ac\rangle}CCZ_{ac\langle ac\rangle}\prod_{\langle bc \rangle}CCZ_{b'c\langle bc\rangle} V_{B,\mathrm{defcet}}, \\
    \mathsf{U}_C &= \prod_{c\in \Lambda_c} X_c\prod_{\langle ab\rangle}CCZ_{ab\langle ab\rangle}\prod_{\langle bc \rangle}CCZ_{b'b\langle bc\rangle}V_{C,\mathrm{defcet}}, 
\end{align}
where $V_{B,\mathrm{defect}}$ and $V_{C,\mathrm{defect}}$ are defined explicitly below, and their illustration is shown in Fig.~\ref{app_B_fig_34}:
\begin{align}
    V_{B,\mathrm{defect}} := \prod_{\langle bc \rangle}CZ_{c\langle bc\rangle}\prod_{\langle ac \rangle}CZ_{c\langle ac\rangle}, \quad V_{C,\mathrm{defect}} := \prod_{\langle ab\rangle}CZ_{b\langle ab\rangle}.
\end{align}
\begin{figure}
  \centering
  \begin{subfigure}{0.48\textwidth}
    \centering
    \resizebox{\linewidth}{!}{
\usetikzlibrary{backgrounds}

\begin{tikzpicture}[scale=1.4, line cap=round, line join=round]
\def\h{0.8660254}
\def\Nx{6}
\def\Ny{4}

\definecolor{Vcolor}{RGB}{0,95,55}
\definecolor{ABcolor}{RGB}{135,55,170}

\begin{scope}[on background layer]
  \fill[gray!20]
    (4,0)
    -- (3.5,{\h})
    -- (4,{2*\h})
    -- (3.5,{3*\h})
    -- (4,{4*\h})
    -- (5,{4*\h})
    -- (4.5,{3*\h})
    -- (5,{2*\h})
    -- (4.5,{\h})
    -- (5,0)
    -- cycle;
\end{scope}

\foreach \j in {0,...,\Ny}{
  \foreach \i in {0,...,\Nx}{
    \pgfmathsetmacro{\x}{\i + 0.5*\j}
    \pgfmathsetmacro{\y}{\h*\j}

    \ifnum\i<\Nx
      \draw[black!55, line width=0.4pt]
        (\x,\y) -- ({\x+1},\y);
    \fi

    \ifnum\j<\Ny
      \draw[black!55, line width=0.4pt]
        (\x,\y) -- ({\x+0.5},{\y+\h});

      \ifnum\i>0
        \draw[black!55, line width=0.4pt]
          (\x,\y) -- ({\x-0.5},{\y+\h});
      \fi
    \fi
  }
}

\draw[blue, line width=0.7pt]
  (5,0)
  -- (4.5,{\h})
  -- (5,{2*\h})
  -- (4.5,{3*\h})
  -- (5,{4*\h});

\draw[red, line width=0.7pt]
  (4,0)
  -- (3.5,{\h})
  -- (4,{2*\h})
  -- (3.5,{3*\h})
  -- (4,{4*\h});

\foreach \j in {1,3}{
  \draw[ABcolor, line width=2.5pt]
    (3.5,{\j*\h}) -- (4,{\j*\h});
}


\foreach \j in {0,2,4}{
  \draw[Vcolor, line width=2.5pt]
    (5,{\j*\h}) -- (4.5,{\j*\h});
}

\foreach \j in {1,3}{
  \draw[Vcolor, line width=2.5pt]
    (3.5,{\j*\h}) -- (4,{\j*\h});
}

\pgfmathtruncatemacro{\NxInner}{\Nx-1}

\foreach \j in {0,...,\Ny}{
  \foreach \i in {1,...,\NxInner}{
    \pgfmathsetmacro{\x}{\i + 0.5*\j}
    \pgfmathsetmacro{\y}{\h*\j}
    \pgfmathtruncatemacro{\sublattice}{mod(\i+2*\j,3)}

    \ifcase\sublattice
      \def\vertexlabel{b}
    \or
      \def\vertexlabel{a}
    \or
      \def\vertexlabel{c}
    \fi

    \fill[black] (\x,\y) circle (1.5pt);
    \node[
      anchor=south west,
      xshift=3pt,
      yshift=2pt,
      font=\large,
      fill=white,
      inner sep=0.7pt
    ] at (\x,\y) {$\vertexlabel$};
  }
}

\node[above=12pt, red, font=\Large]
  at (4,{4*\h}) {$\widetilde{U}_{B'}$};

\node[above=12pt, blue, font=\Large]
  at (5,{4*\h}) {$U_A$};

\end{tikzpicture}}
    \caption{}
  \end{subfigure}%
  \hspace{0.04\textwidth}%
  \begin{subfigure}{0.48\textwidth}
    \centering
    \resizebox{\linewidth}{!}{
\usetikzlibrary{backgrounds}

\begin{tikzpicture}[scale=1.4, line cap=round, line join=round]
\def\h{0.8660254}
\def\Nx{6}
\def\Ny{4}

\definecolor{Vcolor}{RGB}{0,95,55}
\definecolor{ABcolor}{RGB}{135,55,170}

\begin{scope}[on background layer]
  \fill[gray!20]
    (4,0)
    -- (3.5,{\h})
    -- (4,{2*\h})
    -- (3.5,{3*\h})
    -- (4,{4*\h})
    -- (5,{4*\h})
    -- (4.5,{3*\h})
    -- (5,{2*\h})
    -- (4.5,{\h})
    -- (5,0)
    -- cycle;
\end{scope}

\foreach \j in {0,...,\Ny}{
  \foreach \i in {0,...,\Nx}{
    \pgfmathsetmacro{\x}{\i + 0.5*\j}
    \pgfmathsetmacro{\y}{\h*\j}

    \ifnum\i<\Nx
      \draw[black!55, line width=0.4pt]
        (\x,\y) -- ({\x+1},\y);
    \fi

    \ifnum\j<\Ny
      \draw[black!55, line width=0.4pt]
        (\x,\y) -- ({\x+0.5},{\y+\h});

      \ifnum\i>0
        \draw[black!55, line width=0.4pt]
          (\x,\y) -- ({\x-0.5},{\y+\h});
      \fi
    \fi
  }
}

\draw[blue, line width=0.7pt]
  (5,0)
  -- (4.5,{\h})
  -- (5,{2*\h})
  -- (4.5,{3*\h})
  -- (5,{4*\h});

\draw[red, line width=0.7pt]
  (4,0)
  -- (3.5,{\h})
  -- (4,{2*\h})
  -- (3.5,{3*\h})
  -- (4,{4*\h});

\foreach \j in {1,3}{
  \draw[ABcolor, line width=2.5pt]
    (4.25,{(\j+0.5)*\h}) -- (4.5,{\j*\h});

  \draw[ABcolor, line width=2.5pt]
    (4.25,{(\j-0.5)*\h}) -- (4.5,{\j*\h});
}

\pgfmathtruncatemacro{\NxInner}{\Nx-1}

\foreach \j in {0,...,\Ny}{
  \foreach \i in {1,...,\NxInner}{
    \pgfmathsetmacro{\x}{\i + 0.5*\j}
    \pgfmathsetmacro{\y}{\h*\j}
    \pgfmathtruncatemacro{\sublattice}{mod(\i+2*\j,3)}

    \ifcase\sublattice
      \def\vertexlabel{b}
    \or
      \def\vertexlabel{a}
    \or
      \def\vertexlabel{c}
    \fi

    \fill[black] (\x,\y) circle (1.5pt);
    \node[
      anchor=south west,
      xshift=3pt,
      yshift=2pt,
      font=\large,
      fill=white,
      inner sep=0.7pt
    ] at (\x,\y) {$\vertexlabel$};
  }
}

\node[above=12pt, red, font=\Large]
  at (4,{4*\h}) {$\widetilde{U}_{B'}$};

\node[above=12pt, blue, font=\Large]
  at (5,{4*\h}) {$U_A$};

\end{tikzpicture}}
    \caption{}
  \end{subfigure}

  \caption{%
    Action of $V_{B, \mathrm{defect}}$ and $V_{C, \mathrm{defect}}$. The green and purple lines represent $V_{B, \mathrm{defect}}$ and $V_{C, \mathrm{defect}}$, respectively. Each line denotes a $CZ$ operator.
  }
  \label{app_B_fig_34}
\end{figure}
From the same argument of Sec.~\ref{sec.2.1}, around the $\mathsf{U}_{A, B'}$ defect, $\mathsf{U}_B$ and $\mathsf{U}_C$ effectively acts as
\begin{align}
    \mathsf{U}_B|_{\mathrm{defect}} = \prod_{b\in \Lambda_b} X_b
    \prod_{\langle ac\rangle}CZ_{c\langle ac\rangle} 
    \prod_{\langle bc \rangle}CZ_{c\langle bc\rangle}
    V_{B,\mathrm{defect}}, \\
    \mathsf{U}_C|_{\mathrm{defect}} = \prod_{c\in \Lambda_c} X_c
    \prod_{\langle ab\rangle}CZ_{b\langle ab\rangle} 
    \prod_{\langle bc \rangle}CZ_{b\langle bc\rangle}V_{C,\mathrm{defect}}.
\end{align}
From these commutation relations, one can verify the emergence of a $1$-form symmetry localized along the defect, leading to the same projective algebra as~\eqref{p}:
\begin{align}
    \mathsf{U}_B\mathsf{U}_C = \eta_D^{(1)}(\gamma)\mathsf{U}_C\mathsf{U}_B, 
\end{align}
 where $\gamma$ represents the non-contractible loop along the defect. 
This confirms that gauging either an onsite or a non-onsite symmetry in a system with a type-IV anomaly consistently gives rise to a $2$-group symmetry.

\section{Invariance of partition function} \label{invariance}
In this appendix, we derive the invariance of the gauged partition functions, described by \eqref{invariance_1} and \eqref{invariance_2}. 
\subsection{Derivation of \eqref{invariance_1}}
We begin by defining the action of higher gauging on the theory $\mathcal{T}_2$. When restricted to $M_2$, the theory $\mathcal{T}_2$ can be regarded as the theory $\chi$, obtained by gauging the two $0$-form symmetries on $M_2$. Its partition function is given by
\begin{equation}
    Z_{\chi}[\hat{A}^{(1)},\hat{B}^{(1)}] =\# \sum_{\hat{a}^{(1)}, \hat{b}^{(1)}}Z_{\chi_0}[\hat{a}^{(1)}, \hat{b}^{(1)}](-1)^{\int_{M_2}\hat{a}^{(1)}\cup \hat{B}^{(1)} + \hat{b}^{(1)}\cup \hat{A}^{(1)} }, 
\end{equation}
where $\chi_0$ denotes the pregauged theory and $\#$ denotes a normalization factor depending on the topology of the spacetime manifold. We define the topological manipulations $S$ and $T$ on $M_2$ as
\begin{align}
    S:\ &Z_{\chi}[\hat{A}^{(1)},\hat{B}^{(1)}] \rightarrow \sum_{\hat{a}^{(1)}, \hat{b}^{(1)}} Z_{\chi}[\hat{a}^{(1)},\hat{b}^{(1)}](-1)^{\int_{M_2}\hat{a}^{(1)}\cup\hat{B}^{(1)} + \hat{b}^{(1)}\cup\hat{A}^{(1)}} \\
    T:\ &Z_{\chi}[\hat{A}^{(1)},\hat{B}^{(1)}]
    \rightarrow Z_{\chi}[\hat{A}^{(1)},\hat{B}^{(1)}](-1)^{\int_{M_2}\hat{A}^{(1)}\cup\hat{B}^{(1)}}.
\end{align}
Using these operations, we have
\begin{equation}
\begin{aligned}
  &\quad Z_{TST\chi}[\hat{A}^{(1)},\hat{B}^{(1)}] \\
  &=\sum_{\hat{a}^{(1)}, \hat{b}^{(1)}}\sum_{a^{(1)}, b^{(1)}} Z_{\chi_0}[a^{(1)}, b^{(1)}](-1)^{\int_{M_2}a^{(1)}\cup \hat{b}^{(1)} + b^{(1)}\cup \hat{a}^{(1)} + \hat{a}^{(1)}\cup \hat{b}^{(1)} + \hat{a}^{(1)}\cup \hat{B}^{(1)} + \hat{b}^{(1)}\cup \hat{A}^{(1)} + \hat{A}^{(1)}\cup \hat{B}^{(1)}} \\
  &=\sum_{a^{(1)}, b^{(1)}} Z_{\chi_0}[a^{(1)}, b^{(1)}](-1)^{\int_{M_2}a^{(1)}\cup (b^{(1)}+\hat{B}^{(1)}) + (b^{(1)}+\hat{B}^{(1)})\cup \hat{A}^{(1)} + \hat{A}^{(1)}\cup \hat{B}^{(1)}} \\
  &=\sum_{a^{(1)}, b^{(1)}} Z_{\chi_0}[a^{(1)}, b^{(1)}](-1)^{\int_{M_2}a^{(1)}\cup b^{(1)}+a^{(1)}\cup\hat{B}^{(1)} + b^{(1)}\cup \hat{A}^{(1)}}.
\end{aligned}
\end{equation}
In the second equality, we perform the sums over $\hat{a}^{(1)}$ and $\hat{b}^{(1)}$. This implies that the additional term $\int_{M_2}a^{(1)}\cup b^{(1)}$ cancels the variation induced by the action of $g$ on $\mathcal{T}_2$ \eqref{pt2}.
\subsection{Derivation of \eqref{invariance_2}}
From \eqref{S}, \eqref{T} and \eqref{bar_S}, we have
\begin{equation}
\begin{aligned}
    Z_{g\Bar{S}TS\mathcal{T}_3}[\hat{A}^{(2)},\hat{B}^{(2)}, \hat{C}^{(2)}]  
    &= Z_{g\Bar{S}T\mathcal{T}}[\hat{A}^{(2)},\hat{B}^{(2)}, \hat{C}^{(2)}] \\
    &=\sum_{a,b,c} Z_{\mathcal{T}}[a,b,c] (-1)^{\int_{M_3}a\cup b\cup c + a\cup \hat{A}^{(2)} + b \cup \hat{B}^{(2)} + c\cup \hat{C}^{(2)} + a\cup b\cup c} \\
    &=\sum_{a,b,c} Z_{\mathcal{T}}[a,b,c] (-1)^{\int_{M_3} a\cup \hat{A}^{(2)} + b \cup \hat{B}^{(2)} + c\cup \hat{C}^{(2)}} \\
    &=Z_{\mathcal{T}_3}[\hat{A}^{(2)},\hat{B}^{(2)},\hat{C}^{(2)}].
\end{aligned}
\end{equation}
In the first equality, we use the relations $\mathcal{T}_3=\Bar{S}\mathcal{T}$ and $S\Bar{S}=1$. 


\end{appendix}

\bibliography{main}
\bibliographystyle{utphys}
\end{document}